\documentclass[%
 reprint,
 superscriptaddress,
preprintnumbers,
nofootinbib,
 amsmath,amssymb,
 aps,
]{revtex4-1}

\usepackage{tikz}
\usepackage{graphicx}
\usepackage{dcolumn}
\usepackage{bm}
\usepackage{hyperref}
\usepackage{color}
\usepackage{multirow, array}
\hypersetup{
    colorlinks = true
}
\usepackage{aas_macros}

\newcommand{\beq}{\begin{equation}}
\newcommand{\eeq}{\end{equation}}
\newcommand{\beqa}{\begin{eqnarray}}
\newcommand{\eeqa}{\end{eqnarray}}
\newcommand{\be}{\begin{equation}}
\newcommand{\ee}{\end{equation}}
\newcommand{\bea}{\begin{eqnarray}}
\newcommand{\eea}{\end{eqnarray}}
\newcommand{\nn}{\nonumber}

\newcommand{\vpt}{\mbox{\sf VPT}}

\begin{document}

\preprint{TUM-HEP 1557/25}

\title{Vlasov Perturbation Theory and the role of higher cumulants}

\author{Mathias Garny}
\email{mathias.garny@tum.de}
\affiliation{Physik Department T31, Technische Universit\"at M\"unchen, James-Franck-Stra\ss{}e 1, D-85748 Garching, Germany
}%

\author{Rom\'an Scoccimarro}
\email{rs123@nyu.edu}
\affiliation{
 Center for Cosmology and Particle Physics, Department of Physics, New York University, NY 10003, New York, USA
}%

\date{\today}

\begin{abstract} 

We develop a new approach to Vlasov  Perturbation Theory (\vpt{}) that solves for the hierarchy of cumulants of the phase-space distribution function to arbitrarily high truncation order in the context of cosmological structure formation driven by collisionless dark matter. We investigate the impact of higher cumulants on density and velocity power spectra as
well as the bispectrum, and compare to scale-free $N$-body simulations. While there is a strong difference between truncation at the first cumulant, {\it i.e.} standard perturbation theory (SPT), and truncation at the second ({\it i.e.} including the velocity dispersion tensor), the third cumulant has a small quantitative impact and fourth and higher cumulants only have  a minor effect on these summary statistics at weakly non-linear scales. We show that spurious exponential growth is absent in vector and tensor modes if scalar-mode constraints on the non-Gaussianity of the background distribution function that results from shell-crossing are satisfied, guaranteeing the screening of UV modes for all fluctuations of any type, as expected physically. We also show analytically that loop corrections to the power spectrum are finite within \vpt{} for any initial power spectra consistent with hierarchical clustering, unlike SPT. Finally, we discuss the relation to and contrast our predictions with effective field theory (EFT), and discuss how the advantages of \vpt{} and EFT approaches could be combined.

\end{abstract}

\maketitle

\tableofcontents

\section{Introduction}
\label{sec:introduction}

Within the standard paradigm of cold dark matter (CDM), structure formation on large scales is
governed by collisionless dynamics of non-relativistic particles and their gravitational interaction.
On sub-horizon scales, this setup can be described by the collisionless Boltzmann (or Vlasov) equation
for the phase-space distribution function (DF) $f(\tau, \bm x, \bm p)$ together with the Poisson equation for the gravitational potential.
The Vlasov-Poisson equations can be equivalently mapped to a coupled hierarchy for the cumulants of the DF in momentum space, related to the moments $\int d^3p\, p_{i_1}p_{i_2}\cdots p_{i_c}\, f(\tau, \bm x, \bm p)$~\cite{Pueblas:2008uv}.

The lowest-order truncation taking only the zeroth and first cumulant into account, {\it i.e.}~the density and bulk velocity fields,  yields
the pressureless perfect fluid approximation that underlies the framework of Standard Perturbation Theory (SPT), see \emph{e.g.}~\cite{Bernardeau:2001qr}. This approximation is in general
not expected to apply for collisionless dynamics. Within the CDM paradigm of hierarchical structure formation, the rationale leading to
this approach is that the second and higher cumulants are initially small, and the DF is expected to be close to its initial shape on sufficiently large length scales.
Nevertheless, at the non-linear level, perturbation modes on different length scales affect each other.
In particular, small-scale modes for which the ideal fluid picture breaks down may affect larger modes due to non-linear mode coupling, including generating a non-trivial background for the DF.
This leads to well-known limitations of SPT, such as a spurious dependence on highly non-linear scales, see \emph{e.g.}~\cite{BlaGarKon1309}. This is in contradiction to the decoupling of those scales expected from analytical arguments~\cite{peebles1980large} and known from numerical simulations~\cite{NisBerTar1611,NisBerTar1712}.

The advent of Stage-IV surveys such as DESI~\cite{DESI:2016fyo} and Euclid~\cite{Euclid:2024yrr} necessitates the development of an improved framework that goes beyond the limitations of SPT.
Current analyses often employ an effective field theory (EFT) description obtained from complementing SPT with correction terms. The latter depend on free parameters that
encapsulate the most general impact that any conceivable dynamics on small scales could have on the evolution of the density and velocity fields on larger scales~\cite{BauNicSen1207}.
The advantage of this approach is that it applies irrespectively of the actual dynamics on small scales. A disadvantage is the reduction in predictivity, related to the need to fit the 
EFT parameters alongside the cosmological model parameters to observational data. This becomes particularly acute when going to higher orders in the
perturbative expansion, for which more and more EFT parameters need to be added.

The focus of this work is perturbation theory based on collisionless dynamics, instead of an (effective) fluid description. The motivation for this program is avoiding a proliferation of unknown free parameters and overcoming the spurious UV sensitivity of SPT within a first principle approach. 
The framework of Vlasov Perturbation Theory (\vpt{})~\cite{cumPT,cumPT2} takes into account the dynamical evolution of second and higher cumulants, following conceptually from the underlying Vlasov-Poisson equations in a straightforward fashion. 
A novel feature (compared to SPT) is that the velocity dispersion tensor, {\it i.e.}~the second cumulant, has a non-zero average value, which can be described by a time-dependent function and introduces a new (dispersion) scale $k_\sigma(z)$. Its value is sensitive to non-linear scales and the physics of gravitational collapse, halo formation and virialization. 
For given $k_\sigma(z)$, the evolution equations for the fluctuations of the cumulants around the average can be solved perturbatively on large scales. The outcome is that the well-known kernels describing the non-linear evolution of the density field in SPT are replaced by time-dependent kernels $F_n(\bm k_1,\dots,\bm k_n,\eta)$ within \vpt{} that do capture the decoupling of UV modes, \emph{i.e.} are suppressed if one or several wavenumber arguments exceed $k_\sigma(z)$, and thereby overcome a major limitation of SPT. The dispersion scale can itself either be treated as a (single) free parameter, or estimated from halo models or simulations~\cite{cumPT,cumPT2}. 
The resulting power spectra have been shown to be in good agreement with $N$-body simulation results for density and velocity fields as well as for the bispectrum, and capture the generation of vorticity on large scales~\cite{cumPT2}.

A pivotal question within \vpt{} is the truncation of the coupled hierarchy of evolution equations derived from the Vlasov-Poisson equations.
Since going from the first-order truncation (\emph{i.e.} the SPT approximation) to the second-order truncation (\emph{i.e.} including velocity dispersion) does
lead to major qualitative differences, it is a fair question to ask whether third and higher cumulants are relevant. {\em This point is addressed systematically
in this work.} For that purpose, we develop a formalism that allow us to solve the cumulant hierarchy up to, in principle, arbitrarily high truncation order.
We investigate the impact of higher cumulants on the density and velocity fields. Considering their auto- and cross power spectra, the bispectrum as well as the vorticity power spectrum, we quantify the dependence of these summary statistics on
the truncation order of the cumulant hierarchy. For illustration, we focus on cosmologies with scale-free initial power spectra, for which we also compare to $N$-body results following~\cite{cumPT2}. Nevertheless, we expect our main findings to carry
over also to cosmologies featuring an initial power spectrum as predicted in the $\Lambda$CDM model.

This work extends the \vpt{} framework from~\cite{cumPT,cumPT2}, where the second cumulant was fully included, but the third and fourth only partially at non-linear level, while a systematic study of the truncation to arbitrarily high cumulant order was limited to linear theory. Previous works on effects from Vlasov-Poisson dynamics in the context of
perturbation theory for structure formation include~\cite{Pueblas:2008uv}, where the coupled hierarchy of cumulants was discussed
and the generation of velocity dispersion in this context was investigated. The impact of an average velocity dispersion was considered
in an expansion for low wavenumber in~\cite{McD1104}, and in a formulation in Lagrangian coordiantes in~\cite{Aviles_2016,McDVla1801}. A perturbative description involving velocity dispersion, but no higher cumulants, was investigated in~\cite{ErsFlo1906}. Vorticity generation was discussed in this context in~\cite{Pueblas:2008uv,Cusin:2016zvu,cumPT2,Erschfeld:2023aqr}.
A discussion of the coupled hierarchy of cumulants and its relation to the Schrödinger-Poisson method can be found in~\cite{Uhlemann:2018olp}, while~\cite{Nascimento:2024hle} considers perturbation theory using the DF instead of the cumulant hierarchy.

\medskip

This work is structured as follows: after a review of \vpt{} in Sec.~\ref{sec:review}, we introduce the formulation applicable to arbitrarily high truncation order in Sec.~\ref{sec:formalism}.
The impact of the truncation on the non-linear kernels of the density field as well as on power spectra and the bispectrum is analyzed in Sec.~\ref{sec:kernels} and Sec.~\ref{sec:impact}, respectively.
Sec.~\ref{sec:eft} contains a discussion of the relation to the EFT approach, and Sec.~\ref{sec:NLdisp} a general analysis of the scales on which velocity dispersion is expected to become important.
The appendices contain further technical details as well as results for non-linear kernels of the velocity divergence.

In Sec.~\ref{sec:conclusions} we present our conclusions, which are written in a form that can guide the reader through the different results in the paper, and can therefore be read first for those interested in particular aspects of this work.

\section{Review of \vpt}
\label{sec:review}

In this section we briefly review the formalism of \vpt{} introduced in~\cite{cumPT,cumPT2}.
The phase-space distribution function $f(\tau,\bm x,\bm p)$ of the matter density
evolves under gravity as described by the collisionless Boltzmann equation, also known as Vlasov equation,
\be
  0 = \frac{\partial f}{\partial \tau}+\frac{p_i}{a} \frac{\partial f}{\partial x_i}-a(\nabla_i\Phi) \frac{\partial f}{\partial p_i}\,,
\ee
complemented by the Poisson equation $\nabla^2\Phi=\frac{3}{2}{\cal H}^2\Omega_m\delta$.
Here $\tau$ and ${\cal H}=\partial \ln a/\partial\ln\tau$ are conformal time and Hubble rate, respectively, with $a$ the scale factor, $\Omega_m$ is the time-dependent matter density parameter, 
and $\delta$ and $\Phi$ are the density contrast and gravitational potential. The phase-space distribution function can equivalently be characterized by its cumulants in momentum space.
The latter can be obtained from the cumulant generating function (CGF)
\be
  e^{{\cal C}(\tau,\bm x,\bm l)} \equiv \int d^3p\, e^{\bm l\cdot\bm p/a} f(\tau,\bm x,\bm p)\,,
\ee
with the $n$th cumulant obtained by taking the $n$th gradient with respect to the auxiliary vector $\bm l$,
\be
  {\cal C}_{i_1i_2\cdots i_n}(\tau,\bm x) \equiv \nabla_{l_{i_1}}\nabla_{l_{i_2}}\cdots\nabla_{l_{i_n}}{\cal C}(\tau,\bm x,\bm l)\big|_{\bm l=0}\,.
\ee
The zeroth cumulant 
\be
  A\equiv {\cal C}|_{\bm l=0}=\ln(1+\delta)\,,
\ee
is related to the density contrast $\delta$, the first to the peculiar
velocity $v_i$, and the second to the velocity dispersion tensor $\sigma_{ij}$,
\be
  v_i\equiv {\cal C}_i,\quad \sigma_{ij}\equiv {\cal C}_{ij}\,.
\ee
Multiplying the Vlasov equation by $e^{\bm l\cdot\bm p/a}$ and integrating over momentum yields an equation of motion for the CGF~\cite{Pueblas:2008uv},
\be
  \partial_\tau{\cal C}+{\cal H}(\bm l\cdot\nabla){\cal C}+(\nabla{\cal C})\cdot(\nabla_l{\cal C})+(\nabla\cdot\nabla_l){\cal C}=-{\bm l}\cdot\nabla\Phi\,,
\ee
from which the hierarchy of coupled equations for the cumulants can be derived by taking $n$ gradients with respect to ${\bm l}$, and then setting $\bm l=0$.
The zeroth and first gradients yield the familiar continuity and Euler equations,
\bea\label{eq:continuityandEuler}
  \partial_\tau\delta+\nabla_i[(1+\delta)v_i] &=&0\,,\\
  \partial_\tau v_i+{\cal H}v_i+v_j\nabla_jv_i+\nabla_i\Phi &=& -\nabla_j\sigma_{ij}-\sigma_{ij}\nabla_j\ln(1+\delta)\,.\nn
\eea
The canonical SPT approximation corresponds to setting the right-hand side of the Euler equation to zero.
Within \vpt{}, this assumption is lifted, and second and higher cumulants are taken into account as additional
variables. For example, the equations of motion of the second and third cumulant read
\bea
  \partial_\tau \sigma_{ij} &+& 2{\cal H}\sigma_{ij} +v_k\nabla_k\sigma_{ij}+\sigma_{jk}\nabla_kv_i+\sigma_{ik}\nabla_kv_j\nn\\
  &=& -\nabla_k{\cal C}_{ijk}-{\cal C}_{ijk}\nabla_k\ln(1+\delta)\,,\nn\\
  \partial_\tau {\cal C}_{ijk} &+& 3{\cal H}{\cal C}_{ijk} +v_m\nabla_m{\cal C}_{ijk}\nn\\
  &+&\sigma_{km}\nabla_m\sigma_{ij} +\sigma_{im}\nabla_m\sigma_{kj}+\sigma_{jm}\nabla_m\sigma_{ik}\nn\\
  &+&{\cal C}_{jkm}\nabla_mv_i+{\cal C}_{kim}\nabla_mv_j+{\cal C}_{ijm}\nabla_mv_k\nn\\
  &=& -\nabla_m{\cal C}_{ijkm}-{\cal C}_{ijkm}\nabla_m\ln(1+\delta)\,.
\eea
The equations of motion feature the well-known dependence of the evolution of the $n$th cumulant on the one at order $n+1$,
requiring a truncation of the coupled hierarchy of equations. In this work we study the impact of truncation systematically, and develop
a formalism that allow us to describe the cumulant hierarchy to, in principle, arbitrarily high order in the cumulant expansion at the
non-linear level. Before this, we review findings and properties from previous works~\cite{cumPT,cumPT2} that completely (partially) included perturbations
up to cumulant order two (four).

In the following it is convenient to work with the rescaled cumulants
\be
u_i\equiv\frac{v_i}{-f{\cal H}},\ \epsilon_{ij}\equiv\frac{\sigma_{ij}}{(-f{\cal H})^2},\ 
\pi_{ijk}\equiv\frac{{\cal C}_{ijk}}{(-f{\cal H})^3},
\ee
and so on, where $f=d\ln D/d\ln a$ is the linear growth rate and $D(a)$ the growth factor.
In general we denote the rescaled $n$th cumulant by
\be
  \widetilde{\cal C}_{i_1i_2\cdots i_n}(\tau,\bm x) \equiv \frac{{\cal C}_{i_1i_2\cdots i_n}(\tau,\bm x)}{(-f{\cal H})^n}\,,
\ee
with $u_i=\widetilde{\cal C}_i$, $\epsilon_{ij}=\widetilde{\cal C}_{ij}$ and $\pi_{ijk}=\widetilde{\cal C}_{ijk}$.
They can be obtained from the rescaled generating function
\be\label{eq:Ctilde}
  e^{\widetilde{\cal C}(\eta,\bm x,\bm L)} \equiv \int d^3p\ e^{-\bm L\cdot\bm p/(af{\cal H})} f(\tau,\bm x,\bm p)\,,
\ee
by taking gradients with respect to the rescaled auxiliary vector ${\bm L}$, {\it i.e.}
\be\label{eq:Cexpansion}
  \widetilde{\cal C} = \ln(1+\delta)+L_iu_i+\frac12 L_iL_j\epsilon_{ij}+\frac16 L_iL_jL_k\pi_{ijk}+\dots\,.
\ee
 Using also $\eta\equiv\ln D$ instead
of $\tau$ as time variable, its equation of motion reads
\bea\label{eq:eomC}
  \partial_\eta\widetilde{\cal C}+\left(\frac32\frac{\Omega_m}{f^2}-1\right)(\bm L\cdot\nabla)\widetilde{\cal C}&&\nn\\
  -(\nabla\widetilde{\cal C})\cdot(\nabla_L\widetilde{\cal C})-(\nabla\cdot\nabla_L)\widetilde{\cal C}&=&{\bm L}\cdot\nabla\widetilde\Phi\,,
\eea
where $\widetilde\Phi\equiv \Phi/(-f{\cal H})^2$ is the rescaled gravitational potential.

The second cumulant can possess an average value consistent with statistical isotropy and
homogeneity,
\be
  \langle\epsilon_{ij}(\eta,\bm x)\rangle=\delta_{ij}^K\, \epsilon(\eta)\,,
\ee
where $\delta_{ij}^K$ is the Kronecker symbol.
The average of the second cumulant can also be related to the trace of the dispersion tensor,
\be
  \epsilon(\eta) \equiv {\cal E}_2(\eta) = \frac13 \langle \epsilon_{ii}(\eta,\bm x)\rangle\,.
\ee
It plays an important role in \vpt{},
setting a new physical ``dispersion'' scale
\be
  k_\sigma(\eta) \equiv 1/\sqrt{\epsilon(\eta)}\,.
\ee
Even if matter is initially (almost) perfectly cold, it is expected that non-zero dispersion
is generated non-linearly via orbit crossing~\cite{Pueblas:2008uv}.
In~\cite{cumPT2} various estimates of this scale were presented, based on both simulation and
perturbation theory results within \vpt{} and for various cosmologies, generally finding a value comparable to but slightly larger than
the non-linear scale. This new scale enters the evolution of perturbations within~\vpt{}, and is a central input for capturing the
decoupling of small-scale modes~\cite{cumPT,cumPT2}.

Apart from the second cumulant, also higher cumulants can have a non-zero average value.
They can be described by considering the average of the CGF itself,
\be
  {\cal E}(\eta, L^2) \equiv \langle \widetilde{\cal C}(\eta,\bm x,\bm L) \rangle\,.
\ee
Assuming statistical isotropy and homogeneity implies that the average is independent of $\bm x$
and can depend only on $L^2\equiv{\bm L}^2$. Due to analyticity, ${\cal E}$ can be Taylor expanded in
powers of $L^2$,
\be\label{eq:Eexpansion}
  {\cal E}(\eta, L^2) = \sum_{s\geq 0} \frac{L^{2s}}{(2s)!} {\cal E}_{2s}(\eta)\,,
\ee
where ${\cal E}_{2s}(\eta)$ is related to the average of the cumulant of order $2s$,
\be\label{eq:Eaverage}
  {\cal E}_{2s}(\eta) = \frac1{2s+1} \langle \widetilde{\cal C}_{i_1i_1i_2i_2\cdots i_si_s}(\eta,\bm x)\rangle\,.
\ee
Note that only even cumulants can possess an average due to the assumption of statistical isotropy.
At zeroth order ${\cal A}\equiv {\cal E}_0=\langle\ln(1+\delta)\rangle$, which however does not impact
the evolution of $\delta$ and higher cumulants~\cite{cumPT} and is therefore irrelevant in the following.
In contrast, the average of the second cumulant ${\cal E}_2(\eta) \equiv \epsilon(\eta) \equiv 1/k_\sigma(\eta)^2$
is of crucial importantce within \vpt{} as discussed above.
The average values of higher cumulants can be characterized by the dimensionless ratios
\be\label{eq:E2sbar}
  \bar{\cal E}_{2s}(\eta) \equiv {\cal E}_{2s}(\eta)/{\cal E}_2(\eta)^s = {\cal E}_{2s}(\eta)/\epsilon(\eta)^s\,,
\ee
and are generically expected to be of order unity~\cite{cumPT,cumPT2}.
For the fourth cumulant expectation value, we also introduce the notation $\omega\equiv 5{\cal E}_4/3$ and $\bar\omega\equiv\omega/\epsilon^2=5\bar {\cal E}_4/3$
to make contact with~\cite{cumPT,cumPT2}.

An equation of motion for the averaged cumulants can be obtained by taking the expectation value of Eq.\,\eqref{eq:eomC}
and integrating over the direction of $\bm L$,
\be\label{eq:eomE}
  \left[\partial_\eta+\left(\frac32\frac{\Omega_m}{f^2}-1\right)\frac{\partial}{\partial\ln L}\right]{\cal E} = Q_{\cal E}\,,
\ee
with non-linear source term
\be
  Q_{\cal E}\equiv \int\frac{d\Omega_L}{4\pi} \langle (\nabla\widetilde{\cal C})\cdot(\nabla_L\widetilde{\cal C}) \rangle \,.
\ee
Equations of motion for the ${\cal E}_{2s}$ follow by inserting Eq.\,\eqref{eq:Cexpansion} and Taylor expanding in powers of $L^2$~\cite{cumPT}.

The fluctuations of the cumulants around the average can be obtained from the difference of the CGF and its expectation value,
\be
  \delta\widetilde{\cal C}(\eta,\bm x,\bm L) \equiv \widetilde{\cal C}(\eta,\bm x,\bm L) - {\cal E}(\eta,L^2)\,.
\ee
For example, the fluctuation part of the second cumulant $\delta\epsilon_{ij}\equiv\delta\widetilde{\cal C}_{ij}$
obtained from taking two gradients with respect to $\bm L$ reads
\be
  \delta\epsilon_{ij}(\eta,\bm x)=\epsilon_{ij}(\eta,\bm x)-\delta_{ij}^K\epsilon(\eta)\,,
\ee
{\it i.e.} is given by the difference of the dispersion tensor and its average $\langle\epsilon_{ij}\rangle=\delta_{ij}^K\epsilon(\eta)$.
The equation of motion of $\delta\widetilde{\cal C}$ follows from the difference of Eq.\,\eqref{eq:eomC} and Eq.\,\eqref{eq:eomE},
\bea\label{eq:eomdC}
  \Bigg[\partial_\eta &+& \left(\frac32\frac{\Omega_m}{f^2}-1\right)(\bm L\cdot\nabla) -2\frac{\partial{\cal E}}{\partial L^2}{\bm L}\cdot\nabla 
  -(\nabla\cdot\nabla_L) \Bigg]\delta\widetilde{\cal C} \nn\\
  &=& (\nabla\delta\widetilde{\cal C})\cdot(\nabla_L\delta\widetilde{\cal C}) + {\bm L}\cdot\nabla\widetilde\Phi -Q_{\cal E}\,.
\eea
Inserting the expansions from Eq.\,\eqref{eq:Cexpansion} and Eq.\,\eqref{eq:Eexpansion} in terms of the auxiliary vector $\bm L$
yields equations of motion for the fluctuations of the cumulants. Note that they depend on the average values via the
contribution involving $\partial{\cal E}/\partial L^2=\frac12\epsilon+\frac{1}{12}{\cal E}_4+\dots$.
Thus, the evolution of fluctuations and background values are in general coupled with each other.

It is convenient to decompose the cumulants into contributions with different transformation
properties under rotations. The first cumulant,~{\it i.e.} peculiar velocity, can be decomposed into its scalar part, being the
velocity divergence $\theta=\nabla_iu_i$, and its vector part, given by the vorticity $w_i=\varepsilon_{ijk}\nabla_ju_k$,
\be\label{eq:velocitydecomp}
  u_i=u_i^S+u_i^V=\frac{\nabla_i}{\nabla^2}\theta-\frac{\varepsilon_{ijk}\nabla_j}{\nabla^2}w_k\,.
\ee
Here $\varepsilon_{ijk}$ denotes the Levi-Civita symbol.
In Fourier space the scalar and vector part corresponds to
the components of $\bm u$ parallel and perpendicular to the wavevector $\bm k$, respectively,
and the inverse Laplace corresponds to a factor $-1/k^2$. The (fluctuation part of the) second cumulant, {\it i.e.}~the velocity dispersion tensor, can be decomposed into scalar, vector and tensor
parts $\delta\epsilon_{ij}=\delta\epsilon_{ij}^S+\delta\epsilon_{ij}^V+\delta\epsilon_{ij}^T$, with
\bea\label{eq:epsdecomp}
  \delta\epsilon_{ij}^S &=& \delta_{ij}^K\delta\epsilon+\frac{\nabla_i\nabla_j}{\nabla^2}g\,,\nn\\
  \delta\epsilon_{ij}^V &=& -\frac{\varepsilon_{ilk}\nabla_l\nabla_j}{\nabla^2}\nu_k-\frac{\varepsilon_{jlk}\nabla_l\nabla_i}{\nabla^2}\nu_k\,,\nn\\
  \delta\epsilon_{ij}^T &=& t_{ij} \,,
\eea
with two scalar modes denoted by $\delta\epsilon$ and $g$, vector mode $\nu_i$, and tensor mode $t_{ij}$ (see~\cite{cumPT}).
The scalar mode proportional to the unit matrix is called $\delta\epsilon$ to discriminate it from the average value $\epsilon$.
Both tensor and vector parts are traceless. The tensor contribution is in addition transverse to the wavevector with respect to
both of its indices in Fourier space. Each of the two terms contributing to the vector part are transverse with respect to
one of the two indices, and parallel with respect to the other one. In addition, all contributions are symmetric under exchange
of $i$ and $j$ as required. This yields in total two scalar, two vector and two tensor degrees of freedom out of the six
independent components of $\delta\epsilon_{ij}$.

In~\cite{cumPT,cumPT2}, explicit equations of motion for the scalar modes $\delta,\theta,\delta\epsilon,g$, the vector modes $w_i,\nu_i$ and tensor modes $t_{ij}$ have been
worked out, including all non-linear terms resulting from the underlying Vlasov-Poisson equations. In addition, non-linear equations for the scalar modes of the third cumulant perturbations
have been derived, that depend on the background value ${\cal E}_4$ of the fourth cumulant. Furthermore, within the linear approximation, the complete hierarchy of coupled equations
for the scalar perturbation modes has been studied in~\cite{cumPT}, finding convergence when increasing the truncation order provided the background values satisfy certain stability conditions.

In this work, a systematic investigation of the impact of the complete set of higher-order cumulants
at non-linear level is performed, using an implementation that includes all modes and non-linear interactions among them
up to, in principle, arbitrarily high cumulant order.

\section{Formulation of \vpt{} including higher rank tensors}
\label{sec:formalism}

In this section we work out the coupled hierarchy of evolution equations for the cumulants of the matter
phase-space distribution function as obtained from the  Vlasov-Poisson equations, including non-linear terms.
For this task it is important to find a suitable decomposition of the rank-$n$ tensor structure of the $n$th cumulant.
This can be conveniently  organized in terms of irreducible representations of the rotation group by expanding the cumulant generating function in terms
of spherical harmonics, discussed in Sec.~\ref{sec:expansionYlm}. Then we work out the equations of motion in Sec.~\ref{sec:eom}, and discuss
the structure of the non-linear mode-coupling terms for the full hierarchy of cumulants.

\subsection{Expansion in spherical harmonics}\label{sec:expansionYlm}

Our starting point is the generating function $\widetilde{\cal C}(\eta,\bm x,\bm L)=\ln(1+\delta)+L_iu_i+\frac12 L_iL_j\epsilon_{ij}+\dots$ for (rescaled) cumulants of the phase-space disctribution function
as introduced in Eq.~\eqref{eq:Ctilde}, or more precisely the fluctuation part $\delta\widetilde{\cal C}=\widetilde{\cal C}-{\cal E}$ from which the average ${\cal E}=\langle\widetilde{\cal C}\rangle$ has
been subtracted. 
The fluctuations of the $n$th order cumulants follow from Taylor expanding in the auxiliary vector $\bm L$,
\be\label{eq:Ctildeexpansion}
  \delta\widetilde{\cal C}(\eta,\bm x,\bm L) = \sum_{n\geq 0} \frac{1}{n!}L_{i_1}L_{i_2}\cdots L_{i_n}\,\delta\widetilde{\cal C}_{i_1i_2\cdots i_n}(\eta,\bm x)\,,
\ee
with {\it e.g.}~$\delta\widetilde{\cal C}_{i}=u_i$ and $\delta\widetilde{\cal C}_{ij}=\delta\epsilon_{ij}=\epsilon_{ij}-\langle\epsilon_{ij}\rangle$ being the peculiar velocity and fluctuation of the dispersion tensor
for $n=1$ and $n=2$, respectively. The $n=0$ contribution is given by the difference of the log-density field and its average.
The fluctuation of the $n$th cumulant
\be
  \delta\widetilde{\cal C}_{i_1i_2\cdots i_n}=\widetilde{\cal C}_{i_1i_2\cdots i_n}-\langle\widetilde{\cal C}_{i_1i_2\cdots i_n}\rangle\,,
\ee
is given by a tensor of rank $n$ that is totally symmetric under index permutations.
The number of independent components is therefore $(n+1)(n+2)/2$.
In the following, it is convenient to work in Fourier space,
\be
  \delta\widetilde{\cal C}(\eta,\bm x,\bm L) = \int d^3k\, e^{i\bm k\cdot\bm x}\, \delta\widetilde{\cal C}(\eta,\bm k,\bm L)\,,
\ee
with analogous definition of $\delta\widetilde{\cal C}_{i_1i_2\cdots i_n}(\eta,\bm k)$. 

A decomposition of the $n$th cumulant in irreducible representations of the rotation group
can be obtained by separating the dependence of the CGF on the auxiliary vector $\bm L$ into
a dependence on its direction and its magnitude, as given by the unit vector $\hat L\equiv \bm L/L$
and Euclidean norm $L\equiv|\bm L|$. The direction of the unit vector can be parameterized by two angles $\vartheta$ and $\varphi$
as usual. Since the CGF itself is a scalar quantity, its dependence on these angles can be characterized by a multipole expansion.
An analogous decomposition of moments (instead of cumulants) of the distribution function has been used in~\cite{Seljak:2011tx,Okumura:2011pb,Okumura:2012xh,Vlah:2012ni,Vlah:2013lia}
in the context of redshift-space distortions. A decomposition of second or higher rank tensors in a spherical harmonic basis has also been considered in~\cite{Vlah:2019byq,Vlah:2020ovg,Matsubara:2022ohx,Bakx:2023mld} related to tensorial biased tracers such as galaxy shapes. In a different context, spherical basis expansion methods were proposed for three-point function
estimation in~\cite{Slepian:2015qza}.

\begin{figure*}[t]
  \begin{center}
  \includegraphics[width=0.42\textwidth]{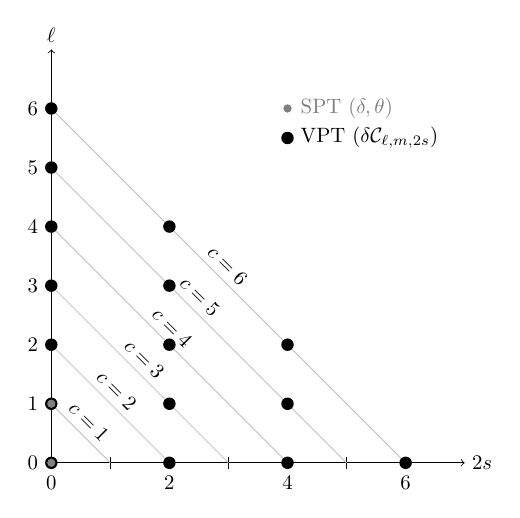}
  \includegraphics[width=0.48\textwidth]{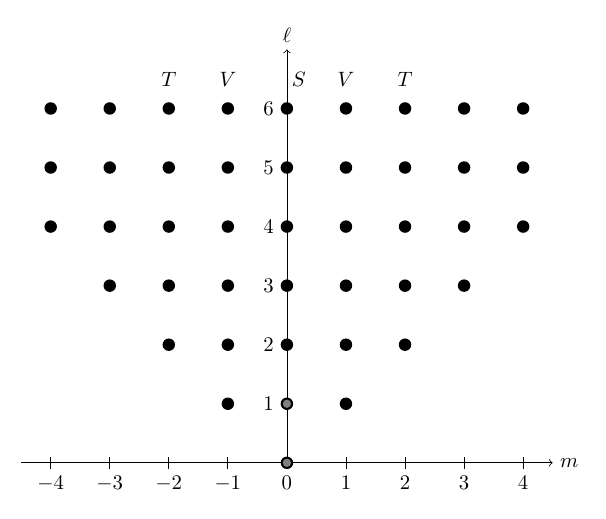}
  \end{center}
  \caption{\label{fig:Clms}
  Illustration of cumulant modes $\delta C_{\ell m ,2s}(\eta,{\bm k})$ with indices $\ell=0,1,2,\dots$, $m=0,\pm1,\pm2,\dots,\pm\ell$ and $2s=0,2,4,\dots$.
  The cumulant order is $c=\ell+2s$. The left figure illustrates the modes of cumulant order $c=0,1,2,\dots,6$ in the $\ell-2s-$plane.
  In the commonly adopted ideal fluid approximation only the modes $\delta C_{000}$ related to the density contrast $\delta$ and $\delta C_{100}=\theta/3$ are taken into account.
  In \vpt{} all modes up to a certain truncation order $\ell+2s\leq c_{\text{max}}$ can be included. The right figure shows the $\ell-m$ plane.
  Perturbations with $m=0$ are scalar modes, with $m=\pm 1$ vector modes, and with $m=\pm2$ tensor modes, and so on. The truncations we consider are
  characterized by $c\leq c_\text{max}$ and $|m|\leq m_\text{max}$, and we obtain numerical results going up to at most $c_\text{max}=6$ and $m_\text{max}=4$.
  }
\end{figure*}

Using some general properties of the decomposition of higher rank tensors in spherical harmonics from~\cite{Matsubara:2022ohx}, we observe that due to analyticity of the CGF in $\bm L$ contributions from the $\ell$-th
multipole involve at least a factor $L^\ell$, and a coefficient that can depend apart from $\eta$ and ${\bm k}$ also on a power series in $L^2$. We therefore make the ansatz
\bea\label{eq:Ylmdecomp}
  \delta\widetilde{\cal C}(\eta,{\bm k},{\bm L}) &=& \sum_{\ell m} \frac{(-iL)^\ell}{k^{a_\ell}} \sqrt{4\pi(2\ell+1)} \, Y_{\ell m}^{\bm k}(\hat L) \nn\\
  && {} \times \delta C_{\ell m}(\eta,{\bm k},L^2)\,.
\eea
Here $Y_{\ell m}^{\bm k}(\hat L)$ are spherical harmonics in a suitable (in general $\bm k$-dependent) basis, see below, and $\delta C_{\ell m}(\eta,{\bm k},L^2)$ are the expansion coefficients that parameterize the
cumulants of the distribution function.\footnote{In our numerical implementation we use a slightly different expansion in terms of real-valued spherical harmonics, see Appendix~\ref{app:real} for details.} Furthermore $a_\ell=0$ for even $\ell$ and $a_\ell=1$ for odd $\ell$. Normalization factors are chosen for later convenience.
If the wavevector $\bm k$ points in $z$-direction, $Y_{\ell m}^{\bm k}(\hat L)\equiv Y_{\ell m}(\hat L)$ are the standard spherical harmonics. Otherwise, they are defined to be ``rotated'' spherical harmonics that coincide with the usual ones in a rotated basis with respect to which $\bm k$ is aligned with the third coordinate axis. In other words, the polar and azimuthal angles are measured with respect to the $\bm k$ vector, see Section~\ref{NLvert} below for more details.

Using
\be\label{eq:Ylmortho}
  \int d^2\hat L\, Y^{\bm k}_{\ell m}(\hat L)\left(Y^{\bm k}_{\ell'm'}(\hat L)\right)^*=\delta_{\ell\ell'}^K\delta_{mm'}^K\,,
\ee
(see below) we can invert the expansion in spherical harmonics and obtain
\bea\label{eq:Ylmdecompinverse}
  \delta C_{\ell m}(\eta,{\bm k},L^2) &=& \frac{1}{\sqrt{4\pi(2\ell+1)}}\frac{k^{a_\ell}}{(-iL)^\ell}\nn\\
  && {} \times \int d^2\hat L \, Y^{{\bm k}*}_{\ell m}(\hat L) \delta\widetilde{\cal C}(\eta,{\bm k},{\bm L})\,.
\eea
Here $d^2\hat L\equiv d\varphi \ d\cos\vartheta$ is the usual measure for angular integration.

To obtain the strict expansion in cumulants, we still have to Taylor expand the $L^2$ dependence,
\be\label{eq:Clms}
  \delta C_{\ell m}(\eta,{\bm k},L^2) = \sum_{s\geq 0} \frac{L^{2s}}{(2s)!}\delta C_{\ell m, 2s}(\eta,{\bm k})\,.
\ee
The $\delta C_{\ell m ,2s}(\eta,{\bm k})$ parameterize the cumulants of the distribution function, contributing at cumulant order
\be
  c=\ell+2s\,.
\ee
For example for $c=0$, there is only a single contribution $\ell=m=s=0$ related to the log-density field,
\be
  \delta C_{000} = \ln(1+\delta) - \langle\ln(1+\delta)\rangle\,.
\ee
For $c>0$, the relation between the spherical harmonic and cartesian bases for cumulants is discussed in Sec.\,\ref{sec:kart} in detail.
Here we provide some examples and general features.
For $c=1$, only $\ell=1,s=0$ is possible, with three contributions $m=0,\pm 1$. The case $m=0$ corresponds to the contribution parallel to the wavevector $\bm k$, being proportional to the scalar part of the velocity field, captured by the
velocity divergence $\theta$. Using the $\ell=1,m=0$ spherical harmonic one finds
\be\label{eq:dC100}
  \delta C_{100} = \frac{i}3 k_iu_i = \frac13 \theta\,.
\ee
The components $\delta C_{1m0}$ with $m=\pm 1$ correspond to fluctuations perpendicular to the wavevector, which are related to the two independent degrees of freedom of the vorticity field $w_i$.
Thus, the scalar part of the peculiar velocity field is associated to $m=0$, and the vector part to $m=\pm 1$.
With our choice of expanding in ``rotated'' spherical harmonics, this association carries on to higher $\ell$ and $m$ (for an analogous discussion for moments instead of cumulants, see~\cite{Seljak:2011tx}):
\begin{itemize}
\item {\it scalar} modes correspond to $m=0$ modes, 
\item {\it vector} modes to $m=\pm 1$, for all $\ell\geq 1$,
\item {\it tensor} modes to $m=\pm 2$, for all $\ell\geq 2$,
\end{itemize}
and so on. {The various modes are illustrated in Fig.~\ref{fig:Clms}.
 
For the second cumulant $c=2$ there are two possibilities, $\ell=2,2s=0$ and $\ell=0,2s=2$.
The latter contributes one scalar mode with $m=0$, and the former one scalar, two vector, and two tensor modes, corresponding to $m=0,\pm 1,\pm 2$. Together this yields precisely the six degrees of freedom of the symmetric
dispersion tensor $\delta\epsilon_{ij}$, with a total of  $2+2+2$ scalar, vector and tensor modes, in accordance with~\cite{cumPT} and Sec.~\ref{sec:review}. The scalar modes are related to the modes $g$ and $\delta\epsilon$
from Eq.~\eqref{eq:epsdecomp},
\bea\label{eq:dC002dC200}
  \delta C_{002} &=& \frac13(\epsilon_{ii}-\langle\epsilon_{ii}\rangle)=\frac{1}{3}g+\delta\epsilon\,,\nn\\
  \delta C_{200} &=& -\frac{1}{10}\left(\frac{k_ik_j}{k^2}-\frac13\delta_{ij}^K\right)(\epsilon_{ij}-\langle\epsilon_{ij}\rangle) = -\frac{1}{15}g\,.\nn\\
\eea
The components $\delta C_{2,\pm1,0}$ are related to the two independent degrees of freedom of the dispersion vector mode $\nu_i$, and $\delta C_{2,\pm2,0}$ to the tensor modes $t_{ij}$.
 
The third cumulant $c=3$ has contributions from $\ell=1,m=0,\pm1,2s=2$ and $\ell=3,m=0,\pm1,\pm2,\pm3,2s=0$, respectively, making up $2+4+2+2$ scalar, vector, tensor and $|m|=3$ modes, respectively, that add up
to the $10$ independent components of the third cumulant $\delta\widetilde{\cal C}_{ijk}$.
In general, for even (odd) cumulant order $c$ there are contributions with $\ell=0,2,\dots,c$ ($\ell=1,3,\dots,c$), and correspondingly $2s=c-\ell$.
Summing up the $2\ell+1$ contributions for each possible $\ell$, one finds $(c+1)(c+2)/2$ contributions independently of whether $c$ is even or odd.
This precisely matches the number of independent components of the symmetric tensor $\delta\widetilde{\cal C}_{i_1i_2\cdots i_c}$ with $c$ indices.
Thus, the set of cumulant variables $\delta C_{\ell m ,2s}(\eta,{\bm k})$ for all $\ell\geq 0, 2s\geq 0, m=-\ell,\dots,+\ell$ is equivalent to the 
set of all $\delta\widetilde{\cal C}_{i_1i_2\cdots i_c}$ with $c\geq 0$, and the number of degrees of freedom at cumulant order $c$ matches between
both representations (see also~\cite{Seljak:2011tx}).

Each of the $\delta C_{\ell m ,2s}(\eta,{\bm k})$ eventually has in addition a perturbative expansion in powers of the initial density contrast $\delta_{\bm k0}$ in Fourier space,\footnote{In the notation of~\cite{cumPT}, see Eq.~(138) therein, the scalar perturbations
in linear approximation were studied, denoted by ${\cal C}_{\ell,2s}$. They are related to the kernels with $n=1$ and $m=0$ via $\delta C_{1,\ell 0,2s}={\cal C}_{\ell,2s}$.}
\bea\label{eq:cumulantkernels}
  \delta C_{\ell m,2 s}(\eta,{\bm k}) &=& \sum_{n\geq 1} \int_{{\bm k}_1\dots{\bm k}_n}\delta C^{(n)}_{\ell m ,2s}({\bm k_1},\dots,\bm k_n,\eta)\nn\\
  && {} \times e^{n\eta}\delta_{\bm k_10}\cdots\delta_{\bm k_n0}\,,
\eea
where $\int_{{\bm k}_1\dots{\bm k}_n} \equiv \int d^3k_1\cdots d^3k_n\, \delta^D(\bm k-\sum\bm k_i)$. The coefficients $\delta C^{(n)}_{\ell m ,2s}({\bm k_1},\dots,\bm k_n,\eta)$
correspond to the perturbation theory kernels for the cumulant characterized by indices $\ell,m,2s$ at order $n$ in perturbation theory, and order $c=\ell+2s$ in the cumulant expansion.
We can also consider the perturbative expansion before the $L^2$ expansion, with an analogous definition of kernels $\delta C^{(n)}_{\ell m }({\bm k_1},\dots,\bm k_n,\eta,L^2)=\sum_s L^{2s}\delta C^{(n)}_{\ell m ,2s}/(2s)!$.

The next step is to derive equations that allow us to compute the perturbative kernels.

\subsection{Equation of motion}\label{sec:eom}

\subsubsection{General structure}

To arrive at the hierarchy of evolution equations for the cumulants in the spherical harmonic basis introduced above,
in a first step, we derive an equation of motion for the coefficient functions $\delta C_{\ell m}(\eta,{\bm k},L^2)$ in the decomposition
of the CGF, see Eq.~\eqref{eq:Ylmdecomp}. Using the equation of motion Eq.~\eqref{eq:eomdC} of the CGF we can project out an equation for a given $\ell m$ by multiplying with the conjugated
spherical hamonic function and averaging over all angles of the auxiliary vector $\bm L$ that the CGF depends on, as in Eq.~\eqref{eq:Ylmdecompinverse}.
After some algebra (see Appendix~\ref{app:relations}), this yields
\bea\label{eq:Clm_raw_eq_of_motion}
   &&  \left[ \partial_\eta+\left(\frac32\frac{\Omega_m}{f^2}-1\right)(\ell+2L^2\partial_{L^2})\right] \delta C_{\ell m}(\eta,{\bm k},L^2)\nn\\
   && {} + \frac{\{1,k^2\}}{2\ell+1}\Bigg[ 2\sqrt{\ell^2-m^2}\left(\frac{\partial{\cal E}}{\partial L^2}+\partial_{L^2}\right)\delta C_{\ell-1, m}(\eta,{\bm k},L^2)\nn\\
   && {} - \sqrt{(\ell+1)^2-m^2}\nn\\
   && {} \times \left(2L^2\frac{\partial{\cal E}}{\partial L^2}+2\ell+3+2L^2\partial_{L^2}\right)\delta C_{\ell+1, m}(\eta,{\bm k},L^2)\Bigg]\nn\\
   &=& -\frac{k^2}{3}\widetilde\Phi(\eta,{\bm k}) \delta^K_{\ell 1}\delta^K_{m 0}\nn\\
   && {} +\frac{1}{\sqrt{4\pi(2\ell+1)}}\frac{k^{a_\ell}}{(-iL)^\ell}\int d^2\hat L \, Y^{{\bm k}*}_{\ell m}(\hat L)  \nn\\
   && {} \qquad \times \int_{\bm p\bm q} [i{\bm p}\delta\widetilde{\cal C}(\eta,{\bm p},{\bm L})\cdot\nabla_L\delta\widetilde{\cal C}(\eta,{\bm q},{\bm L})]\,.
\eea
Here $\{1,k^2\}=1$ for even $\ell$ and $\{1,k^2\}=k^2$ for odd $\ell$. Before working out the structure of this equation more explicitly, and performing the
Taylor expansion in powers of $L^2$ to arrive at equations for the individual cumulants, let us comment on some general features:
\begin{itemize}
\item The terms on the left-hand side originate from the terms in the equation of motion Eq.~\eqref{eq:eomdC} that are linear in the CGF,
and imply that the time-derivative $\partial_\eta\delta C_{\ell m}$ (here represented by its $\eta$-derivative) depends on $\delta C_{\ell m}$, $\delta C_{\ell-1, m}$
and $\delta C_{\ell+1, m}$. In this respect, the structure is analogous to that of the well-known coupled hierarchy of photon or neutrino perturbations at linear
level~\cite{Ma:1995ey}. 
\item The last term
on the right-hand side arises from the non-linear term in the equation of motion Eq.~\eqref{eq:eomdC} for the CGF. In Eq.~\eqref{eq:Clm_raw_eq_of_motion} this
term has not been worked out explicitly yet, which requires to insert the decompositions from Eq.~\eqref{eq:Ylmdecomp} for both $\delta\widetilde{\cal C}(\eta,{\bm p},{\bm L})$
and $\delta\widetilde{\cal C}(\eta,{\bm q},{\bm L})$, see below. This term captures the mode coupling of two perturbation modes of wavevector ${\bm p}$ and ${\bm q}$ into a single mode with
wavevector ${\bm k=\bm p+\bm q}$, {\it i.e.} has a structure that is
in principle analogous to the usual non-linear terms in SPT. However, it gives rise to many more distinct vertex functions that describe the non-linear interactions of arbitrary cumulants.
\item The right-hand side of Eq.~\eqref{eq:Clm_raw_eq_of_motion} contains one term arising from the gravitational potential $\widetilde\Phi$, that captures the impact of gravity and
contributes only to a single equation for $\ell=1$ and $m=0$, {\it i.e.} eventually in the equation for the velocity divergence $\theta$, analogously as in SPT. 
\item The terms on the left-hand side of Eq.~\eqref{eq:Clm_raw_eq_of_motion} do not mix contributions with different values of $m$, only those with different $\ell$. This is consistent with the decoupling of scalar, vector, tensor,$\dots$ modes in the {\it linear} approximation. For example, scalar modes with $m=0$ evolve independently from vector modes ($m=\pm 1$) and tensor modes ($m=\pm 2$) when restricting to the {\it linear} level, even when allowing for contributions of any cumulant order. In contrast, as we will see, the non-linear term on the right-hand side {\it does} lead to cross-talk among cumulants of different $m$ values, and for example in particular captures the generation of vector modes (such as vorticity) from two scalar modes, in line with~\cite{Pueblas:2008uv,cumPT2}.
\item The evolution of the perturbation modes $\delta C_{\ell m}$ does depend on the average background values of the cumulants, that are encoded in the Taylor coefficients with respect to powers of $L^2$ of the function ${\cal E}(\eta, L^2)$. This dependence captures the suppression of small-scale modes relative to the linear growth obtained within SPT. Moreover, as observed in~\cite{cumPT},
the background values of higher cumulants have to satisfy certain conditions to ensure that the system of equations for the perturbations is stable, {\it i.e.} has no exponentially growing solutions in
linear approximation. We comment on how these conditions generalize from the analysis of the coupled hierarchy of cumulants for scalar perturbations studied in~\cite{cumPT} to the general case below.
\end{itemize}

From the structure of Eq.~\eqref{eq:Clm_raw_eq_of_motion} one finds that the coupled hierarchy of equations of motions for the set of cumulants $\delta C_{\ell m, 2s}$
can be brought into the standard form
\bea\label{eq:eommatrixform}
  \partial_\eta\psi_a(\eta,{\bm k})&+&\Omega_{ab}(\eta,k)\psi_b(\eta,{\bm k})\nn\\
  && =\int_{\bm p\bm q} \gamma_{abc}(\bm p,\bm q)\psi_b(\eta,{\bm p})\psi_c(\eta,{\bm q})\,,
\eea
with a generalized ``linear evolution matrix'' $\Omega_{ab}$ and ``vertices'' $\gamma_{abc}(\bm p,\bm q)$ that describe the non-linear mode coupling.
The perturbation modes of the cumulants are all collected into the generalized vector $\psi_a(\eta,{\bm k})$, with a multi-index $a$ that collectively captures all
the individual cumulant modes. In practice we can take $a$ to run over all possible combinations of indices 
\bea
 a &\in& \{(\ell,m,2s),\ \ \ell=0,1,2,\dots, m=0,\pm1,\dots,\pm\ell,\nn\\
 && \qquad 2s=0,2,4,\dots\}\,,
\eea
in order to label all cumulant perturbation modes $\delta C_{\ell m, 2s}$. 
Moreover, due to the appearance of the gravitational potential in Eq.~\eqref{eq:Clm_raw_eq_of_motion}, it is
convenient to include also the density contrast $\delta$ itself as a further variable contained in $\psi_a$. This means we can directly use the Poisson equation $-k^2\widetilde\Phi(\eta,{\bm k})=\frac32\frac{\Omega_m}{f^2}\delta(\eta,{\bm k})$ to express the gravitational potential in terms of $\delta$ itself, and avoid having to expand the log-density field at any point. This strategy has already been implemented in~\cite{cumPT2}.
Altogether, the generalized vector of perturbation modes is given by
\bea\label{eq:psi}
  \psi_a &=& \big(\delta(\eta,{\bm k}),\delta C_{000}(\eta,{\bm k}),\delta C_{1m0}(\eta,{\bm k})|_{m=0,\pm1},\nn\\
  && {} \delta C_{002}(\eta,{\bm k}),\delta C_{2m0}(\eta,{\bm k})|_{m=0,\pm1,\pm2},\nn\\
  && {} \delta C_{1m2}(\eta,{\bm k})|_{m=0,\pm1},\delta C_{3m0}(\eta,{\bm k})|_{m=0,\pm1,\pm2,\pm3},\nn\\
  && {} \delta C_{004}(\eta,{\bm k}),\delta C_{2m2}(\eta,{\bm k})|_{m=0,\pm1,\pm2},\nn\\
  && {} \hspace*{1cm} \delta C_{4m0}(\eta,{\bm k})|_{m=0,\pm1,\pm2,\pm3,\pm4},\dots\big)\,,\nn\\
\eea
where the first line contains apart from $\delta$ the zeroth and first cumulants related to the perturbations of the log-density field $\ln(1+\delta)$ and the peculiar velocity $u_i$, respectively, the second line the second cumulant perturbations related to $\delta\epsilon, g,\nu_i,t_{ij}$, the third line the third cumulant modes, and the last two lines those at fourth cumulant order.
The vector can be systematically extended by perturbation modes $\delta C_{\ell m,2s}$ at higher cumulant orders $c=\ell+2s$ in a straightforward way.

\subsubsection{Linear part}

The explitict form of the matrix $\Omega_{ab}$ can be read off from the Taylor expansion of Eq.~\eqref{eq:Clm_raw_eq_of_motion} in powers of $L^2$
after inserting the $L^2$-expansion Eq.~\eqref{eq:Clms}. For $a=(\ell,m,2s)$ and $b=(\ell',m',2s')$ we find a potentially non-zero result only
if $m=m'$ and $\ell'\in\{\ell-1,\ell,\ell+1\}$,
\be\label{eq:Omegaab}
  \Omega_{ab}=\left(\delta^K_{\ell\ell'}A_{\ell m}^{ss'}+\delta^K_{\ell-1,\ell'}B_{\ell m}^{ss'}+\delta^K_{\ell+1,\ell'}C_{\ell m}^{ss'}\right)\delta^K_{mm'}\,,
\ee
with
\bea\label{eq:ABC}
  A_{\ell m}^{ss'} &=& \left(\frac32\frac{\Omega_m}{f^2}-1\right)(\ell+2s)\delta^K_{ss'}\,,\nn\\
  B_{\ell m}^{ss'} &=& \frac{\{1,k^2\}}{2\ell+1}\sqrt{\ell^2-m^2}\bigg( \frac{\delta^K_{s+1,s'}}{(2s+1)} \nn\\
   && {} +\frac{(2s)!\,{\cal E}_{2s-2s'+2}(\eta)}{(2(s-s')+1)!(2s')!}\Theta_{s-s'} \bigg)\,,\nn\\
  C_{\ell m}^{ss'} &=& -\frac{\{1,k^2\}}{2\ell+1}\sqrt{(\ell+1)^2-m^2}\bigg( \left(2\ell+3+2s\right)\delta^K_{ss'} \nn\\
   && {}  + \frac{(2s)!\,{\cal E}_{2s-2s'}(\eta)}{(2(s-s')-1)!(2s')!}\Theta_{s-1-s'} \bigg)\,,
\eea
where as before $\delta^K$ is the Kronecker symbol, $\{1,k^2\}=1 (k^2)$ for even (odd) $\ell$ and furthermore $\Theta_r=1$ for $r\geq 0$ and zero otherwise. Due to our convention of
including the density contrast $\delta$ as an extra variable, we also need to specify $\Omega_{ab}$ when either one or both
of the indices represent the $\delta$-component of $\psi_a$. The gravitational force term in Eq.~\eqref{eq:Clm_raw_eq_of_motion}
yields
\be
  \Omega_{a\delta} = \left\{\begin{array}{ll}
    -\frac12\frac{\Omega_m}{f^2} & a=(\ell,m,2s)=(1,0,0)\,,\\
    0 & \mbox{else}\,.
    \end{array}\right.
\ee
Furthermore the continuity equation implies $\Omega_{\delta b}=\Omega_{ab}|_{a=(\ell,m,2s)=(0,0,0)}$ and $\Omega_{\delta\delta}=0$.
For scalar modes with $m=m'=0$, we checked that $\Omega_{ab}$ precisely coincides with the linear evolution equations derived in~\cite{cumPT},
specifically Eqs.\,(G1, G2).\footnote{The only difference is the ``+1'' in the square bracket in (G1) of~\cite{cumPT}, which is due to a
relative factor $e^\eta$ in the definition of ${\cal C}_{\ell,2s}$ used in~\cite{cumPT} compared to $\delta C_{\ell 0,2s}$ used here. Note that the prefactors of the terms
involving ${\cal C}_{\ell\pm 1}$ in (G2) of~\cite{cumPT} correspond to $-C_{\ell 0}^{ss'}$ and $-B_{\ell 0}^{ss'}$, respectively.}

Note that the conventional SPT case is contained as a subset of the modes $\psi_a$, specifically the $\delta$-component as well
as $\delta C_{100}=\theta/3$, with {\it e.g.} $\Omega_{100,100}=3\Omega_m/(2f^2)-1$ and $\Omega_{\delta,100}=\Omega_{000,100}=-3$.
The $\Omega_{ab}$ in general depend on the time-dependent background values ${\cal E}_{2}(\eta)\equiv\epsilon(\eta)$, ${\cal E}_4(\eta), {\cal E}_6(\eta),\dots$
of the cumulants of order $2,4,6,\dots$. For example, $\Omega_{100,000}=\frac13 k^2\epsilon(\eta)$ describes a correction to the Euler equation due to the average
background dispersion, and $\Omega_{100,200}=-\frac{10}{3} k^2$ as well as $\Omega_{100,002}=\frac{1}{3} k^2$ capture the impact of scalar
perturbations of the velocity dispersion on the evolution of the velocity divergence. Thus perturbations of different cumulant orders mix with each other under
time-evolution. In contrast, the fact that {\it e.g.} scalar and vector modes can only couple at non-linear level is represented by the property that the ``linear evolution matrix''
$\Omega_{ab}$ is proportional to $\delta^K_{mm'}$, {\it i.e.} is block-diagonal with respect to the sets of perturbation modes with distinct $m$-number.

\subsubsection{Non-linear vertices}
\label{NLvert}

Let us now turn to the vertices $\gamma_{abc}(\bm p,\bm q)$. They originate from the last term in the equation of motion  Eq.~\eqref{eq:Clm_raw_eq_of_motion}.
The vertices for individual components $a=(\ell,m,2s), b=(\ell_1,m_1,2s_1)$ and $c=(\ell_2,m_2,2s_2)$ can be extracted by inserting the decompositions Eq.~\eqref{eq:Ylmdecomp} into each of the two factors of the CGF appearing in the last term of  Eq.~\eqref{eq:Clm_raw_eq_of_motion}, performing the angular integral $\int d^2\hat L$ with respect to the direction of the auxiliary vector $\bm L$,
and then Taylor expanding in powers of $L^2$ using Eq.~\eqref{eq:Clms}. The pivotal step in this derivation is the angular integral over a product of three generalized spherical harmonic functions.
Since each of them is defined with respect to a basis related to a different wavevector, we cannot simply use the standard Gaunt integral. Instead, at this point, we need to specify the generalized $Y_{\ell m}^{\bm k}$ more precisely. We want to define them such that they are given by the usual $Y_{\ell m}$ with respect to a basis in which $\bm k$ points in the third coordiante direction. In particular, we introduce an orthogonal matrix $R^{\bm k}$ for each wavevector $\bm k$, that describes a change of our three-dimensional cartesian coordinate system spanned by unit vectors $\bm e_i$ ($i=x,y,z$) to a primed basis $\bm e_i'=R^{\bm k}_{ij}\bm e_j$. We require $R^{\bm k}$ to be chosen such that, within the primed basis, the vector $\bm k$ is aligned with the third coordinate axis, {\it i.e.} $\bm k=k\bm e_z'$. Using the transformation $k_i'=k_jR^{\bm k}_{ji}$ of the components of $\bm k=k_i\bm e_i=k_i'\bm e_i'$, this requires that the third column of $R^{\bm k}$ is given by the unit vector $\hat k=\bm k/k$, while the first and second columns, denoted by $R^{\bm k}_1$ and $R^{\bm k}_2$, span a basis in the plane orthogonal to $\bm k$,
\be\label{eq:rotationmatrix}
  R^{\bm k} = (R^{\bm k}_1, R^{\bm k}_2, \hat k)\,.
\ee
Under coordinate rotations the spherical harmonics transform as $Y_{\ell m}(\vartheta',\varphi')=Y_{\ell m'}(\vartheta,\varphi)D^{m'}_{\ell m}(R^{\bm k})$, where $D^{m'}_{\ell m}$ are the Wigner rotation matrices and $\vartheta',\varphi'$ the angles measured with respect to the primed basis. We therefore define
\be\label{eq:Ylmk}
  Y_{\ell m}^{\bm k}(\vartheta,\varphi) \equiv Y_{\ell m'}(\vartheta,\varphi)D^{m'}_{\ell m}(R^{\bm k})\,,
\ee
such that $Y_{\ell m}^{\bm k}(\vartheta,\varphi)=Y_{\ell m}(\vartheta',\varphi')$ when expressed with respect to the angles in the rotated basis.
As above, for brevity, we use the shorthand $Y_{\ell m}^{\bm k}(\hat L)=Y_{\ell m}^{\bm k}(\vartheta,\varphi)$ to denote the dependence on the direction of the auxiliary vector $\hat L=\bm L/L$.

The generalized spherical harmonics can be understood as modified spherical harmonic functions that would correspond to the standard ones in a basis relative to which $\bm k$ points along the third coordinate direction.
Due to rotational covariance, the $Y_{\ell m}^{\bm k}$ inherit all properties that can be written in covariant form from the $Y_{\ell m}$.
For example, using $D^{m'}_{\ell m}(R^{-1})=D^{m*}_{\ell m'}(R)$ we see that the orthogonality relation Eq.\,\eqref{eq:Ylmortho} for the $Y_{\ell m}^{\bm k}$
follows directly from the one of the $Y_{\ell m}$. Furthermore, the inverse relation is
\be\label{eq:Ylmkinv}
  Y_{\ell m}(\hat L)= Y_{\ell m'}^{\bm k}(\hat L)D^{m'}_{\ell m}((R^{\bm k})^{-1})\,.
\ee
The property $Y_{\ell m}^*(\hat L)=(-1)^mY_{\ell, -m}(\hat L)$ implies $D^{m*}_{\ell m'}(R)=(-1)^{m-m'}D^{-m}_{\ell, -m'}(R)$, which in turn
means that also 
\be
  Y_{\ell m}^{\bm k*}(\hat L)=(-1)^mY_{\ell, -m}^{\bm k}(\hat L)\,.
\ee
Due to rotational invariance the Gaunt integral for a product of three generalized spherical harmonics all featuring an \emph{identical} $\bm k$-vector
\bea\label{eq:Gaunt}
  \lefteqn{ \int d^2\hat L\ Y^{\bm k}_{\ell m}(\hat L)^*\, Y^{\bm k}_{\ell_1 m_1}(\hat L)\, Y^{\bm k}_{\ell_2 m_2}(\hat L) }\nn\\
  &=& \sqrt{\frac{(2\ell+1)(2\ell_1+1)(2\ell_2+1)}{4\pi}}\left(\begin{array}{ccc}\ell&\ell_1&\ell_2\\ 0&0&0\end{array}\right)\nn\\
 && {} \times \left(\begin{array}{ccc}\ell&\ell_1&\ell_2\\ -m&m_1&m_2\end{array}\right)(-1)^m\,,
\eea
is independent of ${\bm k}$. This can be shown easily by changing integration variables to the primed basis for the angles characterizing the unit vector $\hat L$.
Here the round brackets stand for the usual Wigner-3j symbols, being non-zero for $m=m_1+m_2$ and $|\ell_1-\ell_2|\leq\ell\leq\ell_1+\ell_2$, {\it i.e.} imposing the rules known from coupling of two angular momenta, as well as
the condition that $\ell_1+\ell_2+\ell$ is even. For our purpose we need the Gaunt integral evaluated, in general, with three spherical harmonic
functions corresponding to different reference vectors,
\bea\label{eq:Gauntkpq}
  \lefteqn{ \int d^2\hat L\ Y^{\bm k}_{\ell m}(\hat L)^*\, Y^{\bm p}_{\ell_1 m_1}(\hat L)\, Y^{\bm q}_{\ell_2 m_2}(\hat L) }\nn\\
  &=& \sqrt{\frac{(2\ell+1)(2\ell_1+1)(2\ell_2+1)}{4\pi}}\left(\begin{array}{ccc}\ell&\ell_1&\ell_2\\ 0&0&0\end{array}\right)\nn\\
 && {} \times \left(\begin{array}{ccc}\ell&\ell_1&\ell_2\\ -m'&m_1&m_2'\end{array}\right)(-1)^{m'}\nn\\
 && {} \times [D^{m'}_{\ell m}((R^{\bm p})^{-1}R^{\bm k})]^* D^{m_2'}_{\ell_2 m_2}((R^{\bm p})^{-1}R^{\bm q})\,,
\eea
where summation over $m'=-\ell,\dots,+\ell$ and $m_2'=-\ell_2,\dots,+\ell_2$ is implied. To derive this expression we expressed all spherical harmonics
in terms of those with respect to $\bm p$ using Eqs.~\eqref{eq:Ylmk} and \eqref{eq:Ylmkinv}, and exploited the group property of the Wigner rotation matrices (see Appendix~\ref{app:relations}).
Note that the result depends on the relative orientation of the wave-vectors $\bm p$, $\bm q$ and $\bm k$ via the arguments of the Wigner rotation matrices
and the corresponding matrices $R^{\bm k}, R^{\bm p}$ and $R^{\bm q}$. There are equivalent ways of writing this generalized Gaunt integral, but we find the
form above most convenient. Note that the Wigner-3j symbols impose $m'=m_1+m_2'$, such that effectively only a single summation over either $m'$ or $m_2'$ occurs.
However, note that the generalized Gaunt integral can be non-zero also for $m\not= m_1+m_2$, while the conditions on the $\ell$'s are identical as for the
usual case quoted above.

\begin{figure}[t!]
  \begin{center}
  \includegraphics[width=\columnwidth]{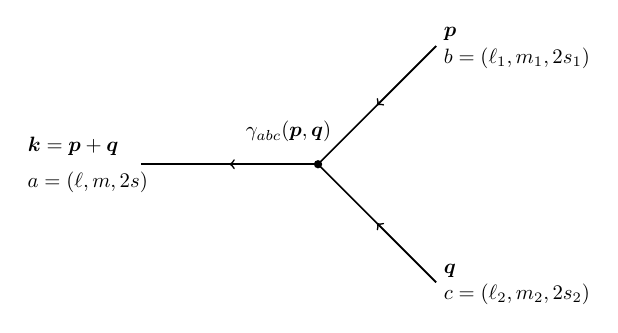}
  \end{center}
  \caption{\label{fig:Gammaabc}
  Illustration of the general mode-coupling function (``vertex'') $\gamma_{abc}(\bm p,\bm q)$ that describes how two (incoming) cumulant perturbation modes $\delta C_{\ell_1m_1,2s_1}(\eta,{\bm p})$ and $\delta C_{\ell_2m_2,2s_2}(\eta,{\bm q})$
  couple into one (outgoing) mode $\delta C_{\ell m,2s}(\eta,{\bm k})$ with $\bm k=\bm p+\bm q$. The modes are characterized by multi-indices $a=(\ell,m,2s),  b=(\ell_1,m_1,2s_1)$ and $c=(\ell_2,m_2,2s_2)$. The respective order in the
  cumulant expansion are given by $c_{\bm p}=\ell_1+2s_1, c_{\bm q}=\ell_2+2s_2$ and $c_{\bm k}=\ell+2s$. The vertex is in general non-zero provided the selection rules $c_{\bm k}=c_{\bm p}+c_{\bm q}-1$ as well as
  $\text{min}(|\ell_1-\ell_2-1|,|\ell_1-\ell_2+1|)\leq\ell\leq\ell_1+\ell_2+1$ are satisifed.
  }
\end{figure}

Using Eq.~\eqref{eq:Gauntkpq} to evaluate the non-linear term in Eq.~\eqref{eq:Clm_raw_eq_of_motion} and expanding in powers of $L^2$, we finally arrive
at the general form of the vertex (see Appendix~\ref{app:relations})
\be\label{eq:gammaabcexplicit}
  \gamma_{abc}(\bm p,\bm q) = \frac12 \left( \tilde\gamma_{abc}(\bm p,\bm q) + \tilde\gamma_{acb}(\bm q, \bm p)\right)\,,
\ee
with
\begin{widetext}
\bea\label{eq:gammaunsymmabcexplicit}
 \tilde\gamma_{abc}(\bm p,\bm q) &=& (-1)^\frac{\ell_1+\ell_2-1-\ell}{2} \frac{(2s)!}{(2s_1)!(2s_2)!}\delta^K_{\ell_1+2s_1+\ell_2+2s_2,\ell+2s+1} \nn\\
 && {} \times \frac{pk^{a_\ell}}{p^{a_{\ell_1}} q^{a_{\ell_2}}} (2\ell_1+1) (-1)^{m'}[D^{m'}_{\ell m}((R^{\bm p})^{-1}R^{\bm k})]^* D^{m_2'}_{\ell_2 m_2}((R^{\bm p})^{-1}R^{\bm q})\nn\\
 && {} \times \Bigg[(2\ell_2+1+2s_2)  \left(\begin{array}{ccc}\ell&\ell_1&\ell_2-1\\ 0&0&0\end{array}\right)
  \left(\begin{array}{ccc}\ell&\ell_1&\ell_2-1\\ -m'&m_1&m_2'\end{array}\right) \sqrt{\ell_2^2-(m_2')^2}\nn\\
 && {} + 2s_2 \left(\begin{array}{ccc}\ell&\ell_1&\ell_2+1\\ 0&0&0\end{array}\right)
  \left(\begin{array}{ccc}\ell&\ell_1&\ell_2+1\\ -m'&m_1&m_2'\end{array}\right) \sqrt{(\ell_2+1)^2-(m_2')^2} \Bigg]\,.
\eea
\end{widetext}
This vertex is one of the main results of this work, and describes the non-linear mode coupling
of the two modes $\delta C_{\ell_1m_1,2s_1}(\eta,{\bm p})$ and $\delta C_{\ell_2m_2,2s_2}(\eta,{\bm q})$ into one mode
$\delta C_{\ell m,2s}(\eta,{\bm k})$, {\it i.e.} of modes with cumulant perturbations characterized by
multi-indices $a=(\ell,m,2s), b=(\ell_1,m_1,2s_1)$ and $c=(\ell_2,m_2,2s_2)$, and wave-numbers $\bm k=\bm p+\bm q$.
We also recall that $a_\ell = 0 (1) $ for even (odd) $\ell$, $\delta^K$ is the Kronecker symbol, and that summation over $m'$ and $m_2'$ within the ranges
$-\ell\leq m'\leq \ell$ and $-\ell_2\leq m_2'\leq \ell_2$ is implied. The vertex is schematically illustrated in Fig.~\ref{fig:Gammaabc}.

Before discussing the properties of this general mode-coupling vertex, we also need to provide the mode-coupling terms
containing the extra perturbation mode $\delta$ for completeness. They follow from the usual continuity equation, with
the only non-zero vertices involving $\delta$ being of the form $\tilde\gamma_{\delta\delta c}(\bm p,\bm q)$ with $c=(1,m_2,0)$.
They follow from the vertices $\gamma_{\delta\delta\theta}$, being the usual $\alpha$-vertex, and the property
$\gamma_{\delta\delta w_i}=\gamma_{AA w_i}$, see Eq.\,(C3) in~\cite{cumPT},
\bea\label{eq:gammawithdelta}
  \tilde\gamma_{\delta,\delta ,100}(\bm p,\bm q) &=& 3\alpha(\bm q,\bm p) = 3\frac{(\bm p+\bm q)\cdot \bm q}{q^2}\,,\nn\\
  \tilde\gamma_{\delta,\delta ,1\pm1 0}(\bm p,\bm q) &=& \tilde\gamma_{100,100 ,1\pm1 0}(\bm p,\bm q)\,,
\eea
where we used $\delta C_{100}=\theta/3$ and $\delta C_{000}=\delta A=\ln(1+\delta)-\langle\ln(1+\delta)\rangle$.

Let us now discuss some general properties of the mode-coupling vertex Eq.~\eqref{eq:gammaabcexplicit}.
\begin{itemize}
\item The Kronecker delta $\delta^K_{\ell_1+2s_1+\ell_2+2s_2,\ell+2s+1}$ implies that modes of cumulant order $c_{\bm p}=\ell_1+2s_1$
and $c_{\bm q}=\ell_2+2s_2$ can couple non-linearly to generate a mode of cumulant order $c_{\bm k}=\ell+2s$ if
\be\label{eq:selectionrule1}
  c_{\bm k} = c_{\bm p}+c_{\bm q}-1\,.
\ee
Thus, in a diagrammatic interpretation of the mode-coupling vertex, the cumulant order of the outgoing line is one order lower than the sum of the cumulant orders of the incoming lines.
This can be easily understood from the structure of the non-linear term on the right-hand side of the equation of motion Eq.~\eqref{eq:eomdC} of the CGF. It contains a derivative with
respect to the auxiliary vector $\bm L$, being equivalent to a reduction of the cumulant order by one unit within the mode-coupling term.
\item The Wigner-3j symbols impose the condition
\bea\label{eq:selectionrule2}
\ \ \ \ \   \text{min}(|\ell_1-\ell_2-1|,|\ell_1-\ell_2+1|)\leq\ell\leq\ell_1+\ell_2+1\,,\nn\\
\eea
on the ``angular momentum'' numbers of the cumulant perturbation modes, arising from the two summands in Eq.~\eqref{eq:gammaunsymmabcexplicit}, {\it i.e.} from the triangle inequality required to
allow for a coupling of angular momenta $\ell_1$ and either $\ell_2+1$ or $\ell_2-1$ into $\ell$. The additional condition that $\ell_1+\ell_2\pm1-\ell$ has to be even is automatically satisfied
by virtue of Eq.~\eqref{eq:selectionrule1}, {\it i.e.} does not impose an additional constraint.
\item Due to the summation over $m'$ and $m_2'$ in Eq.~\eqref{eq:gammaunsymmabcexplicit}, the constraint that $m'=m_1+m_2'$ can always be satisfied, and in practice just implies that
the double sum collapses to a single one. Since the off-diagonal entries of the Wigner rotation matrices $D^{m'}_{\ell m}$ for $m'\not= m$ are in general non-zero, there is in practice
no selection rule on the $m$-numbers of the perturbation modes. That is, modes with all possible values $m_1=0,\pm 1,\dots,\pm \ell_1$ and $m_2=0,\pm 1,\dots,\pm \ell_2$ can couple
to any $m=0,\pm 1,\dots,\pm \ell$. This for example implies that a pair of scalar modes, {\it i.e.} modes with $m_1=m_2=0$, can generate modes of any $m$, provided the angular momentum
and cumulant selection rules from Eqs.~(\ref{eq:selectionrule1},\,\ref{eq:selectionrule2}) are satisfied.
\end{itemize}

\subsubsection{Examples}

For the evaluation of the vertices, it is convenient to express the Wigner rotation matrices in terms of the three Euler angles $\alpha,\beta,\gamma$ that characterize the orthogonal
rotation matrix it depends on. To gain some intuition we can consider the special case for which the wavevectors $\bm p$ and $\bm q$ lie in the $xz$-plane. Then also $\bm k=\bm p+\bm q$
is in this plane. We can further choose coordinates such that $\bm p=p(0,0,1)^T$ and $\bm q=q(\sin(\theta_{qp}),0,\cos(\theta_{qp}))^T$. Then $\bm k=k(\sin(\theta_{kp}),0,\cos(\theta_{kp}))^T$
with $\theta_{kp}$ being fixed by $\bm k=\bm p+\bm q$. It is also convenient to set $\theta_{kq}\equiv\theta_{kp}-\theta_{qp}$ and $\theta_{pq}\equiv-\theta_{qp}$, etc. Then all angles can be interpreted
as the rotation angle in the $xz$-plane between the respective wave-vectors, with counter-clockwise orientation. For this choice, the Euler angles $\alpha$ and $\gamma$ vanish, and the Wigner rotations
can be evaluated via
\bea\label{eq:WignerSimplified}
  D^{m'}_{\ell m}((R^{\bm p})^{-1} R^{\bm k}) &=& D^{m'}_{\ell m}(\theta_{kp})\,,\nn\\
  D^{m'}_{\ell m}((R^{\bm p})^{-1} R^{\bm q}) &=& D^{m'}_{\ell m}(\theta_{qp})\,,
\eea
and analogously for the case with $\bm p$ and $\bm q$ interchanged. The notation on the
right-hand side corresponds to a Wigner rotation evaluated with Euler angles $\alpha=\gamma=0$ and $\beta=\theta_{kp}$ or $\beta=\theta_{qp}$ in the first or second line, respectively.\footnote{Note that in notation of {\sc mathematica} (version~12.3), for $R=${\tt EulerMatrix}$[\{\alpha,\beta,\gamma\}]$, 
one has $D^{m'}_{\ell m}(R)=${\tt WignerD}$[\{\ell,m',m\},-\alpha,-\beta,-\gamma]$.}
For example, one trivially has $D^{m'}_{0m}(\theta)=1$ for $\ell=0$, and
\be
  D^{m'}_{1m}(\theta) = \left(\begin{array}{ccc}
  \cos^2\frac{\theta}{2} & \frac{\sin(\theta)}{\sqrt{2}} & \sin^2\frac{\theta}{2} \\
  -\frac{\sin(\theta)}{\sqrt{2}} & \cos\theta & \frac{\sin(\theta)}{\sqrt{2}} \\
  \sin^2\frac{\theta}{2} & -\frac{\sin(\theta)}{\sqrt{2}} & \cos^2\frac{\theta}{2} 
  \end{array}\right)\,,
\ee
with $m'=-1,0,1$ in the first, second and third row, respectively, and similarly
$m=-1,0,1$ in the three columns.

As a first sanity check, we verify that the standard SPT vertices are recovered.
For the $\alpha$-vertex, this is trivially guaranteed by the first line of Eq.~\eqref{eq:gammawithdelta} together with Eq.~\eqref{eq:gammaabcexplicit}.
The corresponding vertex for the log-density field, that enters via $\delta C_{000}$, is given by
the case $a=c=(000)$ and $b=(100)$. Eqs.~(\ref{eq:gammaabcexplicit},\,\ref{eq:gammaunsymmabcexplicit}) yield $\gamma_{000,100,000}(\bm p,\bm q)=3q\cos(\theta_{qp})/(2p)=3{\bm p}\cdot{\bm q}/(2p^2)=3\gamma_{A\theta A}(\bm p,\bm q)$,
where the latter refers to the notation from~\cite{cumPT}, specifically Eq.~(C1) therein. Given that $\delta C_{100}=\theta/3$, we recover the known form of this vertex.
The $\beta$-vertex within SPT corresponds to the vertex with $a=b=c=(100)$. From the general vertex given in Eqs.\,(\ref{eq:gammaabcexplicit},\,\ref{eq:gammaunsymmabcexplicit}) we find
\bea
  \gamma_{100,100,100}(\bm p,\bm q)&=&3k(p\cos(\theta_{kp})+q\cos(\theta_{kq})) \frac{\cos(\theta_{qp})}{2pq}\nn\\
  &=&\frac{3k^2{\bm p}\cdot{\bm q}}{2p^2q^2}=3\gamma_{\theta\theta \theta}(\bm p,\bm q)=3\beta(\bm p,\bm q)\,,\nn\\
\eea
with
the latter being the standard $\beta$-vertex known from SPT, and the extra factor three accounting for the rescaling $\delta C_{100}=\theta/3$ as above.
As an example for vorticity generation from scalar modes we consider the vertex
\be
  \gamma_{1\pm 1 0,000,002}(\bm p,\bm q)=\mp k p \frac{\sin(\theta_{kp})}{6\sqrt{2}}\,,
\ee
being non-zero if the cross product $\bm k\times\bm p=\bm q\times \bm p$ is non-vanishing, as observed in~\cite{cumPT2}.
This vertex corresponds to a non-linear interaction of a (log-)density mode with a scalar mode of the velocity dispersion that generate a vorticity perturbation.

The configuration of wavevectors within the $xz$-plane considered for the examples above is in fact already sufficient for computing one-loop power spectra or tree-level
bispectra. The reason is that only two linearly independent wavevectors appear in this case, and thus all possible linear combinations that potentially
enter the non-linear kernels can always be chosen to lie in a single plane. Moreover, since only the relative orientation enters in Eq.\,\eqref{eq:WignerSimplified},
it remains valid also when $\bm p$ is not along the $z$-axis but takes any direction in the $xz$-plane, when taking {\it e.g.} $\theta_{kp}$ as the angle between $\bm k$
and $\bm p$. 

Thus, in summary, the simplified expressions Eq.\,\eqref{eq:WignerSimplified} are sufficient to compute arbitrary non-linear interaction vertices Eqs.\,(\ref{eq:gammaabcexplicit},\,\ref{eq:gammaunsymmabcexplicit}) among
any cumulant modes entering one-loop power spectra or tree-level bispectra. The most general case is needed only when three or more linearly independent
wavevectors are present, such as for two-loop power spectra or one-loop bispectra. The general rotation matrices Eq.\,\eqref{eq:rotationmatrix} can be
constructed using the choice of basis vectors as described in App.\,D.2 in~\cite{cumPT2} in that case.\footnote{In our implementation we specifically choose $R^{\bm k}_1\equiv -b_{{\bm k}2}$ and $R^{\bm k}_2\equiv b_{{\bm k}1}$ in the notation of App.\,D.2 in~\cite{cumPT2}.}

\subsection{Comparison to the cartesian basis}\label{sec:kart}

In order to be able to compare the vertices Eqs.\,(\ref{eq:gammaabcexplicit},\,\ref{eq:gammaunsymmabcexplicit}) obtained within the spherical harmonic decomposition with those
derived in~\cite{cumPT,cumPT2} based on a cartesian basis, we note that both can be related 
using~\cite{Matsubara:2022ohx}
\be\label{eq:YlmPolynomialRepresentation}
  Y_{\ell m}(\vartheta,\varphi) = Y^{(m)*}_{i_1i_2\cdots i_\ell}\hat L_{i_1}\hat L_{i_2}\cdots\hat L_{i_\ell}\,,
\ee
where $\hat L=(\sin(\vartheta)\cos(\varphi),\sin(\vartheta)\sin(\varphi),\cos(\vartheta))^T$ and the $Y^{(m)*}_{i_1i_2\cdots i_\ell}$
are completely traceless tensors of rank $\ell$. They can be constructed using the spherical basis
$\bm e^0=\hat z$, $\bm e^\pm = \mp \frac{\hat x\mp i\hat y}{\sqrt{2}}$, that satisfies
$\bm e^\pm\cdot(\bm e^\pm)^*=\bm e^0\cdot(\bm e^0)^*=1, \bm e^\pm\cdot(\bm e^\mp)^*=0, \bm e^\pm\cdot (\bm e^0)^*=\bm e^0\cdot (\bm e^\pm)^*=0$.
The tensors for the first few multipoles are
\bea\label{eq:Ytensors}
  Y^{(0)} &=& \frac{1}{\sqrt{4\pi}}\,,\nn\\
  Y^{(m)}_j &=& \sqrt{\frac{3}{4\pi}} \bm e^m_j\,,\nn\\
  Y^{(0)}_{ij} &=& \sqrt{\frac{5}{16\pi}}(3\bm e^0_i\bm e^0_j-\delta^K_{ij})\,,\nn\\
  Y^{(\pm 1)}_{ij}&=&\sqrt{\frac{15}{16\pi}}(\bm e^0_i\bm e^\pm_j+\bm e^0_j\bm e^\pm_i)\,,\nn\\
  Y^{(\pm 2)}_{ij}&=&\sqrt{\frac{15}{8\pi}}(\bm e^\pm_i\bm e^\pm_j)\,,
\eea
and satisfy
\bea\label{eq:Ytensororth}
  Y^{(m)}_{i}Y^{(m)*}_{i}&=&\frac{3}{4\pi}\delta^K_{mm'}, \quad Y^{(m)}_{ij}Y^{(m)*}_{ji}=\frac{15}{8\pi}\delta^K_{mm'}\,,\nn\\
  \delta^K_{ij}Y^{(m)*}_{ji}&=&0\,.
\eea
The expansion can be simply generalized to the rotated spherical harmonics $Y^{\bm k}_{\ell m}$ by replacing the cartesian basis vectors $\hat x,\hat y,\hat z$ by $\hat x',\hat y',\hat z'$ with 
components $\hat z'_j=R^{\bm k}_{j3}=\hat k_j$, $\hat x'_j=R^{\bm k}_{j1}, \hat y'_j=R^{\bm k}_{j2}$. This means we obtain the tensor expansion of the rotated spherical harmonics
by using
\be
 \bm e^0=\hat k,\quad \bm e^\pm = \mp \frac{1}{\sqrt{2}}\left(R^{\bm k}_1\mp i R^{\bm k}_2\right)\,,
\ee
where $R^{\bm k}_i$ denotes the $i$th column of the rotation matrix $R^{\bm k}$, see Eq.~\eqref{eq:rotationmatrix}.
Comparing Eq.\,\eqref{eq:Ylmdecomp} to Eq.\,\eqref{eq:Ctildeexpansion}, this yields for example for the peculiar velocity
\bea\label{eq:velocity}
  \bm u_j &=& {\cal N}_m\frac{-i}{k}\sqrt{3\cdot 4\pi} Y^{(m)*}_j\,\delta C_{1m0} \\
  &=& \frac{-3i}{k}\left(\bm e^0_j\,\delta C_{100} + (\bm e^+_j)^*\,\delta C_{1,+1,0} + (\bm e^-_j)^*\,\delta C_{1,-1,0}\right)  \,.\nn
\eea
The part aligned with the wavevector is given by $\delta C_{100}$, and the ones perpendicular by $\delta C_{1,\pm 1,0}$, being related
to velocity divergence and vorticity, respectively, as claimed above. Indeed, using Eq.\,\eqref{eq:velocitydecomp} yields
$\theta=3\delta C_{100}$, in line with Eq.\,\eqref{eq:dC100}, as well as
\bea
  \delta C_{1,\pm 1,0} &=& \mp \frac{1}{3i}\bm e^\pm\cdot \bm w\,,
\eea
for the relation to vorticity $\bm w$. 
Similarly, for the perturbation of the dispersion tensor one obtains
\be
  \delta\epsilon_{ij} = \sqrt{4\pi} \delta_{ij}^K Y^{(0)*} \,\delta C_{002} -2 \sqrt{5\cdot 4\pi} Y^{(m)*}_{ij}\,\delta C_{2m0} \,.
\ee
Using Eq.\,\eqref{eq:Ytensororth} we can project out the components via
\be
  \delta C_{002} = \frac13 \delta^K_{ij} \delta\epsilon_{ij},\quad \delta C_{2m0} = -\frac{1}{30}\sqrt{\frac{16\pi}{5}} Y^{(m)}_{ij}\delta\epsilon_{ij}\,.
\ee
Comparing to Eq.\,\eqref{eq:epsdecomp} this yields the relations given in Eq.\,\eqref{eq:dC002dC200} between the scalar components 
$\delta C_{002}$, $\delta C_{200}$ and the corresponding cartesian parameterization $g$, $\delta\epsilon$. Furthermore, we obtain that the vector mode $\nu_i$ of the
dispersion tensor defined in Eq.\,\eqref{eq:epsdecomp} is related to $\delta C_{2,m,0}$ with $m=\pm 1$, as claimed above. Specifically,
\be
  \delta C_{2,\pm1,0} = \pm\frac{1}{5\sqrt{3}i}\bm e^\pm\cdot \nu\,.
\ee
Analogously, $\delta C_{2m0}$ for $m=\pm 2$ is related to the tensor modes $t_{ij}$, completing the set of modes at second cumulant order.
The mapping between cartesian and spherical harmonic bases can be extended to arbitrary cumulant order using Eq.\,\eqref{eq:YlmPolynomialRepresentation}.
For example, the two scalar modes of the third cumulant parameterized by $\pi$ and $\chi$ in~\cite{cumPT,cumPT2} are related via
\be
  \delta C_{102} = -\pi/15,\quad \delta C_{300} = (\pi-\chi)/105\,.
\ee

Using this mapping we checked agreement between the general result Eqs.~(\ref{eq:gammaabcexplicit},\,\ref{eq:gammaunsymmabcexplicit}) for the vertex $\gamma_{abc}(\bm p,\bm q)$ 
as well as the explicit results provided in~\cite{cumPT}. In particular, this includes 22 vertices involving three scalar modes up to third order in cumulants given in Eqs.\,(C1, C2, D3, D4, D5) of~\cite{cumPT}. 
Furthermore we checked that all vertices involving one, two or three vector modes (either vorticity $w$ or $\nu$) agree with those given in App.\,C of~\cite{cumPT}.
These are 13+8+2 vertices with either a single, two or three vector modes, respectively, given in Eqs.\,(C3-C6), Eqs.\,(C8-C9) and Eq.\,(C11) in~\cite{cumPT}.

\subsection{Stability conditions}\label{sec:stability}

The equations of motion for the perturbation modes of the cumulants in general lead to a strong suppression at scales $k\gg k_\sigma$,
or equivalently for $k^2\epsilon(\eta)\gg 1$. This suppression can be interpreted as a decoupling of small-scale modes, being an
expected physical behaviour that is captured by \vpt{} but not by SPT.

As noted in~\cite{cumPT}, the asymptotic behaviour of the cumulant perturbations for $k^2\epsilon(\eta)\gg 1$ depends on the
size of the cumulant expectation values ${\cal E}_{2s}$. More precisely, by studying the linear set of coupled evolution equations for
scalar modes it was found that the dimensionless, normalized quantities $\bar {\cal E}_{2s}$ defined in Eq.~\eqref{eq:E2sbar}
have to satisfy a set of ``stability conditions''. If these conditions are violated, the scalar perturbations feature exponentially
growing solutions for $k^2\epsilon(\eta)\gg 1$. If the stability conditions are satisfied, the perturbations are strongly suppressed
in the limit $k^2\epsilon(\eta)\gg 1$, as stated above. We therefore interprete these conditions as constraints that the expectation
values of higher cumulants have to satisfy in order to yield physically viable solutions for the cumulant perturbations.

The set of conditions depends in general on the cumulant order $c_{\text{max}}$ at
which the hierarchy of equations for cumulant perturbations is truncated. For example, for $c_{\text{max}}=3$ and $c_{\text{max}}=4$ only a single stability condition
on $\bar{\cal E}_4\equiv {\cal E}_4/\epsilon^2$ arises, being~\cite{cumPT}
\bea\label{eq:stability_scalar_cmax34}
  -6\leq \bar{\cal E}_4\leq 3 &\quad& (c_{\text{max}}=3)\,,\nn\\
  -2\leq \bar{\cal E}_4\leq 3 &\quad& (c_{\text{max}}=4)  \,.
\eea
For $c_{\text{max}}\leq 2$ no stability conditions arise, while for $c_{\text{max}}\geq 5$ the conditions
involve in addition $\bar{\cal E}_6$, and for $c_{\text{max}}\geq 7$ additionally $\bar{\cal E}_8$, etc.
For example, for $c_{\text{max}}=6$ the conditions read~\cite{cumPT}
\bea
  -6/5 &\leq& \bar{\cal E}_4\leq 3\,,\nn\\
  15(\bar{\cal E}_4-1) &\leq& \bar{\cal E}_6\leq 20 + 10\bar{\cal E}_4/3\,,\nn\\
  0 &\leq& 20(216+324\bar{\cal E}_4+90\bar{\cal E}_4^2+175\bar{\cal E}_4^3)\nn\\
  && -216\bar{\cal E}_6(2+5\bar{\cal E}_4)-27\bar{\cal E}_6^2\,.
\eea
In addition, the conditions obtained for higher even (odd) $c_{\text{max}}$ encompass (and potentially strengthen)
those for lower even (odd) $c_{\text{max}}$. That is, if the conditions for {\it e.g.} $c_{\text{max}}=8$ are satisfied,
also those for $c_{\text{max}}=4$ and $6$ are automatically fulfilled, as shown in~\cite{cumPT}.

Here we extend this analysis to non-scalar cumulant perturbation modes. To derive stability conditions, it is sufficient to consider the linear part of the evolution equation Eq.~\eqref{eq:eommatrixform},
described by the $\Omega_{ab}$ matrix. The linear equations do mix modes of different cumulant order, but only those sharing the same value of the $m$-number.
For the stability analysis, it is thus sufficient to consider only modes of a given $m$. For $m=0$, we precisely recover the linear evolution of scalar
modes studied in~\cite{cumPT}, and correspondingly the stability conditions derived there. To investigate non-scalar modes, we denote the subset of
cumulant perturbation modes with fixed $m$ by $\psi^{(m)}_a(\eta,\bm k)$, with index $a$ running over all modes $\delta C_{\ell m,2s}(\eta,\bm k)$
up to cumulant order $c_{\text{max}}$, {\it i.e.} comprising all $\ell$ and $s$ values satisfying $\ell+2s\leq c_{\text{max}}$. In addition, for given $m$ we have $\ell\geq|m|$. For example, for
$m=+1$ and $c_{\text{max}}=4$, 
\be\label{eq:psiVcmax4}
  \psi^{(+1)}_a=(\delta C_{110},\delta C_{210},\delta C_{112},\delta C_{310},\delta C_{212},\delta C_{410})\,.
\ee
Here $\delta C_{110}$ is related to vorticity $w_i$, and $\delta C_{210}$ to the vector mode $\nu_i$ of the velocity dispersion tensor.
Furthermore $\delta C_{112}$ and $\delta C_{310}$ are the $m=1$ vector modes at third cumulant order, and $\delta C_{212},\delta C_{410}$
those at fourth cumulant order. Modes with $m=+2,+3,+4$ up to $c_{\text{max}}=4$ are, respectively,
\bea
  \psi^{(+2)}_a &=& (\delta C_{220},\delta C_{320},\delta C_{222},\delta C_{420})\,,\nn\\
  \psi^{(+3)}_a &=& (\delta C_{330},\delta C_{430})\,,\nn\\
  \psi^{(+4)}_a &=& (\delta C_{440})\,.
\eea
The cases $m=-1,-2,-3,-4$ are completely analogous. The decoupling of subsets of modes with distinct $m$ at linear level is manifested by the
block-diagonal structure of the evolution matrix $\Omega_{ab}$ with respect to blocks of given $m$, see Eq.~\eqref{eq:Omegaab}.
Denoting the corresponding sub-matrix by $\Omega^{(m)}_{ab}$, the linear parts of the evolution equations take the form
\be
  \partial_\eta\psi_a^{(m)}+\Omega^{(m)}_{ab}\psi_b^{(m)} = \ \text{non-linear\ terms}.
\ee
Here $a$ and $b$ run over the subsets of cumulant perturbations with fixed $m$, as illustrated above.
It is convenient to switch to rescaled, dimensionless variables
\be
  \delta\bar C_{\ell m, 2s}(\eta,\bm k)\equiv \delta C_{\ell m, 2s}(\eta,\bm k)/\epsilon(\eta)^{[(\ell+2s)/2]}\,,
\ee
where $[\cdot]$ stands for the integer part. We denote the collection of rescaled cumulant perturbations by $\bar\psi^{(m)}_a(\eta,\bm k)$ accordingly.
Then the equation of motion can be written in the form 
\be\label{eq:eomlinearscaled}
  \partial_\eta\bar\psi^{(m)}+(\Omega^{(m)}_0+\epsilon k^2\Omega^{(m)}_1)\bar\psi^{(m)} = 0\,.
\ee
Here we suppress the indices $a,b$ for brevity, using vector-matrix notation, and set the non-linear terms to
zero, as appropriate for the linear stability analysis we are pursuing in this section. The scalar case $m=0$
corresponds to Eq.~(146) of~\cite{cumPT}. Here we consider $m\not=0$, and generalize the derivation of
stability conditions performed for the case $m=0$ in~\cite{cumPT}. From Eq.~\eqref{eq:Omegaab} we find\footnote{These expressions also hold for $m=0$, if the contribution arising from the gravitational force
term is included in addition. In linear approximation (for which $\ln(1+\delta)$ and $\delta$ are equivalent), this amounts to adding a term
$-\frac12\frac{\Omega_m}{f^2}\delta^K_{\ell,1}\delta^K_{\ell',0}\delta^K_{s,0}\delta^K_{s',0}\delta^K_{m,0}$ to the right-hand side of the first line in Eq.\,\eqref{eq:OmegaRescaled}.}
\bea\label{eq:OmegaRescaled}
  \left(\Omega^{(m)}_0\right)_{ab} &=& \delta^K_{\ell\ell'}\bar A^{ss'}_{\ell m} + \delta^K_{\ell,\text{even}}\left(\delta^K_{\ell-1,\ell'}\bar B^{ss'}_{\ell m} + \delta^K_{\ell+1,\ell'}\bar C^{ss'}_{\ell m}\right)\,,\nn\\
  \left(\Omega^{(m)}_1\right)_{ab} &=& \delta^K_{\ell,\text{odd}}\left(\delta^K_{\ell-1,\ell'}\bar B^{ss'}_{\ell m} + \delta^K_{\ell+1,\ell'}\bar C^{ss'}_{\ell m}\right)\,,
\eea
where $a=(\ell,m,2s), b=(\ell',m',2s')$ with fixed $m=m'$ and $\ell,s$ as well as $\ell',s'$ labelling the modes for a given value of $m$ up to cumulant order $c_{\text{max}}$.
Furthermore, $\delta^K_{\ell,\text{even}}$ is a generalization of the Kronecker symbol, being equal to $1$ if $\ell$ is even and $0$ otherwise, and correspondingly
for $\delta^K_{\ell,\text{odd}}$. Moreover $\bar A^{ss'}_{\ell m}\equiv A^{ss'}_{\ell m}+\delta^K_{ss'}[(\ell+2s)/2]\partial_\eta\ln\epsilon$, and $\bar B^{ss'}_{\ell m}$ is obtained from $B^{ss'}_{\ell m}$
defined in Eq.~\eqref{eq:ABC} by replacing $\{1,k^2\}\mapsto 1$, {\it i.e.} omitting the factor $k^2$ for odd $\ell$, and ${\cal E}\mapsto\bar{\cal E}$, {\it i.e.} replacing the cumulant expectation values by their normalized version, see Eq.~\eqref{eq:E2sbar}.
Finally, $\bar C^{ss'}_{\ell m}$ is obtained from $C^{ss'}_{\ell m}$ defined in Eq.~\eqref{eq:ABC} by analogous replacements. For example, for the vector modes with $m=+1$
and $c_{\text{max}}=4$ as shown in Eq.~\eqref{eq:psiVcmax4} one has
\bea\label{eq:Omega01cmax3m1}
 \Omega^{(+1)}_0 &=& \left(\begin{array}{cccccc}
 \frac12 & 0 & 0 & 0 & 0 & 0 \\
 \frac{\sqrt{3}}{5} & 1+\alpha & \frac{\sqrt{3}}{5} & -\frac{14\sqrt{2}}{5} & 0 & 0 \\
 0 & 0 & \frac32+\alpha & 0 & 0 & 0 \\
 0 & 0 & 0 & \frac32+\alpha  & 0 & 0 \\
 \frac{\bar{\cal E}_4}{5\sqrt{3}} & 0 & \frac{\sqrt{3}}{5} & -\frac{4\sqrt{2}}{5} & 2+2\alpha & 0 \\
 0 & 0 & 0 & \frac{\sqrt{5}}{3\sqrt{3}} & 0 & 2+2\alpha  
 \end{array}\right)\,,
 \nn\\
 \Omega^{(+1)}_1 &=& \left(\begin{array}{cccccc}
 0 & -\frac{5}{\sqrt{3}} & 0 & 0 & 0 & 0 \\
 0 & 0 & 0 & 0 & 0 & 0 \\
 0 & -\frac{2}{\sqrt{3}} & 0 & 0 & -\frac{7}{\sqrt{3}} & 0 \\
 0 & \frac{2\sqrt{2}}{7} & 0 & 0 & \frac{2\sqrt{2}}{7} & -\frac{9\sqrt{15}}{7} \\
 0 & 0 & 0 & 0 & 0 & 0 \\
 0 & 0 & 0 & 0 & 0 & 0 
 \end{array}\right)\,,
\eea
where we have set $\alpha\equiv\partial_\eta\ln\epsilon$ and used the EdS approximation $\Omega_m/f^2\to 1$.

For simplicity, we assume in the remainder of this section that $\alpha$ as well as the normalized
cumulant expectation values $\bar{\cal E}_{2s}$ are constant in time, following~\cite{cumPT}.
This is strictly justified in a scaling universe, and expected to be approximately satisfied in
realistic $\Lambda$CDM cosmologies. Then the matrices $\Omega^{(m)}_0$ and $\Omega^{(m)}_1$ are constant
in time, and the only source of time-dependence is the factor $\epsilon(\eta)k^2$.
We furthermore observe that for all $m$ and $c_{\text{max}}$ the matrices $\Omega^{(m)}_1$ are nilpotent. Specifically,
we find that the matrix product $\Omega^{(m)}_1\Omega^{(m)}_1=0$  (no summation over $m$)  yields the zero-matrix. This is due to
the property that only modes with odd $\ell$ \emph{and} even $\ell'$ give rise to non-zero entries in $\Omega^{(m)}_1$.
These properties allow us to generalize the analysis from Sec.\,VII\,D in~\cite{cumPT} to non-scalar modes.
In particular, the asymptotic behaviour for $k^2\epsilon \gg 1$ is determined by the eigenvalues $\lambda$
of the matrix
\be\label{eq:Mm}
  M^{(m)}_{c_\text{max}} \equiv \frac13\left(\Omega^{(m)}_0\Omega^{(m)}_1+\Omega^{(m)}_1\Omega^{(m)}_0-\frac{\alpha}{2}\Omega^{(m)}_1\right)\,,
\ee
with eigenmodes scaling as
\be
  \bar\psi^{(m)}\propto \exp\left(\pm\sqrt{\lambda}\int^{\eta}d\eta'\,\sqrt{3\epsilon(\eta')k^2}\right)\,,
\ee
with a prefactor that can depend polynomially on $\epsilon(\eta)$. Here we indicated the dependence on the cumulant
truncation order explicitly in the subscript, and $M^{(m)}_{c_\text{max}}$ reduces to Eq.\,(153) of~\cite{cumPT} for
the case of scalar modes ($m=0$).

Stability requires \emph{all} eigenvalues $\lambda$ of $M^{(m)}_{c_\text{max}}$ to be zero or real and negative, since otherwise an
exponentially growing solution exists. For the case of scalar modes with $m=0$, this requirement was used in~\cite{cumPT} to derive conditions on
the parameters $\bar{\cal E}_{2s}$ entering the matrix $M^{(m)}_{c_\text{max}}$. Following the same strategy,
we can now derive corresponding conditions for each $m\not=0$. Since $m$ enters only via $m^2$, see Eqs.~(\ref{eq:OmegaRescaled}, \ref{eq:ABC}),
the resulting conditions are identical for positive and negative $m$, {\it i.e.} depend only on $|m|$. It is thus sufficient to consider $m\geq 0$.

Let us start with the example $c_{\text{max}}=4$ and consider vector modes, {\it i.e.} $m=+1$. Using Eq.\,\eqref{eq:Omega01cmax3m1} we find that the matrix $M^{(+1)}_4$
from Eq.\,\eqref{eq:Mm} has the three distinct eigenvalues $-1/3$ and $-1\pm\sqrt{6+\bar{\cal E}_4}/3$.
They are real and less or equal to zero for 
\be
 -6\leq\bar{\cal E}_4\leq 3 \quad (c_{\text{max}}=4,\ \text{vector modes})\,.
 \ee
Remarkably, we find that this stability condition for \emph{vector}
modes within \emph{fourth} cumulant truncation precisely agrees with the one for \emph{scalar} modes in \emph{third} cumulant truncation, see Eq.\,\eqref{eq:stability_scalar_cmax34}.
In fact, all eigenvalues of $M^{(+1)}_4$ are identical to those of $M^{(0)}_3$.

We find that this pattern continues to higher truncation orders $c_{\text{max}}$. The characteristic polynomials of the matrices $M^{(\pm 1)}_{c_\text{max}}$ are
identical to those of $M^{(0)}_{c_\text{max}-1}$, and thus also the eigenvalues and consquently the stability conditions derived from them agree. We checked this
statement analytically up to $c_\text{max}\leq 9$.
For tensor modes ($|m|=2$) we find a similar correspondence between $M^{(\pm 2)}_{c_\text{max}}$ and $M^{(0)}_{c_\text{max}-2}$.
Even more generally, we find that the characteristic polynomials of $M^{(m)}_{c_\text{max}}$ are identical to those of $M^{(0)}_{c_\text{max}-|m|}$ for
$c_\text{max}\geq |m|$. Analytically, we checked this for all $c_\text{max}\leq 10$ and $|m|\leq 3$. Consequently, the stability conditions derived
from the evolution equations of modes with a certain $m$ and cumulants up to order $c_\text{max}$ are identical to those derived from scalar modes
up to cumulant order $c_\text{max}-|m|$.

Thus, the set of stability conditions on the normalized background cumulant expectation values $\bar{\cal E}_{2s}$ that were derived in~\cite{cumPT}
for scalar modes can be easily generalized when including vector, tensor and even higher $|m|$ modes. 
In practice, for an approximation that includes all modes up to a certain cumulant order $c_\text{max}$ and for all $|m|\leq m_\text{max}$, the stability conditions
for the scalar modes are thus the most restrictive ones. Those for vector, tensor and higher $|m|$ modes are then automatically satisfied as well, since they
correspond to the (less restrictive) set of stability conditions obtained for scalar modes at lower values of $c_\text{max}$.

Overall, we find a remarkable universality of the stability conditions. They are not only independent of the time-dependence of $\epsilon(\eta)$ as
found in~\cite{cumPT}, but also (up to a simple shift) independent of $|m|$.

In the following examples, we assume background values of the cumulants in accordance with stability conditions. We recall that truncating the hierarchy of
cumulants at the perturbation level at order $c_\text{max}$, the background cumulants ${\cal E}_{2s}$ up to order $2s\leq c_\text{max}$ ($2s\leq c_\text{max}+1$)
affect the \vpt{} kernels for even (odd) $c_\text{max}$. Furthermore, the trivial choice $0={\cal E}_4={\cal E}_6=\dots$ satisfies the stability conditions
for any truncation order.

\section{Impact of truncation on non-linear kernels}
\label{sec:kernels}

In this section we first discuss the practical implementation of \vpt{} as well as the
various truncation schemes of the cumulant expansion that we consider in this work, and
then present results for the impact of higher cumulants on the kernels furnishing the perturbative expansion
of the density field. The latter enter the power- and bispectra at linear and loop level that we discuss
in the following section.

\subsection{Truncation scheme and computation of \vpt{} kernels}

The solution of the coupled hierarchy of evolution equations for the perturbations in the cumulants of the phase-space distribution
function leads to a back-reaction on the density contrast $\delta$. Consider the generic perturbative Taylor
expansion of $\delta(\eta,\bm k)$ in terms of the usual initial density field $\delta_{\bm k0}$ (see Eq.~\eqref{eq:cumulantkernels} for notation),
\bea\label{eq:Fn}
  \delta (\eta,{\bm k}) &=& \sum_{n\geq 1} \int_{{\bm k}_1\dots{\bm k}_n}F_n({\bm k}_1,\dots,{\bm k}_n,\eta)\nn\\
  && {} \times e^{n\eta}\delta_{\bm k_10}\cdots\delta_{\bm k_n0}\,.
\eea
Regarding the density field, the difference between \vpt{} and SPT is encapsulated in the non-linear kernels $F_n({\bm k}_1,\dots,{\bm k}_n,\eta)$.
Within SPT, and assuming the common EdS-SPT approximation, they are independent of $\eta$ and can be constructed based on the well-known
algebraic recursion relations, see {\it e.g.}~\cite{Bernardeau:2001qr}.

Within \vpt{}, the kernels can also be computed recursively, but replacing the algebraic solution by a numerical solution of a set of coupled ordinary differential
equations with respect to the time-variable $\eta$~\cite{cumPT2}. This set of variables comprises the density contrast $\delta(\eta,\bm k)$ as well as all cumulant
perturbation modes $\delta C_{\ell m,2 s}(\eta,{\bm k})$ included in a given truncation. These quantities are collected in the perturbation vector $\psi_a(\eta,\bm k)$
from Eq.~\eqref{eq:psi}. Generalizing the perturbative expansion of the density contrast from Eq.~\eqref{eq:Fn} yields kernels for each of the cumulant perturbations given by
\bea\label{eq:Fan}
  \psi_a (\eta,{\bm k}) &=& \sum_{n\geq 1} \int_{{\bm k}_1\dots{\bm k}_n}F_{n,a}({\bm k}_1,\dots,{\bm k}_n,\eta)\nn\\
  && {} \times e^{n\eta}\delta_{\bm k_10}\cdots\delta_{\bm k_n0}\,,
\eea
with $F_n\equiv F_{n,a=\delta}$. The kernels for the
velocity divergence $\theta=3\delta C_{100}$ are given by $G_n\equiv 3F_{n,a=(100)}$, see Eq.~\eqref{eq:dC100}.
More generally, for values $a=(\ell m,2s)$ of the multi-index $a$, the kernels are related to those from Eq.~\eqref{eq:cumulantkernels} via $F_{n,a}=\delta C^{(n)}_{\ell m ,2s}$.

Using the evolution equation for the density field coupled to the perturbations of the cumulants Eq.~\eqref{eq:eommatrixform} yields
a coupled set of ordinary differential equations for the non-linear kernels
\bea\label{eq:kerneleom}
 \lefteqn{ (\partial_\eta+n)F_{n,a}({\bm k_1},\dots,{\bm k}_n,\eta) } \nn\\
 &+& \Omega_{ab}(\eta,k)F_{n,b}({\bm k}_1,\dots,{\bm k}_n,\eta)  \nn\\
 &=& \sum_{m=1}^{n-1} \Big\{ \gamma_{abc}(\bm p,\bm q) F_{m,b}({\bm q}_1,\dots,{\bm q}_m,\eta)\nn\\
 && \qquad \qquad F_{n-m,c}({\bm q}_{m+1},\dots,{\bm q}_n,\eta)\Big\}^s\,,
\eea
where summation over $b,c$ is implied, $k\equiv |\sum_{i=1}^n{\bm k}_i|$, $\bm p\equiv {\bm q}_1+\cdots +{\bm q}_m$, $\bm q\equiv {\bm q}_{m+1}+\cdots +{\bm q}_n$.
Furthermore, the wavenumbers $({\bm q}_1,\dots,{\bm q}_n)$ represent a permutation of $({\bm k}_1,\dots,{\bm k}_n)$,
and the superscript $s$ stands for \emph{averaging} over all $n!/m!/(n-m)!$ possibilities to select the subset of wavenumbers $\{{\bm q}_1,\dots,{\bm q}_m\}$
out of $\{{\bm k_1},\dots,{\bm k}_n\}$.

\begin{figure*}[t]
  \begin{center}
  \includegraphics[width=0.48\textwidth]{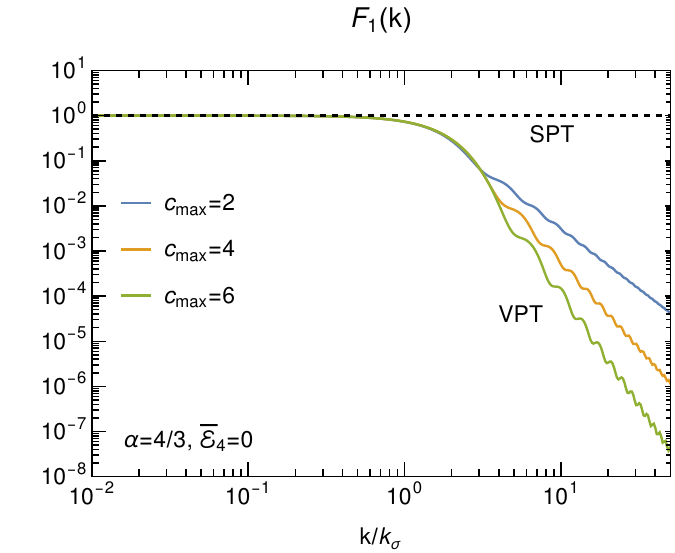}
  \includegraphics[width=0.48\textwidth]{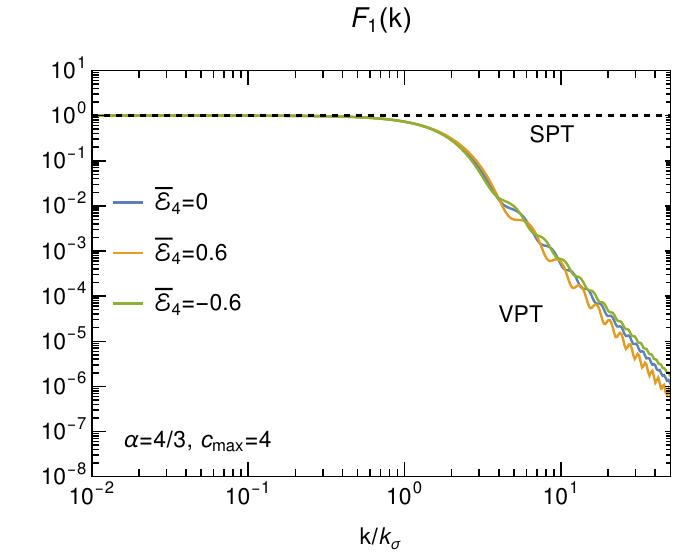}
  \end{center}
  \caption{\label{fig:F1}
  Linear kernel $F_1(k,\eta=0)$ of the density contrast $\delta$ within \vpt{} for truncations of the
  cumulant hierarchy at order $c_\text{max}=2,4,6$ (left), and for $\alpha=4/3$, $\bar{\cal E}_4=0$.
  The suppression relative to SPT (black dashed) sets in for $k\gtrsim k_\sigma$. The right panel
  shows the dependence on the normalized average value $\bar{\cal E}_4$ of the fourth cumulant for $\alpha=4/3, c_\text{max}=4$.
  }
\end{figure*}

We consider solutions of these differential equations for various truncations of the coupled hierarchy of equations for the cumulant perturbations, characterized by $c_\text{max}$ and $m_\text{max}$.
All truncations include the density field $\delta$ as well as a subset of $\delta C_{\ell m,2s}$ within the vector $\psi_a$ as follows:
\begin{itemize}
\item Truncation at order $c_\text{max}$ in the cumulant expansion of the phase-space distribution function: we include $\delta C_{\ell m,2s}$ with $c=\ell+2s\leq c_\text{max}$.
\item Truncation in the decomposition in scalar, vector, tensor and higher-rank modes of order $m_\text{max}$: we include $\delta C_{\ell m,2s}$ with $|m|\leq m_\text{max}$. Note that this is a restriction only for modes with $\ell>m_\text{max}$.
\end{itemize}
The truncation $c_\text{max}=1, m_\text{max}=0$ corresponds to SPT.
We do not consider the case $c_\text{max}=1, m_\text{max}=1$, which would correspond to SPT plus vorticity modes, but neglecting velocity dispersion, since in this truncation vorticity has only decaying modes
and no non-linear source term~\cite{Bernardeau:2001qr}.

All non-trivial truncations we consider thus have $c_\text{max}\geq 2$. In this case, vorticity is generated, and we consider only
truncations that do include vector modes, \emph{i.e.} with $m_\text{max}\geq 1$. This ensures that momentum conservation is respected in any truncation we consider~\cite{cumPT2}.
For all $c_\text{max}\geq 2$, the background value $\epsilon(\eta)=1/k_\sigma(\eta)^2$ of the velocity dispersion enters the evolution equations for perturbations via the matrix $\Omega_{ab}$.
Furthermore, for $c_\text{max}\geq 3$, also the background value ${\cal E}_4(\eta)$ of the fourth cumulant enters, that we parameterize by the dimensionless ratio
$\bar{\cal E}_4={\cal E}_4/\epsilon^2$. Similarly, for $c_\text{max}\geq 5$, $\bar{\cal E}_6={\cal E}_6/\epsilon^3$ enters in $\Omega_{ab}$.

The evolution of the \emph{average} values ${\cal E}_{2s}(\eta)$ is highly sensitive to the dynamics on small, highly non-linear scales on which shell-crossing occurs. While they can be estimated via
a halo model description~\cite{cumPT}, we take these quantities as (given) input in the following, and explore the impact on the evolution of the \emph{perturbations} of the cumulants on large, perturbative scales.
In practice, this means we assume a certain ansatz for ${\cal E}_{2s}(\eta)$, and compute the resulting kernels as described above. 
We find that the most important role for the impact of higher cumulants on the density and velocity divergence kernels is played by the dispersion scale $k_\sigma(\eta)=1/\epsilon(\eta)^{1/2}$ related
to the average value $\epsilon(\eta)={\cal E}_2(\eta)$ of the second cumulant. We expect that the non-linear shell-crossing dynamics on small scales leads to an increase of $\epsilon(\eta)$ with time,
and parameterize it as
\be\label{eq:epsilonparam}
  \epsilon(\eta)=\epsilon_0\,D(z)^\alpha = \epsilon_0\,e^{\alpha\eta}\,,
\ee
with some power-law index $\alpha$. Note that for a scaling universe with scale-free initial power spectrum $P_0\propto k^{n_s}$, one has $\alpha=4/(3+n_s)$.
The prefactor is related to the dispersion scale 
\be
  k_\sigma\equiv 1/\epsilon_0^{1/2}\,,
\ee
today ($\eta=0$). For $c_\text{max}\geq 3$ we furthermore assume that $\bar{\cal E}_4$ is constant in time, which is satisfied in a scaling universe.
We did not find a significant dependence of the density kernels on $\bar{\cal E}_6$ for $c_\text{max}\geq 5$, and therefore set it to zero for simplicity in the following.

We solve Eq.~\eqref{eq:kerneleom} numerically using Eq.~\eqref{eq:Omegaab} and Eqs.~(\ref{eq:gammaabcexplicit},\ref{eq:gammaunsymmabcexplicit})
for $\Omega_{ab}$ and $\gamma_{abc}$, respectively, starting at some initial time $\eta_\text{ini}$ for which the dispersion scale $k_\sigma(\eta_\text{ini})=1/\epsilon(\eta_\text{ini})^{1/2}$ is much
larger than the wavenumbers of interest entering the arguments of the kernels. This means the impact of velocity dispersion and higher cumulants is suppressed at the initial time, and
we initialize the kernels with the corresponding EdS-SPT kernels $F_n^\text{SPT}$ for $F_{n,\delta}$, $F_{n,a=(000)}$ and $(1/3)G_n^\text{SPT}$ for $F_{n,a=(100)}$, while all other kernels are
initially set to zero, following~\cite{cumPT2}. We choose $\eta_\text{ini}$ early enough such that all results are independent of its choice ($\eta_\text{ini}=-20$).
In practice, we first solve for the linear kernels $F_{n=1,a}(k,\eta)$, for which the right-hand side in Eq.~\eqref{eq:kerneleom} vanishes. The structure of $\Omega_{ab}$ and the average cumulants
entering it leads to a non-trivial time- and scale-dependence already at the linear level. Next, we use these solutions as well as the result Eqs.~(\ref{eq:gammaabcexplicit},\ref{eq:gammaunsymmabcexplicit}) for $\gamma_{abc}$ to compute the second-order kernels $F_{n=2,a}(\bm k_1,\bm k_2,\eta)$, and so on for higher $n$.

\subsection{Dependence of \vpt{} kernels on truncation order}

The kernels $F_n(\bm k_1,\dots,\bm k_n,\eta)$ obtained from the perturbative expansion Eq.~\eqref{eq:Fn}
of the density field within \vpt{} differ from the corresponding standard EdS-SPT kernels mainly due to
the presence of the dispersion scale $k_\sigma$. In the following we show results for $n=1,2,3$ and discuss the
dependence on the truncation scheme characterized by $c_\text{max}$ and $m_\text{max}$. Corresponding figures for
the kernels $G_n(\bm k_1,\dots,\bm k_n,\eta)$ are shown in Appendix~\ref{app:Gkernels}.

\begin{figure*}[t]
  \begin{center}
  \includegraphics[width=0.48\textwidth]{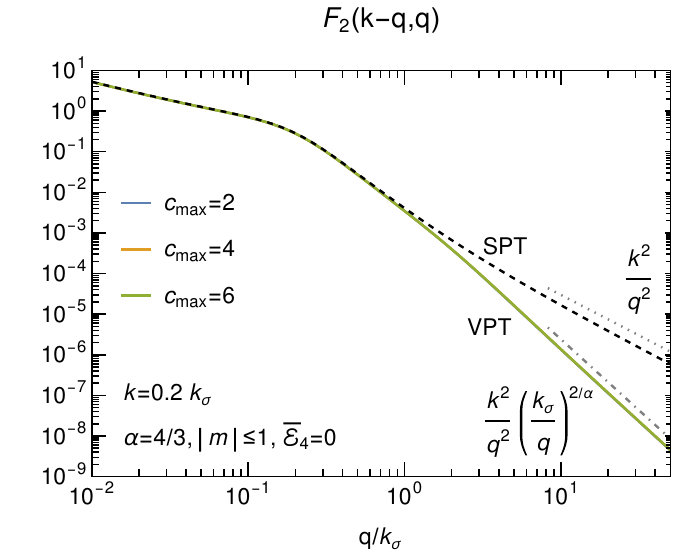}
  \includegraphics[width=0.48\textwidth]{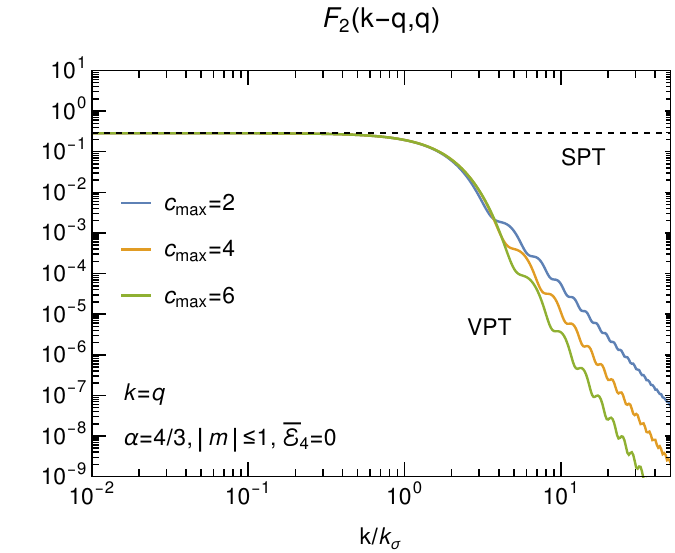}
  \\[1.5ex]
  \includegraphics[width=0.48\textwidth]{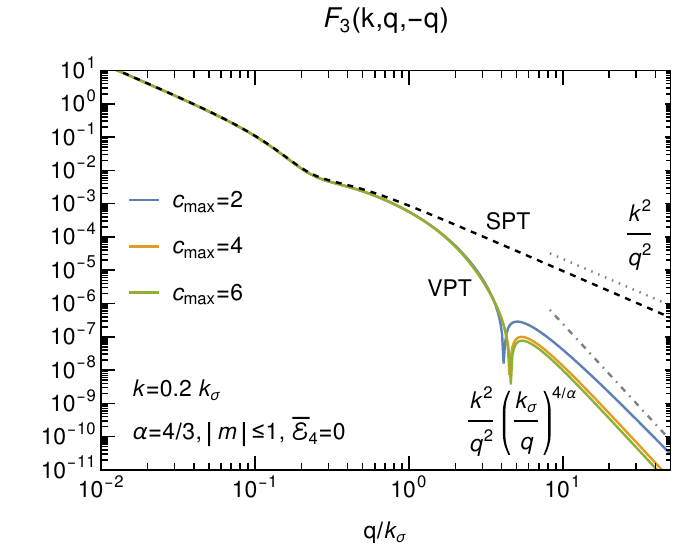}
  \includegraphics[width=0.48\textwidth]{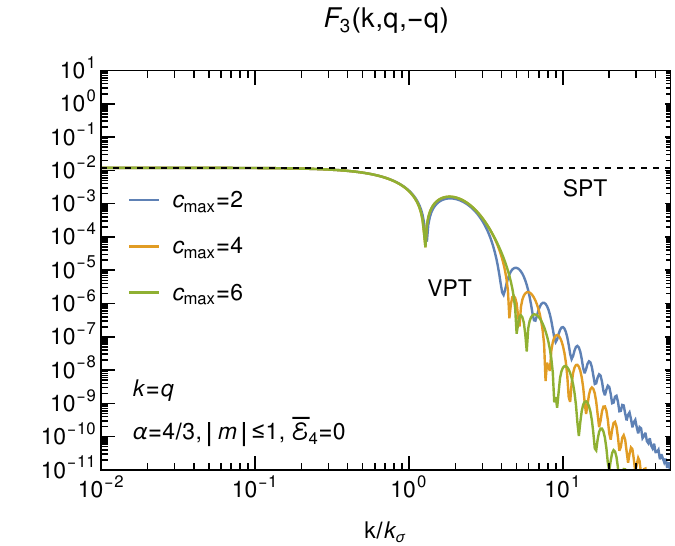}
  \end{center}
  \caption{\label{fig:F23_c246}
  Non-linear kernels $F_2(\bm k-\bm q,\bm q,\eta=0)$ (upper row) and $F_3(\bm k,\bm q,-\bm q,\eta=0)$ (lower row) of the density contrast $\delta$ within \vpt{} for truncations of the
  cumulant hierarchy at order $c_\text{max}=2,4,6$, respectively, and for $\alpha=4/3$, $\bar{\cal E}_4=0$, $m_\text{max}=1$, $\bm k\cdot\bm q/(k q)=0.5$. The left panels show the dependence on $q=|\bm q|$ for fixed $k=|\bm k|$, and the right panels the dependence when varying all arguments jointly with $k=q$. The SPT kernels are shown as well for comparison (black dashed). The universal parametric scaling in the limit $k\ll k_\sigma\ll q$ from Eq.~\eqref{eq:F23scaling} is indicated by the grey dot-dashed lines, and captures the expected screening of the backreaction of small-scale modes onto larger scales within loop integrals, as compared to SPT (grey dotted) where this effect is absent due to the underlying fluid approximation.
  }
\end{figure*}

\begin{figure*}[t]
  \begin{center}
  \includegraphics[width=0.48\textwidth]{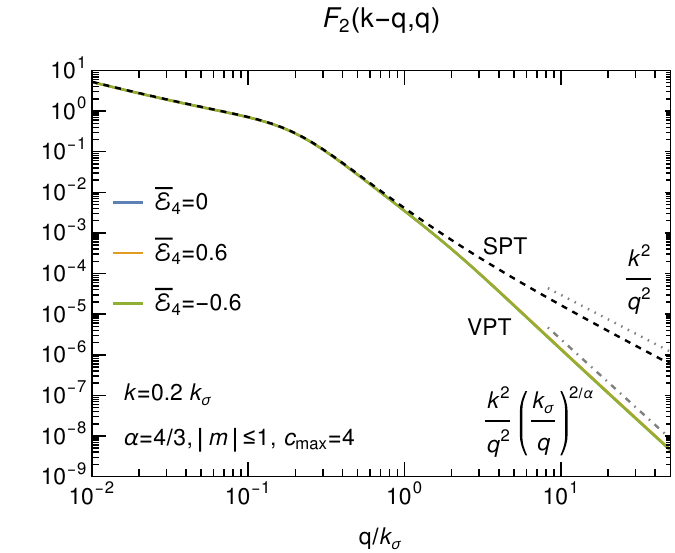}
  \includegraphics[width=0.48\textwidth]{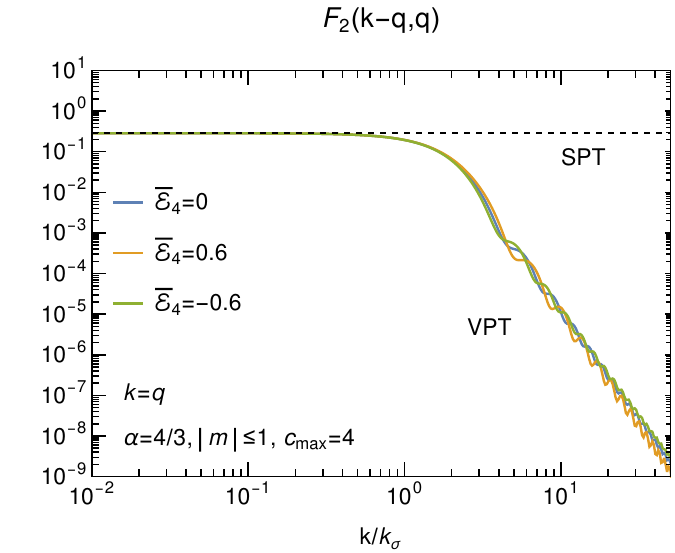}
  \\[1.5ex]
  \includegraphics[width=0.48\textwidth]{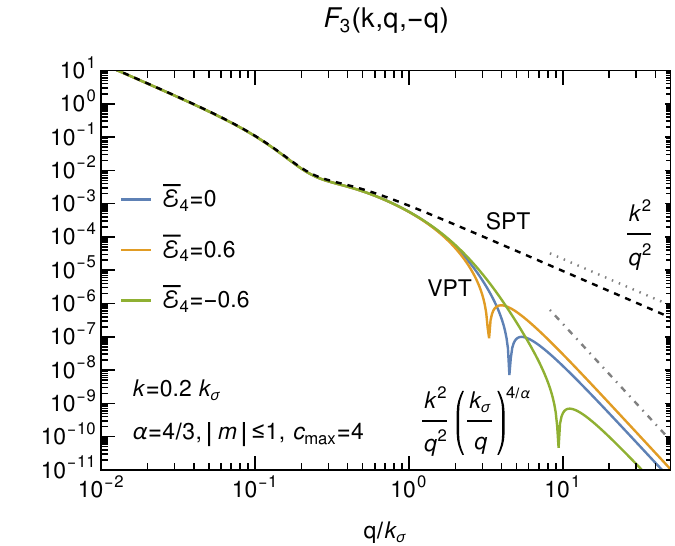}
  \includegraphics[width=0.48\textwidth]{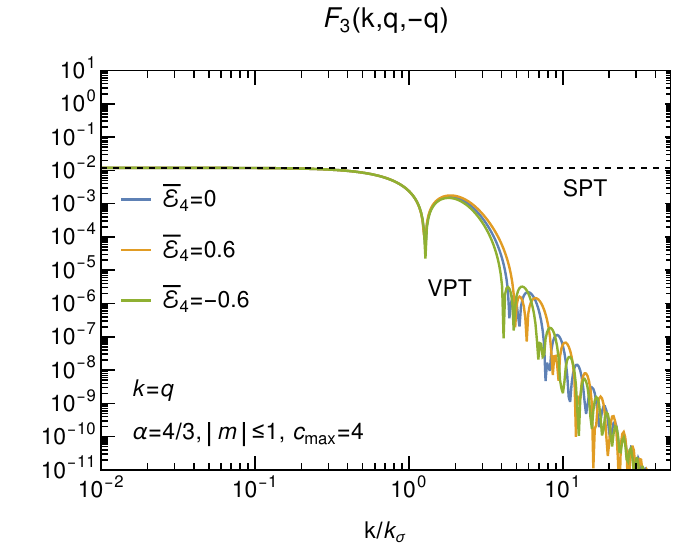}
  \end{center}
  \caption{\label{fig:F23_e4}
  As Fig.~\ref{fig:F23_c246}, but showing the dependence on the average value $\bar{\cal E}_4$ of the fourth cumulant for fixed truncation order $c_\text{max}=4$.
  }
\end{figure*}

Already the linear density kernel $F_1(k,\eta)$ differs significantly from SPT, $F_1^\text{SPT}=1$.
The $k$-dependence for $\eta=0$ is shown in Fig.~\ref{fig:F1}, for $c_\text{max}=2,4,6$ and choosing
$\alpha=4/3, \bar{\cal E}_4=0$ for illustration (left panel). Note that at linear level no mixing of scalar modes with vector, tensor
or other higher-rank modes occurs, such that $F_1$ is independent of $m_\text{max}$. This linear result has already been obtained in~\cite{cumPT}, and
its properties are reviewed for completeness here. We observe a pronounced suppression of $F_1$ for $k>k_\sigma$ relative to SPT. This is expected, since velocity dispersion
leads to a slow-down of linear growth, and modes with higher $k$ are affected by the dispersion scale $k_\sigma(\eta)=k_\sigma e^{-\alpha\eta/2}$
already starting at earlier times $\eta<0$. Furthermore, the \emph{linear} kernel features a pronounced dependence on the truncation order $c_\text{max}$.
Nevertheless, for given $k$ (and $\eta$), the increasing level of suppression of $F_1(k,\eta)$ for higher $c_\text{max}$ saturates at some point. For example, at the value of $k$ for which
the kernel $F_1(k,0)\simeq 0.1$ is suppressed by an order of magnitude with respect to SPT, the second cumulant truncation $c_\text{max}=2$ agrees already very well with those at higher $c_\text{max}$.
Similarly, all scales for which $F_1(k,0)\geq {\cal O}(0.01)$ are well captured by the truncation $c_\text{max}=4$. The dependence on the normalized average value $\bar{\cal E}_4$ of the fourth cumulant
is shown in the right panel of Fig.~\ref{fig:F1}, and is rather mild. Note that these background values are consistent with the stability conditions discussed in Sec.~\ref{sec:stability} for all
truncations considered here. The qualitative behavior is independent of the precise choice of the power-law index $\alpha$, see Eq.~\eqref{eq:epsilonparam}, and we therefore show results only for $\alpha=4/3$.

\begin{figure*}[t]
  \begin{center}
  \includegraphics[width=0.48\textwidth]{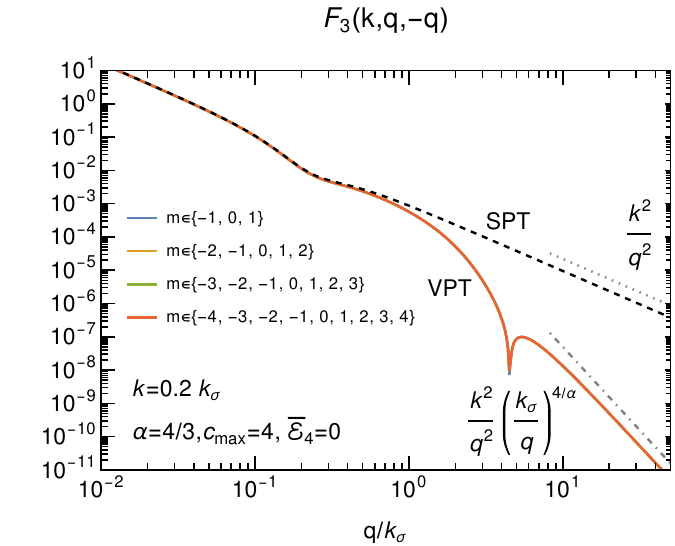}
  \includegraphics[width=0.48\textwidth]{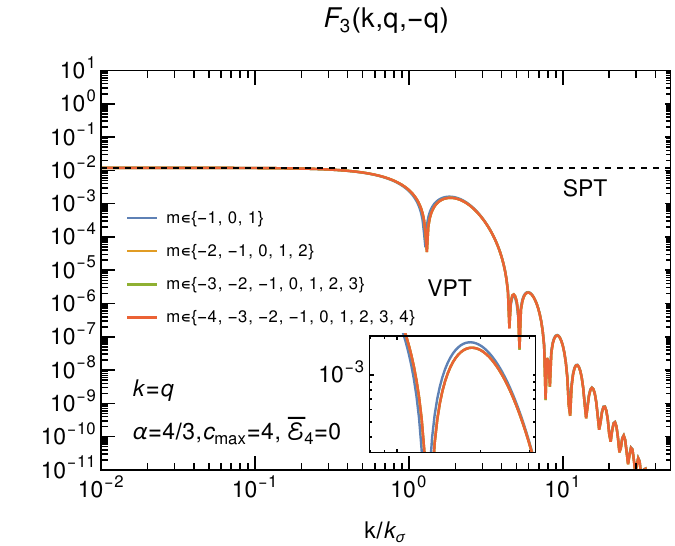}
  \end{center}
  \caption{\label{fig:F3_m}
  As lower row in Fig.~\ref{fig:F23_c246}, but showing the dependence on $m_\text{max}=1,2,3,4$, respectively, for fixed truncation order $c_\text{max}=4$. Note that all lines are almost on top of each other, showing that backreaction of tensor- or even higher-$|m|$ modes on the density contrast is very small.
  }
\end{figure*}

In Fig.~\ref{fig:F23_c246}, the dependence of the kernels $F_2(\bm k-\bm q,\bm q,\eta)$ (upper row) and $F_3(\bm k,\bm q,-\bm q,\eta)$ (lower row) on the truncation order $c_\text{max}$ is shown.
The left panels show the dependence on $q=|\bm q|$ for fixed $\bm k$. This is the limit that is relevant for the sensitivity of the one-loop matter power spectrum to small scales, \emph{i.e.} to large loop
wavenumbers $q$. Within SPT, both kernels are parametrically suppressed as $k^2/q^2$ in that limit. The reason is that the SPT kernel itself is a rational function, such that the $k^2$-dependence resulting from
momentum conservation leads to a $1/q^2$-dependence when $k\ll q$. For \vpt{}, momentum conservation implies a $k^2$-dependence for small $k$ as well~\cite{cumPT2}\footnote{Note that only the kernels $F_n$ are required to scale as $k^2$ due to momentum conservation in the limit when the sum of wavenumbers becomes small. For the $G_n$ kernels, this is only true in SPT due to some accidental cancellation, while in the most general case a scaling proportional to $k$ can occur, see~\cite{cumPT2}.}. However, the kernel depends in addition on the dimensionful scale $k_\sigma$. Thus, the scaling of the kernels in the limit of large $q$ differs from the SPT case. Interestingly, the scaling in the limit $k\ll k_\sigma\ll q$
is independent of the truncation, and given by~\cite{cumPT2}
\bea\label{eq:F23scaling}
  F_{2}(\bm k-\bm q,\bm q,\eta) &\propto& \frac{k^2}{q^2}\left(\frac{k_\sigma}{q}\right)^{2/\alpha}\,,\nn\\
  F_{3}(\bm k,\bm q,-\bm q,\eta) &\propto& \frac{k^2}{q^2}\left(\frac{k_\sigma}{q}\right)^{4/\alpha}\,,
\eea
as indicated by the grey dot-dashed lines in the left panels of Fig.~\ref{fig:F23_c246}. This suppression
of the non-linear kernels at large $q$ is a central result within \vpt{}, and captures the expected screening
of the backreaction of small-scale perturbations ($q$) onto larger scales ($k$). It is not captured in the ideal fluid description underlying SPT,
for which the kernels are proportional to $k^2/q^2$ in that limit (grey dotted lines in Fig.~\ref{fig:F23_c246}).
This difference between \vpt{} and SPT implies that loop integrals are much less sensitive to small-scale modes
in \vpt{}, in agreement with the expectation from mode coupling measured in $N$-body simulations~\cite{NisBerTar1611,NisBerTar1712}
as well as analytical arguments~\cite{peebles1980large}.

The reason why the asymptotic scaling of the non-linear kernels in Eq.~\eqref{eq:F23scaling} is independent of $c_\text{max}$, even though the asymptotic behaviour of the
linear kernel $F_1$ does depend on $c_\text{max}$, is the following~\cite{cumPT2}: for $k\ll k_\sigma\ll q$, the small-scale modes at scale $q$ essentially stop growing
once they enter the ``dispersion horizon'', {\it i.e.} for times $\eta>\eta_q$, where $\eta_q$ is defined by the condition
\be
  q=\epsilon(\eta)^{-1/2}\Big|_{\eta=\eta_q}\,.
\ee
Thus $k_\sigma(\eta)<q$ for $\eta<\eta_q$ and $k_\sigma(\eta)>q$ for $\eta>\eta_q$.
Consider $n$ modes with wave-numbers $\bm k_i$, $i=1,\dots,n$, that are all of the order of a common scale $q\gg k_\sigma$, but sum up to $\bm k=\sum_i\bm k_i$ with
$k=|\bm k|\ll k_\sigma$. This is a typical configuration entering an $n$th order kernel that appears when computing loop corrections to the power spectrum $P(k,\eta)$ on large, perturbative scales $k$, and
with loop wavenumber in the ``hard'' limit $q\gg k$.
In the computation of non-linear corrections to the density field, see Eq.~\eqref{eq:Fn}, the $n$th order contribution grows as $e^{n\eta}=D(z)^n$ within SPT. The same is true in \vpt{} as long as $\eta\ll\eta_q$. However, once $\eta\gg\eta_q$, the sourcing of the growth stops. This means that the $n$th order contribution to the density field is
parametrically of order $e^{\eta-\eta_q}\times e^{n\eta_q}$ within \vpt{}. Here the first factor arises from linear growth of the final, joint mode of wavenumber $k\ll k_\sigma$, while
the second factor arises from the stalling of linear growth of small-scale modes at scale $q$. Using Eq.~\eqref{eq:epsilonparam} this yields for the kernels Eq.~\eqref{eq:Fn}
\be\label{eq:Fnscaling}
  F_{n,\delta}(\bm k_1,\dots,\bm k_n,\eta) \sim \frac{k^2}{q^2}e^{(n-1)\eta_q}\sim \frac{k^2}{q^2} \left(\frac{k_\sigma}{q}\right)^{2(n-1)/\alpha}\,,
\ee
for $k\equiv|\sum\bm k_i|\ll k_\sigma\ll q\sim |\bm k_i|$, generalizing Eq.~\eqref{eq:F23scaling} for $n=2,3$\footnote{This scaling holds if \emph{all} arguments $|\bm k_i|\propto q$ become large as $q\to\infty$. In addition, it also holds if \emph{one} of the arguments, {\it e.g.} $\bm k_1$, has magnitude $|\bm k_1|=k$, such that $\sum_{i=2}^n\bm k_i=0$. The latter case is relevant {\it e.g.} for $F_{3}(\bm k,\bm q,-\bm q,\eta)$ or $F_{5}(\bm k,\bm q_1,-\bm q_1,\bm q_2,-\bm q_2,\eta)$ when both $|\bm q_1|$ and $|\bm q_2|$ become large. See also Eq.~\eqref{eq:F4scaling} for a generalization to kernels entering the bispectrum.}. The truncation order $c_\text{max}$ quantitatively influences the way how linear growth
stalls at $\eta\sim \eta_q$. However, Eq.~\eqref{eq:Fnscaling} is independent of this, and only relies on the fact that the $n$ modes sourcing the $n$th order kernel do stop growing at $\eta>\eta_q$. This explains the universality of the power-law scaling of the kernels in Eq.~\eqref{eq:Fnscaling}, and the independence of the truncation order, as long as the velocity dispersion scale is taken into account, {\it i.e.} for all $c_\text{max}\geq 2$.

The right panels in Fig.~\ref{fig:F23_c246} show the dependence of $F_2$ and $F_3$ when \emph{all} wavenumbers are varied proportionally to each other. This case is more similar to the linear kernel $F_1$ shown in Fig.~\ref{fig:F1}, with a dependence on $c_\text{max}$ that sets in at successively larger wavenumbers the larger $c_\text{max}$. However, since also the external wavenumber $k$ is increased in this case alongside with $q$, this dependence is relevant only for $k\gtrsim {\cal O}(k_\sigma)$. At these small scales, at the edge of validity of the perturbative expansion, a dependence on the truncation is expected.

Fig.~\ref{fig:F23_e4} shows the dependence of $F_2$ and $F_3$ on $\bar{\cal E}_4$. Similarly as for the linear kernel (see right panel of Fig.~\ref{fig:F1}), this dependence is rather mild.

The dependence on $m_\text{max}$ is illustrated in Fig.~\ref{fig:F3_m}. The various lines can hardly be distinguished in this figure, meaning that the impact of tensor- and higher-rank modes
on the density kernels is rather small. We therefore use $m_\text{max}=1$ in all results (except for Fig.~\ref{fig:F3_m}) in this work, {\it i.e.} take scalar and vector modes into account, but neglect the backreaction of tensor modes ($m=\pm 2$) as well as modes with even higher $|m|$ on the density field. Note that the kernel $F_2$ for the density contrast is independent of $m_\text{max}$, since it is determined by a single non-linear interaction of two scalar modes sourced by the initial density field. For $F_3$, modes with $m\not=0$ contribute in the ``internal'' propagator connecting the two vertices $\gamma_{abc}$ entering this kernel, in a diagrammatic language following {\it e.g.}~\cite{Bernardeau:2001qr}.

In summary, we find that the non-linear kernels $F_n$ furnishing the perturbative expansion of the density contrast within \vpt{} feature a suppression for large wavenumber that encodes the
physically expected screening of the backreaction of small-scale modes onto larger scales. The main difference compared to SPT arises already when incorporating the velocity dispersion, while third- and higher cumulants have a certain quantitative impact for scalar ($m=0$) and vector ($m=\pm 1$) modes. The backreaction of tensor ($m=\pm 2$) or higher modes on the density field is negligible on weakly non-linear scales. Importantly, the parametric scaling of the kernels Eq.~\eqref{eq:Fnscaling} within the screening regime is independent of the truncation order $c_\text{max}$.

\section{Impact of truncation on power spectra}
\label{sec:impact}

In this section we discuss the impact of the truncation order $c_\text{max}$ within \vpt{} on power spectra of the density and velocity field
as well as on the matter bispectrum, and compare to numerical $N$-body simulations.\footnote{See~\cite{cumPT2} for details on the $N$-body simulations and results.} 

\subsection{Setup and UV behavior: \vpt{} vs SPT}\label{sptvpt}

The perturbative expansion of power spectra within \vpt{} is
analogous to SPT. For power spectra involving the density or velocity divergence in real space, the only change is to replace SPT by \vpt{} kernels,
and taking into account that already the linear kernels are non-trivial.
For example, the linear and one-loop contributions to the matter power spectrum $P_{\delta\delta}(k,\eta)$ read
\bea
  P_\text{lin}(k,\eta) &=& e^{2\eta}F_1(k,\eta)^2P_0(k)\nn\,,\\
  P_{1L}(k,\eta) &=& e^{4\eta}\int d^3q\,P_0(q)\Big(6F_3(\bm k,\bm q,-\bm q,\eta)F_1(k,\eta)\nn\\
  && {} P_0(k) + 2F_2(\bm k-\bm q,\bm q,\eta)^2P_0(|\bm k-\bm q|)\Big)\,,\nn\\
\eea
with \vpt{} kernels $F_n$ for the density field.

For illustration, we consider in this work a scale-free, Gaussian initial density field with power
spectrum
\be
  P_0(k)\propto k^{n_s}\,,
\ee
characterized by spectral index $n_s$, and an EdS background cosmology. 
The scaling symmetry of a scale-free Universe requires the average velocity dispersion $\epsilon(\eta)$ to depend on time as in Eq.~\eqref{eq:epsilonparam} with power-law index 
\be
  \alpha=\frac{4}{n_s+3}\,,
\ee
such that $k_\sigma(\eta)=\epsilon(\eta)^{-1/2}$ has the same time-dependence as the non-linear scale $k_\text{nl}(\eta)$, defined as usual via $4\pi k^3e^{2\eta}P_0(k)|_{k=k_\text{nl}(\eta)}=1$.
Thus the average dispersion is fully characterized by the constant ratio 
\be
  \frac{k_\sigma}{k_\text{nl}} = \frac{k_\sigma(\eta)}{k_\text{nl}(\eta)}\,.
\ee
We consider values $n_s=-1,0,1,2$ for which the naive SPT one-loop correction is UV divergent. 
Notably, the screening captured by \vpt{}, see Eq.~\eqref{eq:F23scaling}, renders the one-loop power spectrum UV \emph{finite} within \vpt{}. 
Consider for example the contribution to the one-loop result from the region of large wavenumber, such that $q\gg k_\sigma\gg k$.
Using Eq.~\eqref{eq:F23scaling}, as well as $P_0(q)=Aq^{n_s}$ with $A=1/(4\pi k_\text{nl}^{n_s+3})$, yields for the scaling of the loop integrand involving the $F_3$-kernel (the ``$P_{13}$'' part) 
\bea\label{eq:P13UVscaling}
  && \int d^3q \, F_3(\bm k,\bm q,-\bm q,\eta)\, P_0(q)\nn\\
  &\to& \int d\ln(q)\,4\pi q^3\,\frac{k^2}{q^2}\,\left(\frac{ k_\sigma}{q}\right)^{4/\alpha}\, Aq^{n_s}\nn\\
  &=& \int d\ln(q)\,4\pi q^3\,\frac{k^2}{q^2}\,\left(\frac{ k_\sigma}{q}\right)^{n_s+3}\, Aq^{n_s} \nn\\
  &=& \left(\frac{k_\sigma}{k_\text{nl}}\right)^{n_s+3}\int d\ln(q)\,\frac{k^2}{q^2}\,,
\eea
where the arrow in the second line indicates that we consider the scaling of the integrand in the UV domain $q\gg k_\sigma\gg k$.
In addition, the expression on the right of the arrow is understood to represent the UV contribution to the loop integral up to some overall, multiplicative proportionality factor.

Thus, the loop integral is rendered UV finite by the screening of the $F_3$-kernel captured in \vpt{}.
More importantly, this implies the one-loop result is dominated by modes that are below the scale $k_\sigma$. Since $k_\sigma$ is close to $k_\text{nl}$~\cite{cumPT2} (see also below),
this means the result of the \emph{loop integral is mostly sensitive to modes under perturbative control, in agreement with expectations}. In contrast, in SPT, the ``screening'' factor
$(k_\sigma/q)^{4/\alpha}$ in Eq.~\eqref{eq:P13UVscaling} is absent, leading to the well-known spurious UV sensitivity of SPT and resulting in a formal UV divergence for $n_s\geq -1$.

\begin{figure*}[t]
  \begin{center}
  \includegraphics[width=0.48\textwidth]{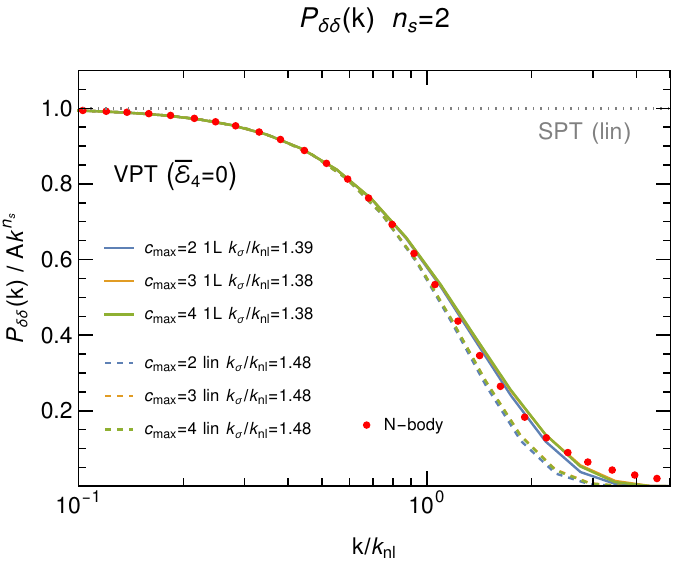}
  \includegraphics[width=0.48\textwidth]{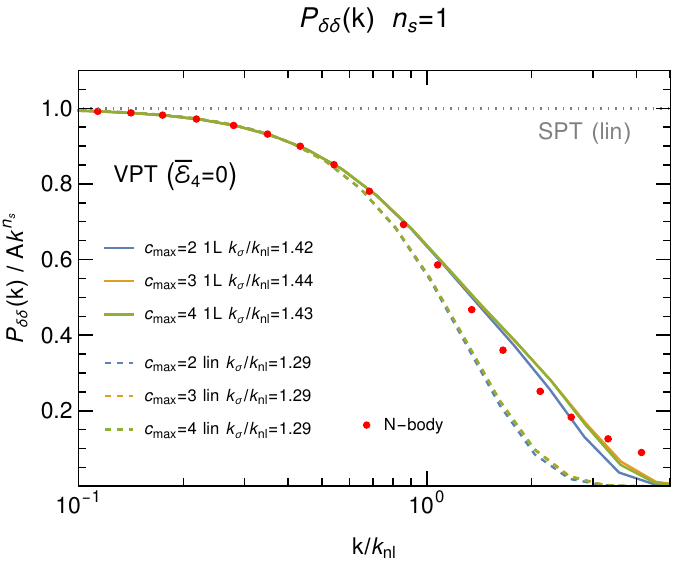}
  \\[1.5ex]
  \includegraphics[width=0.48\textwidth]{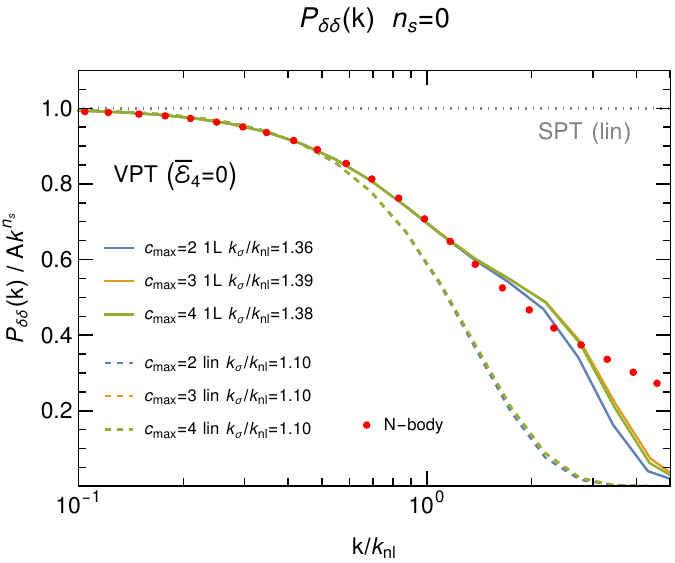}
  \includegraphics[width=0.48\textwidth]{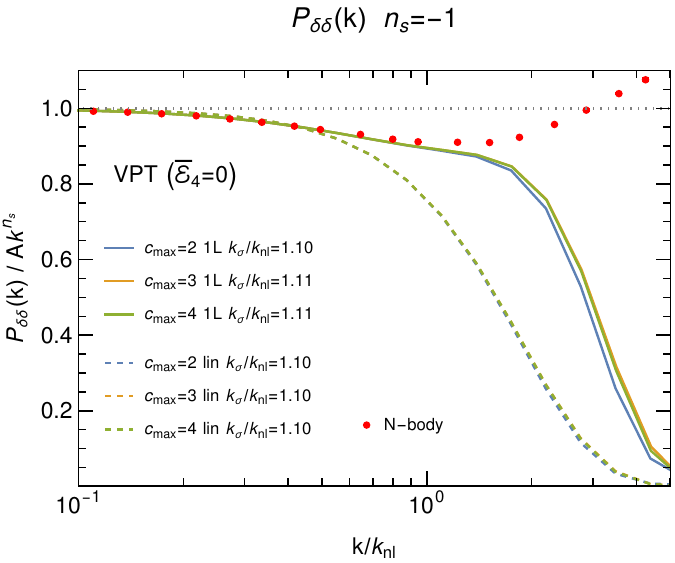}
  \end{center}
  \caption{\label{fig:Pdd_c234}
  Matter density power spectrum $P_{\delta\delta}(k)$ for a scale-free initial spectrum $P_0\propto k^{n_s}$, for spectral indices $n_s=2,1,0,-1$, respectively. Each panel shows the linear (dashed) as well as one-loop (solid) \vpt{} results when using a truncation of the coupled cumulant hierarchy obtained from the Vlasov-Poisson equations at orders $c_\text{max}=2,3,4$. Red points show $N$-body simulation results as detailed in~\cite{cumPT2}. The \vpt{} prediction depends on one free parameter, $k_\sigma/k_\text{nl}$, that is fixed by fitting to the $N$-body result up to $k_\text{max}=0.5k_\text{nl}$. Here $k_\text{nl}$ is the usual non-linear scale, and $k_\sigma=1/\epsilon^{1/2}$ is the scale related to the average velocity dispersion $\epsilon$. We set the average value of the fourth cumulant to $\bar{\cal E}_4=0$ here (see Fig.~\ref{fig:Pdd_e4} for the dependence on $\bar{\cal E}_4$).
  }
\end{figure*}

\begin{figure*}[t]
  \begin{center}
  \includegraphics[width=0.48\textwidth]{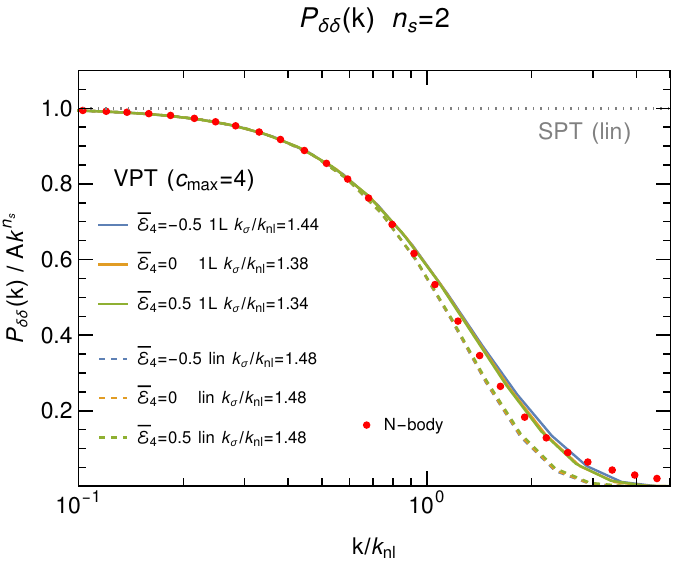}
  \includegraphics[width=0.48\textwidth]{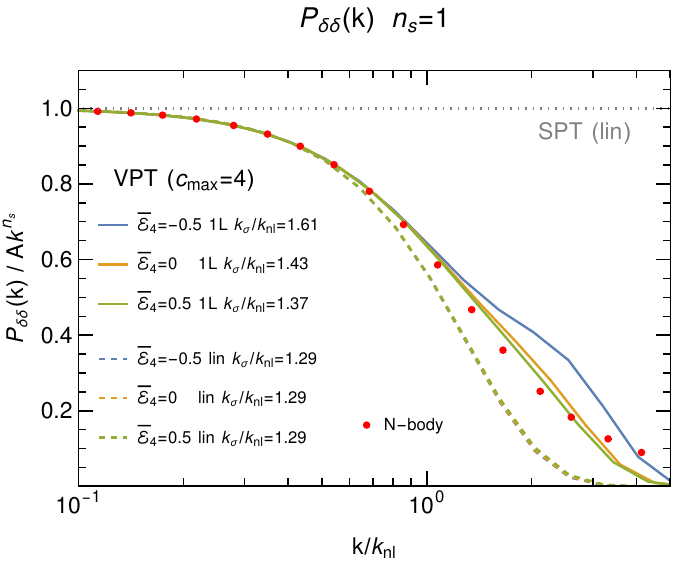}
  \\[1.5ex]
  \includegraphics[width=0.48\textwidth]{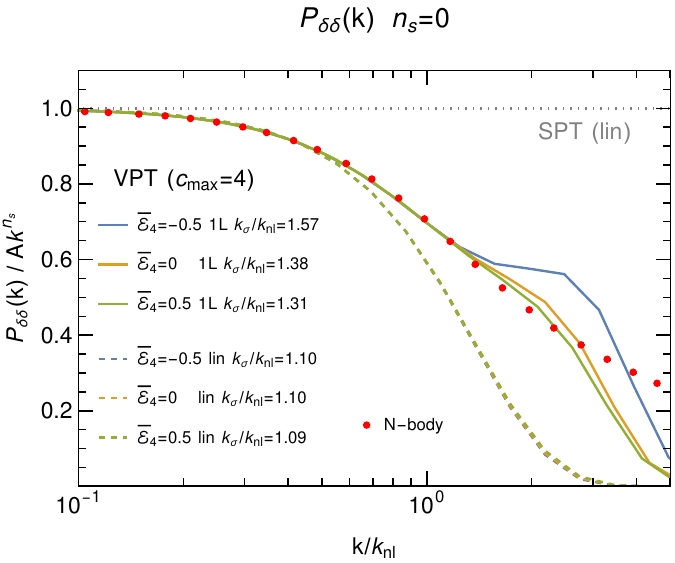}
  \includegraphics[width=0.48\textwidth]{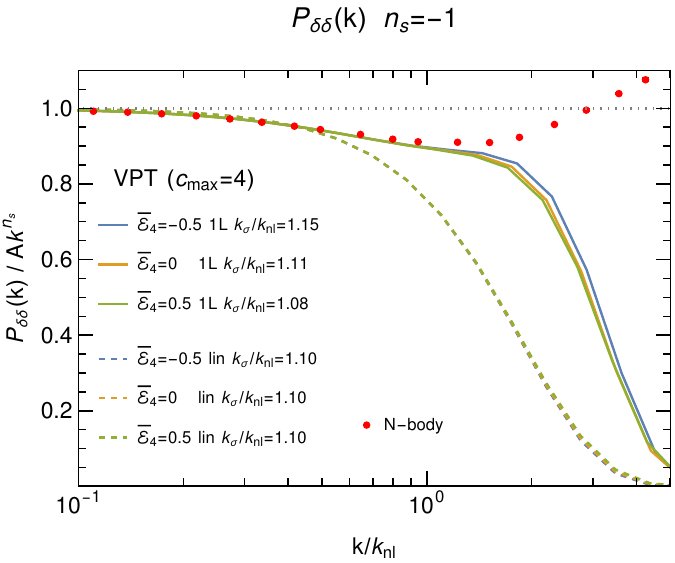}
  \end{center}
  \caption{\label{fig:Pdd_e4}
  As Fig.~\ref{fig:Pdd_c234}, but showing the dependence of the matter density power spectrum $P_{\delta\delta}(k)$ on the average value $\bar{\cal E}_4$ of the fourth cumulant of the matter phase-space distribution function for $c_\text{max}=4$ within \vpt{}.
  }
\end{figure*}

Notably, in \vpt{}, the value of $n_s$ drops out in the scaling of the integrand in Eq.~\eqref{eq:P13UVscaling}. Thus, the integral is UV convergent even for arbitrarily large values of $n_s$~\cite{cumPT2}.
Moreover, from Eq.~\eqref{eq:Fnscaling}, one easily checks that this remains true at arbitrary loop order, for all $L$-loop contributions to the power spectrum of the form $P_{1,2L+1}$, involving
a product of the $F_{2L+1}$ and $F_1$ kernels. Those are UV convergent in \vpt{} for arbitrary $n_s$ and all $L$.
For comparison, within SPT, the $P_{1,2L+1}$ contributions to the perturbative expansion of the power spectrum are the most UV sensitive ones, and become UV divergent for $n_s\geq -3+2/L$.

Using Eq.~\eqref{eq:F23scaling}, the contribution to the one-loop power spectrum involving $F_2$ (the ``$P_{22}$'' part) can be shown to be UV finite in \vpt{} as long as the rather weak
condition $n_s<4$ is met. Interestingly, based on Eq.~\eqref{eq:Fnscaling} this can be generalized to all $L$-loop contributions $P_{NM}$, featuring a product of kernels $F_N$ and $F_M$,
with $N,M>1$. They are UV finite in \vpt{} if $n_s<4$ for all loop orders $L=(N+M)/2-1$. Within SPT, $P_{NM}$ with $N,M>1$ are UV divergent for $n_s\geq (4-3L)/(L+1)$.

In summary we find that \emph{all} loop corrections to the matter power spectrum are UV finite in \vpt{} for $n_s<4$.
We note that the IR limit of small loop wavenumber $q$ is analogous to SPT, as dictated by Galilean invariance, being finite for $n_s>-3$ when summing all contributions at a given loop order (see {\it e.g.}~\cite{peebles1980large,ScoFri9607,Blas:2013bpa,BlaGarKon1309}).
Therefore, loop corrections to the power spectrum are absolutely convergent within \vpt{} for
\be\label{eq:nsVPT}
  -3 < n_s < 4\qquad\text{(\vpt{})}\,.
\ee
Importantly, this statement formally holds at all orders in the perturbative expansion within \vpt{}.
We stress that this covers the entire range of $n_s$ that is theoretically consistent with the paradigm of hierarchical structure formation (see {\it e.g.}~§26 and §28E in~\cite{peebles1980large}),
that is, a non-linear power spectrum $P_{\delta\delta}(k,\eta)$ that exhibits scaling symmetry and approaches the linear spectrum $e^{2\eta}Ak^{n_s}$ for $k\ll k_\text{nl}(\eta)$.
Thus, \vpt{} can be used to compute non-linear corrections at (in principle) all orders of peraturbation theory whenever the dynamics leads to the existence of a weakly non-linear regime, at wavenumbers around $k_\text{nl}$.
In SPT, the corresponding range is 
\be
  -3<n_s<-3+2/L\qquad\text{(SPT)}\,,
\ee
and shrinks to zero for large loop order $L$, as a consequence of the spurious UV sensitivity of SPT. As stressed above, the main virtue of \vpt{} is that the UV screening limits the sensitivity of the loop
integral to modes within the physically expected range, such that the result is dominated by scales that are under perturbative control.
This is true independently of the precise shape of the input power spectrum, but manifests itself in particular in the extended convergence range with respect to the
spectral index $n_s$.

\subsection{Comparing \vpt{} to Simulations}\label{fit}

Following~\cite{cumPT2}, we treat $k_\sigma/k_\text{nl}$ as a single free parameter within the \vpt{} setup (for each $n_s$ and each choice of $c_\text{max}$, respectively). Its value has been estimated via a halo model approach as well as by self-consistently solving coupled equations for perturbations and average values within \vpt{} in~\cite{cumPT2}, finding good agreement with a determination of $k_\sigma/k_\text{nl}$ obtained from fitting the one-loop prediction of the matter power spectrum in \vpt{} to simulation results. Here we thus follow the latter approach. We stress that once $k_\sigma/k_\text{nl}$ is fixed from the fit to the $N$-body matter power spectrum, no more free parameters remain. Thus, the velocity power spectra as well as the bispectrum are fixed, without any free parameters that can be tuned to $N$-body measurements within~\vpt{}.

Compared to~\cite{cumPT2}, we significantly enlarge the number of cumulant perturbation modes that are taken into account, including all scalar and vector modes up to cumulant order $c_\text{max}=2,3,4$, respectively, as well as all non-linear vertices $\gamma_{abc}$. In~\cite{cumPT2}, only scalar modes of the third cumulant have been included, and for vector modes only those vertices that contain exclusively cumulants up to second order.

Let us now discuss the effects of truncation and compare to simulation measurements for each statistic in turn. 

\subsection{Matter power spectrum}\label{eq:mpk}

\begin{figure*}[t]
  \begin{center}
  \includegraphics[width=0.48\textwidth]{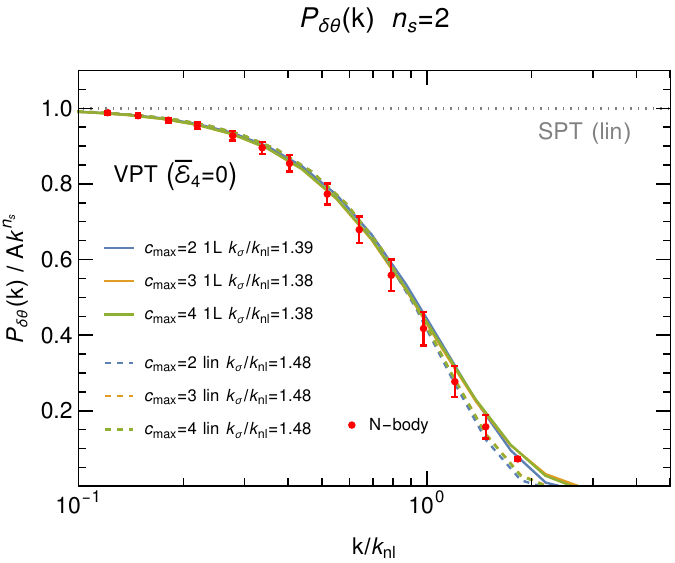}
  \includegraphics[width=0.48\textwidth]{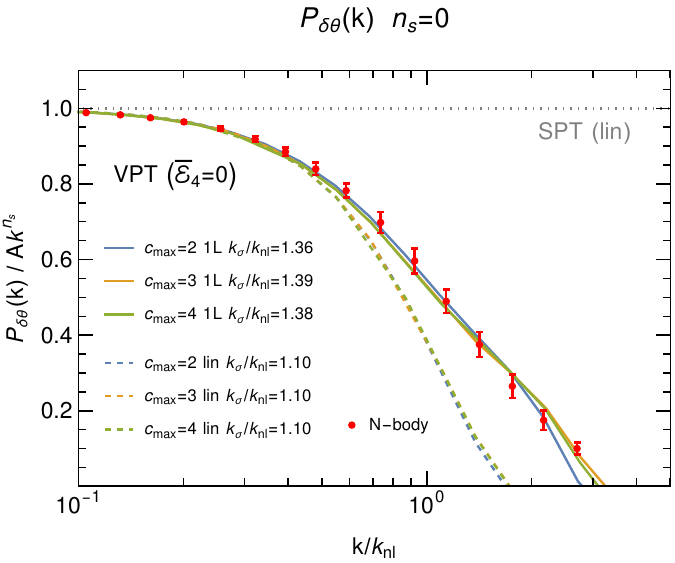}
  \\[1.5ex]
  \includegraphics[width=0.48\textwidth]{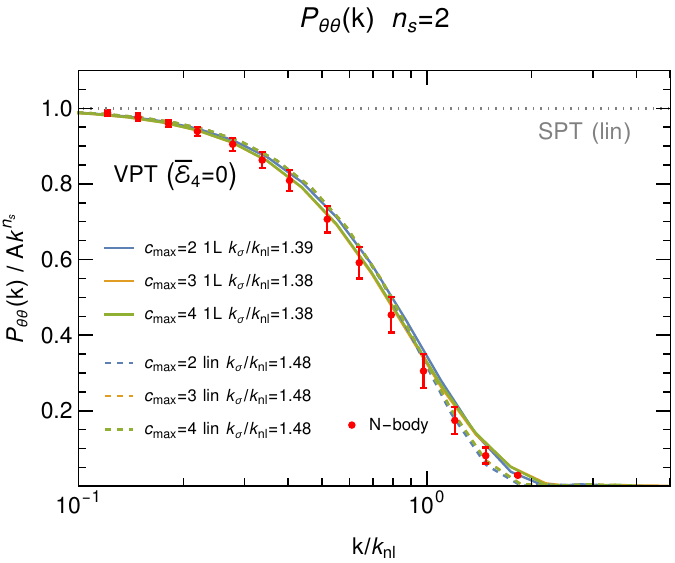}
  \includegraphics[width=0.48\textwidth]{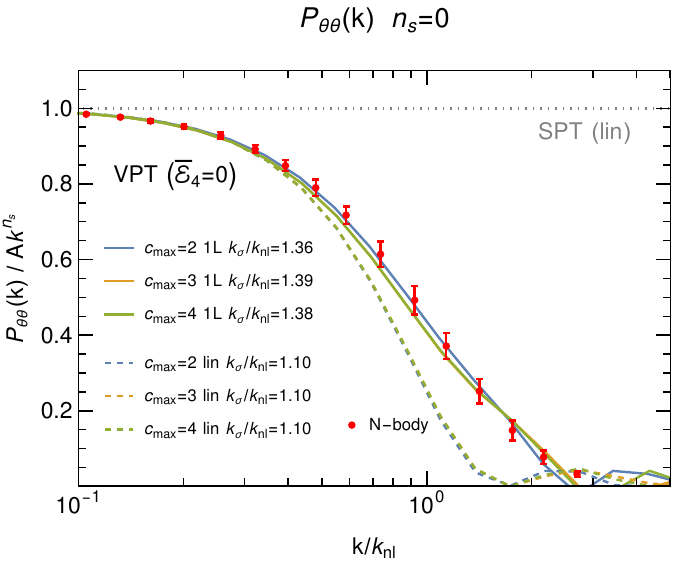}
  \end{center}
  \caption{\label{fig:Pdttt_c234}
  Cross- and auto power spectra $P_{\delta\theta}(k)$ (upper row) and $P_{\theta\theta}(k)$ (lower row) for $n_s=2$ (left) and $n_s=0$ (right).
  Each panel shows the dependence of \vpt{} predictions on the truncation order $c_\text{max}=2,3,4$
  of the Vlasov-Poisson cumulant hierarchy, at linear (dashed) and one-loop (solid) order in perturbation theory, respectively.
  $N$-body results are shown with red error bars, see~\cite{cumPT2} for details.
  We stress that these \vpt{} results
  do not involve any free parameters, but employ the respective values of $k_\sigma/k_\text{nl}$ that are already fixed via $P_{\delta\delta}$ (see Fig.~\ref{fig:Pdd_c234}).
  }
\end{figure*}

The matter power spectrum $P_{\delta\delta}(k)$ in \vpt{} is shown in Fig.~\ref{fig:Pdd_c234},
for $n_s=-1,0,1,2$ in the four panels, respectively. Each panel shows results when truncating the
cumulant hierarchy at orders $c_\text{max}=2,3,4$. Solid lines show the one-loop result, and
dashed lines the linear approximation within \vpt{}. The power spectrum is normalized to the linear input power
spectrum, and plotted against the ratio $k/k_\text{nl}$, such that the result is time-independent due to
scaling symmetry. Red symbols show $N$-body simulation results. As discussed above, the ratio $k_\sigma/k_\text{nl}$
is treated as a free parameter, determined by fitting the \vpt{} prediction for $P_{\delta\delta}(k)$ to the $N$-body result
up to $k_\text{max}=0.5k_\text{nl}$. The corresponding best-fit values are also indicated in the legend in Fig.~\ref{fig:Pdd_c234}.

We observe a rather mild dependence of the \vpt{} result for $P_{\delta\delta}(k)$ on the truncation order $c_\text{max}$ for scales $k\lesssim {\cal O}(k_\text{nl})$.
The results for $c_\text{max}=3$ and $4$ are almost on top of each other, with a small shift compared to $c_\text{max}=2$. Also the best-fit values for $k_\sigma/k_\text{nl}$
are rather stable against variation of $c_\text{max}$, and in good agreement with those obtained in~\cite{cumPT2} for each $n_s$. The one-loop approximation is slightly more sensitive to the truncation compared to the linear result. In Fig.~\ref{fig:Pdd_e4} we also show the dependence on the average value $\bar{\cal E}_4$ of the fourth cumulant, for $c_\text{max}=4$. In this case, there is a stronger dependence on the value of $\bar{\cal E}_4$ for one-loop corrections {\em at fixed $k_\sigma$}, which however can be absorbed into a small but noticeable shift in $k_\sigma$ (note the labels in Fig.~\ref{fig:Pdd_e4}). Once this is done through our standard procedure that fits for $k_\sigma/k_\text{nl}$, there is as seen in Fig.~\ref{fig:Pdd_e4} a  residual small dependence for $k\lesssim {\cal O}(k_\text{nl})$, with a slightly larger shift between $\bar{\cal E}_4=-0.5$ and $0$, compared to the difference to $\bar{\cal E}_4=0.5$.

We stress that for the very blue spectral indices considered here, the UV screening captured by \vpt{} is instrumental for obtaining finite one-loop results. Thus, the (in-)sensitivity to the truncation is a finding that could not have a priori be expected. We find that the inclusion of velocity dispersion, {\it i.e.} the step of going from SPT to \vpt{} with $c_\text{max}=2$, captures already the essential differences between SPT and \vpt{}. A reason for this finding might be that the screening is captured for all $c_\text{max}\geq 2$ and that the scaling of the
kernel in the configuration of Eq.~\eqref{eq:F23scaling} is independent of the truncation order. In addition, when expanding \vpt{} in the limit $\epsilon\to 0$, perturbations of higher cumulants are suppressed by higher powers of $\epsilon$, {\it e.g.} ${\cal O}(\epsilon)$ for vorticity and perturbations of the velocity dispersion, ${\cal O}(\epsilon^2)$ for third and fourth cumulants, and so on (see Sec.~VII in~\cite{cumPT}). Even though \emph{no} expansion in $\epsilon$ is perfomed when numerically computing \vpt{} kernels entering the power spectrum, this argument still indicates that the impact of perturbations of higher cumulants on the density field becomes suppressed on sufficiently large scales.

Overall, we confirm a good agreement between \vpt{} and $N$-body results for $P_{\delta\delta}(k)$ on weakly non-linear scales, even for initial power spectra for which SPT is UV divergent, in accord with~\cite{cumPT2}. Interestingly, we find only a mild sensitivity of \vpt{} to the truncation order of the cumulant hierarchy for $c_\text{max}\geq 2$.

\subsection{Velocity and cross power spectra}

\begin{figure*}[t]
  \begin{center}
  \includegraphics[width=0.48\textwidth]{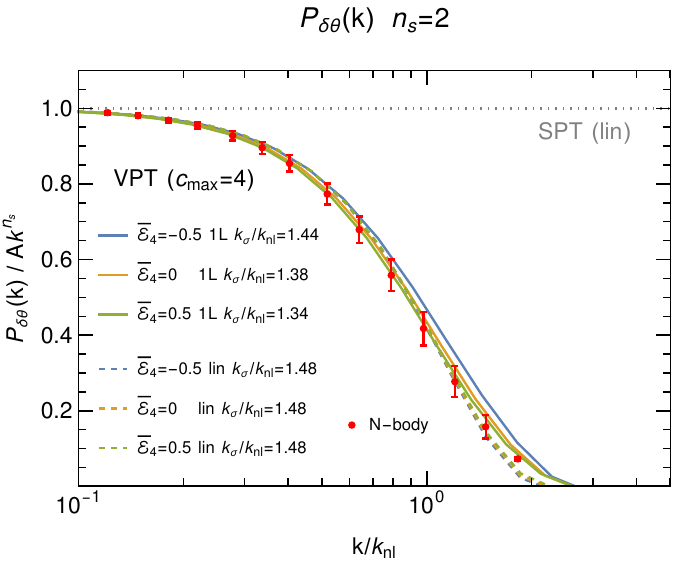}
  \includegraphics[width=0.48\textwidth]{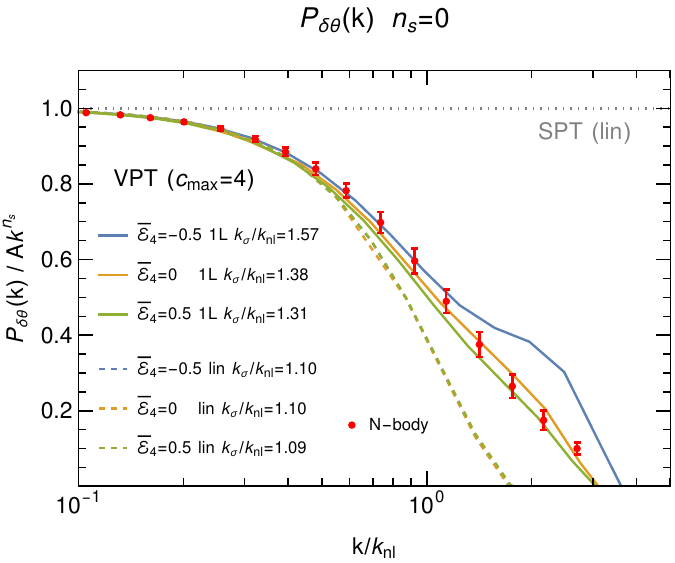}
  \\[1.5ex]
  \includegraphics[width=0.48\textwidth]{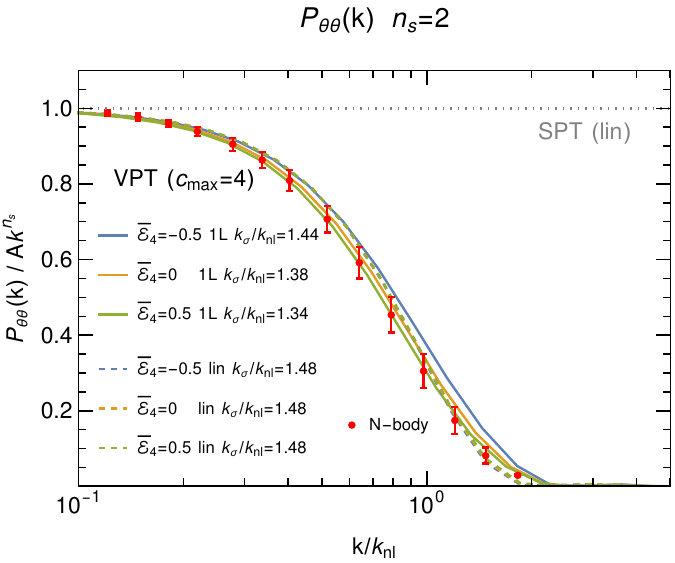}
  \includegraphics[width=0.48\textwidth]{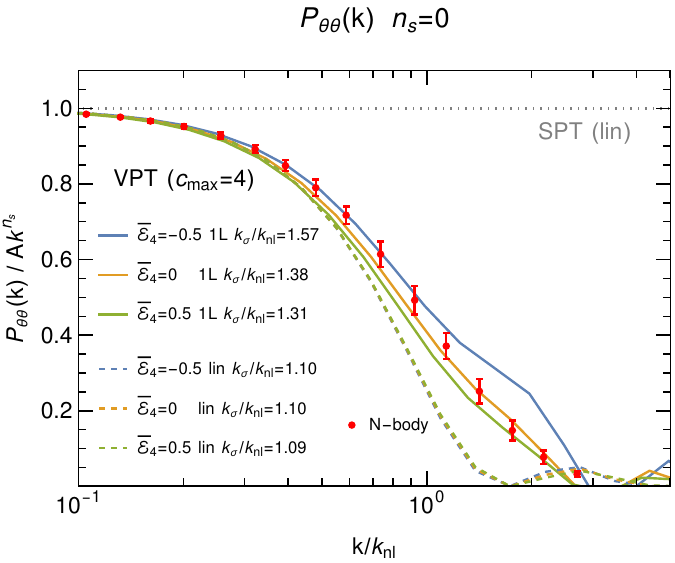}
  \end{center}
  \caption{\label{fig:Pdttt_e4}
  As Fig.~\ref{fig:Pdttt_c234}, but showing the dependence of $P_{\delta\theta}(k)$ and $P_{\theta\theta}(k)$ on the average value $\bar{\cal E}_4$ of the fourth cumulant, for $c_\text{max}=4$.
  The \vpt{} results do not contain free parameters, but employ the values of $k_\sigma/k_\text{nl}$ fixed via $P_{\delta\delta}$ (see Fig.~\ref{fig:Pdd_e4}).
  }
\end{figure*}

\begin{figure}[t]
  \begin{center}
  \includegraphics[width=0.48\textwidth]{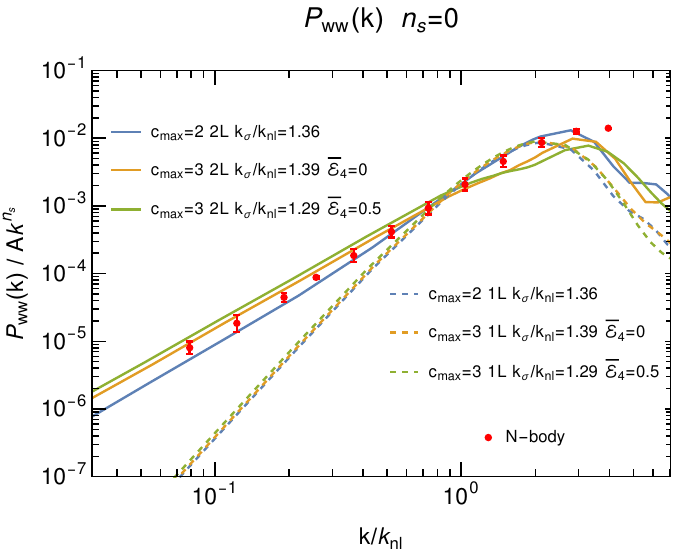}
  \end{center}
  \caption{\label{fig:Pww}
  Vorticity power spectrum $P_{w_iw_i}(k)$ within \vpt{}, in one-loop (dashed) and two-loop (solid) approximation for $n_s=0$. We show \vpt{} results for $c_\text{max}=2$ and
  $c_\text{max}=3$ as well as $\bar{\cal E}_4=0,0.5$ for the latter. \vpt{} results feature no free parameters, and we fix $k_\sigma/k_\text{nl}$ determined from $P_{\delta\delta}$ (see Figs.~\ref{fig:Pdd_c234},~\ref{fig:Pdd_e4}). $N$-body results are taken from~\cite{cumPT2}.
  }
\end{figure}

We now turn to power spectra involving the velocity field, including the velocity divergence $\theta$ as well
as the vorticity $w_i$. The generation of the latter is not captured in SPT, while $\theta$ auto- and cross power spectra
would be UV divergent in SPT for the spectral indices we consider here, similarly as for the density spectrum. Within \vpt{}, all
power spectra shown here are UV finite, due to the screening mechanism captured by \vpt{}. We stress that those results correspond
to predictions of \vpt{} without any free parameters. Specifically, we use the fixed value of the ratio $k_\sigma/k_\text{nl}$ determined from
the density power spectrum $P_{\delta\delta}(k)$ as described above when computing \vpt{} results for velocity spectra.

In Fig.~\ref{fig:Pdttt_c234} we show results for $P_{\delta\theta}(k)$ (upper row) and $P_{\theta\theta}(k)$ (lower row), for $n_s=2$ (left column)
and $n_s=0$ (right column), respectively. In each case, linear (dashed) and one-loop (solid) \vpt{} results are shown, when using a truncation of the
cumulant hierarchy at orders $c_\text{max}=2,3,4$. As for the density power spectrum, the dependence on the truncation is rather mild, and almost invisible in the figure for $c_\text{max}=3$ and $4$. This is remarkable, since the velocity field, being a first-order cumulant, could have potentially been more sensitive to the impact of higher cumulants compared to the density contrast.
The dependence on the average value $\bar{\cal E}_4$ of the fourth cumulant is shown in Fig.~\ref{fig:Pdttt_e4}, and is found to be slightly more pronounced as compared to the density field (see Fig.~\ref{fig:Pdd_e4}). We observe a good agreement of \vpt{} with $N$-body results on weakly non-linear scales for both $P_{\delta\theta}(k)$ and $P_{\theta\theta}(k)$ (see~\cite{cumPT2} for details on the velocity reconstruction).

Finally, we also consider the generation of \emph{vorticity}, an effect beyond the ideal fluid aproximation underlying SPT, see {\it e.g.}~\cite{PicBer9903,Pueblas:2008uv,Cusin:2016zvu,Jelic-Cizmek:2018gdp,cumPT2}. We compute the vorticity power spectrum $P_{w_iw_i}(k)$
within \vpt{} and compare to $N$-body results, see Fig.~\ref{fig:Pww}. In particular, we consider the one- and also two-loop approximation,
since only the latter captures the leading behaviour $P_{w_iw_i}(k)\propto k^2$ in the limit $k\to 0$~\cite{cumPT2}. We note that, in this work, we obtain $P_{w_iw_i}(k)$
for $c_\text{max}=3$  at two-loop order in perturbation theory including \emph{all} scalar and vector perturbation modes up to the third cumulant order as well as \emph{all} vertices $\gamma_{abc}$ among them (as was also the case for the one-loop results discussed above). For comparison, the two-loop results for $P_{w_iw_i}(k)$ presented in~\cite{cumPT2} were limited to scalar modes of the third cumulant, and did not include vertices of third-cumulant modes beyond scalar ones.
For the truncation $c_\text{max}=2$, on the other hand, the results shown here are based on the identical set of modes and vertices as in~\cite{cumPT2}. 

Fig.~\ref{fig:Pww} shows the one- (dashed) and two-loop (solid) vorticity power spectrum, for $c_\text{max}=2$ as well as for $c_\text{max}=3$ and two choices of the average $\bar{\cal E}_4$ of the fourth cumulant for illustration. The slope follows the scaling $P_{w_iw_i}(k)\propto k^2$ obtained from the $N$-body result at low $k$, requiring the inclusion of the contribution featuring two third-order kernels arising at two-loop order, as first noticed in~\cite{cumPT2}. Despite of the inclusion of additional modes and vertices, our results for $c_\text{max}=3$ are comparable to those in~\cite{cumPT2}. 

We find that the dependence of $P_{w_iw_i}(k)$ on the truncation order $c_\text{max}$ as well as on $\bar{\cal E}_4$ is significantly more pronounced than for $P_{\delta\delta}, P_{\delta\theta}$ and $P_{\theta\theta}$. This is expected, since vorticity generation is intimately linked to the small-scale shell crossing dynamics that is responsible for generating the average cumulants and consequently also the cumulant perturbations. In principle, the vorticity power spectrum could even be used to determine $\bar{\cal E}_4$ by fitting \vpt{} to $N$-body results at low values of $k$. Nevertheless, this is impractical, since it requires going to two-loop order on the \vpt{} side, and since extracting the vorticity field at low $k$ from $N$-body simulations with high accuracy is challenging.

Overall, we confirm the good agreement of \vpt{} predictions with $N$-body results for power spectra involving the velocity fields observed in~\cite{cumPT2}. In addition, we find only a mild dependence on the truncation order for $P_{\delta\theta}$ and $P_{\theta\theta}$, while the vorticity power spectrum $P_{w_iw_i}$ is more sensitive to $c_\text{max}$ and $\bar{\cal E}_4$.

\subsection{Bispectrum}

\begin{figure*}[t]
  \begin{center}
  \includegraphics[width=0.48\textwidth]{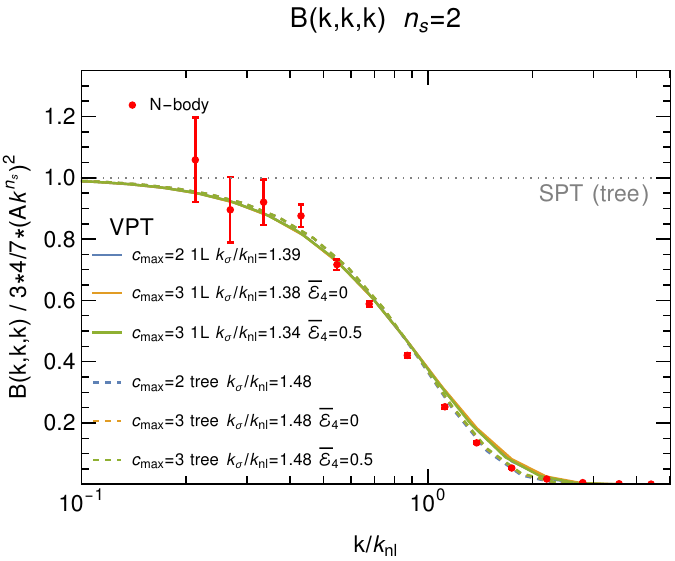}
  \includegraphics[width=0.48\textwidth]{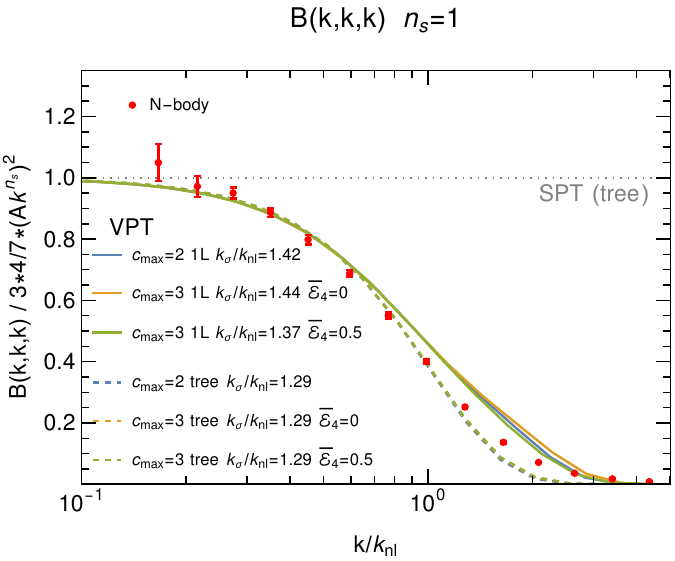}
  \\[1.5ex]
  \includegraphics[width=0.48\textwidth]{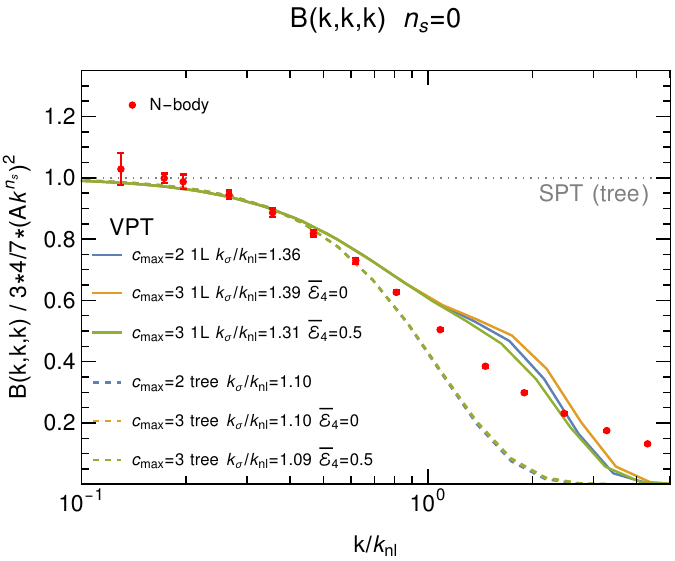}
  \includegraphics[width=0.48\textwidth]{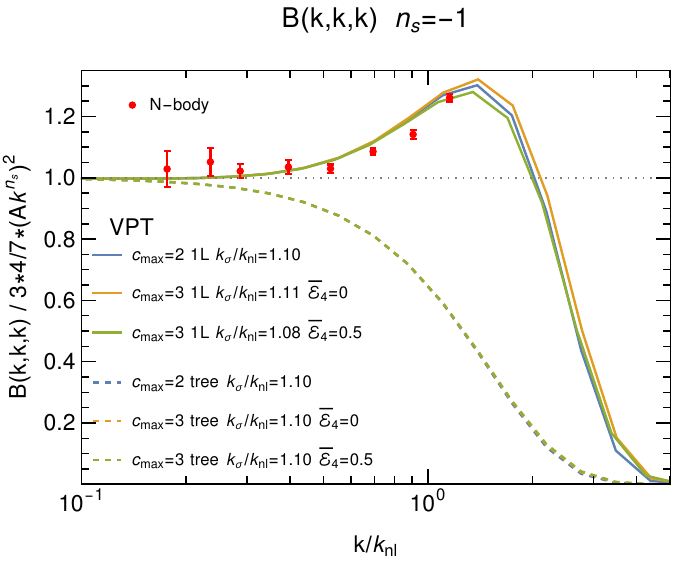}
  \end{center}
  \caption{\label{fig:Bddd}
    Tree- (dashed) and one-loop (solid) bispectrum $B(k,k,k)$ in the equilateral configuration in \vpt{}, for $c_\text{max}=2$ as well as $c_\text{max}=3$ and $\bar{\cal E}_4=0$ and $0.5$, respectively. Also shown are $N$-body results~\cite{cumPT2}. The vertical axis is normalized to the tree-level SPT bispectrum given by $3\times2\times 2/7\times P_0(k)^2$. \vpt{} results have no free parameters, as we fix $k_\sigma/k_\text{nl}$ to those obtained from $P_{\delta\delta}$.
  }
\end{figure*}

The bispectrum can be computed in \vpt{} in close analogy to SPT, by replacing the
common SPT kernels $F_n^\text{SPT}$ for the density contrast by the corresponding \vpt{} kernels $F_n$ from Eq.~\eqref{eq:Fn}. 
Noting that also the linear kernel ($n=1$) is non-trivial, the lowest order contribution (``tree'') reads
\bea
  B^\text{tree}(k_1,k_2,k_3,\eta) &=& 2e^{4\eta}F_2(\bm k_1,\bm k_2,\eta)F_1(k_1,\eta)\nn\\
  && {} \times F_1(k_2,\eta)P_0(k_1)P_0(k_2)\nn\\
  && {} + 2 \ \text{permutations}\,.
\eea
The one-loop correction to the bispectrum within \vpt{}
features the usual loop integrals involving kernels up to $F_4$, see {\it e.g.} Eq.~(125) in~\cite{cumPT2}.
Notably, the UV screening of the $F_n$ kernels renders also the one-loop bispectrum finite even for the
blue spectral indices $n_s\geq -1$ considered here, for which the SPT result would be UV divergent, analogously to the power
spectrum. For example, the contribution involving the $F_4$ kernel (the ``$B_{411}$'' term) contains a loop integral of the form
\be\label{eq:B411}
  \int d^3q \, F_4(\bm k_1,\bm k_2,\bm q,-\bm q,\eta)\, P_0(q)\,.
\ee
The scaling of the kernel in the limit $|\bm k_i|\ll k_\sigma\ll q$ can be determined similarly as in Eq.~\eqref{eq:Fnscaling}, giving
\be\label{eq:F4scaling}
  F_4(\bm k_1,\bm k_2,\bm q,-\bm q,\eta) \sim \frac{k^2}{q^2} \left(\frac{k_\sigma}{q}\right)^{4/\alpha}\,,
\ee
where $k$ is parametrically of order of the $k_i$. From this, we find a scaling of the loop integrand that is parametrically
identical to that of the $P_{13}$ contribution Eq.~\eqref{eq:P13UVscaling} to the power spectrum. In particular, $B_{411}$ is UV convergent
for any value of $n_s$. A similar analysis shows that, in \vpt{}, the two contributions to $B_{321}$ are UV finite for either all $n_s$ or $n_s<4$, respectively,
and $B_{222}$ for $n_s<5$, using Eq.~\eqref{eq:F23scaling} as well as $F_3(\bm k_1,\bm q,-\bm q+\bm k_2)\sim k^2/q^2(k_\sigma/q)^{2/\alpha}$.
Overall, this means the one-loop bispectrum in \vpt{} is absolutely convergent for spectral indices $-3<n_s<4$, {\it i.e.} within the full range that
is theoretically compatible with hierarchical structure formation as discussed in Sec.~\ref{sec:impact} above.
In SPT this is only the case for $-3<n_s<-1$ at one-loop order.

We show \vpt{} results for the tree-level (dashed) and one-loop (solid) bispectrum in the equilateral configuration $k_1=k_2=k_3\equiv k$ in Fig.~\ref{fig:Bddd}.
The various lines in each panel correspond to the truncations $c_\text{max}=2$ and $c_\text{max}=3$, respectively, as well as two values of $\bar{\cal E}_4$ for
the latter. The value of $k_\sigma/k_\text{nl}$ is fixed by the one determined from $P_{\delta\delta}$, {\it i.e.} the \vpt{} result for the bispectrum
does not feature any free parameters. Also shown are $N$-body measurements from~\cite{cumPT2}, finding good agreement with \vpt{} below the non-linear scale.
Remarkably, the bispectrum is highly insensitive to the truncation order $c_\text{max}$ as well as the average value $\bar{\cal E}_4$ of the fourth cumulant.

\section{Comparison to effective theory approach}
\label{sec:eft}

In this section we first discuss the relation of \vpt{} to the effective field theory (EFT) approach,
and then compare \vpt{} predictions for $P_{\delta\delta}$ at one-loop order with those obtained from
the EFT approach for scale-free cosmologies in~\cite{Pajer:2013jj}.

Let us first very briefly recall the general idea and setup of the EFT framework.
It provides a systematic approach for complementing SPT predictions by an (in principle infinite) series of correction terms that
encapsulate the impact of small-scale dynamics on the evolution of the density and velocity fields on large scales~\cite{BauNicSen1207}. These EFT corrections are constrained only by the
symmetries of the theory, involving generalized Galilean symmetry, mass and momentum conservation, the equivalence principle, causality, as well as rotational and shift symmetries~\cite{MerPaj1403}.
Specifically, the most general EFT that follows from these principles can be expressed in terms of coarse-grained density and velocity fields that obey the continuity and Euler equations
Eq.~\eqref{eq:continuityandEuler}, with the velocity dispersion tensor being replaced by an effective stress tensor~\cite{Mirbabayi:2015,2016JCAP...05..063A} $\sigma_{ij}(\tau,\bm x)\mapsto \sigma_{ij}^\text{eff}(\tau,\bm x)$ given in general
by (see {\it e.g.}~\cite{Baldauf:2021zlt})
\bea\label{eq:sigmaeff}
  \sigma_{ij}^\text{eff}(\tau,\bm x) &=& \sum_{\cal O} \int^\tau d\tau_1 \cdots \int^\tau d\tau_{n_{\cal O}} \, c_{\cal O}(\tau;\tau_1,\dots,\tau_{n_{\cal O}})\nn\\
  && {} \times {\cal O}_{ij}(\tau,\bm x;\tau_1,\dots,\tau_{n_{\cal O}})\,.
\eea
Here $c_{\cal O}$ are (a priori unknown) ``Wilson coefficients'' (also referred to as ``counterterms'' within the literature related to structure formation) that parameterize the most general
impact of small-scale modes onto large scales, and ${\cal O}_{ij}$ are ``operators''.\footnote{In addition, when considering products of coarse-grained density or velocity fields (such as when computing power spectra), the EFT construction requires to include ``contact
terms'' in position space, that give rise to noise terms in Fourier space, see {\it e.g.}~\cite{Desjacques:2018}.}
 The EFT construction requires to consider the most general structure allowed by the symmetries of the system,
being non-local in time for non-relativistic theories. Specifically, the operators ${\cal O}_{ij}$ are themselves given by a product of $n_{\cal O}\geq 1$ Galilean invariant ``building blocks'' ${\cal O}^{(a)}(\tau,\bm x)$,
that are in turn given by (spatial gradients of) either $\nabla_k\nabla_l\Phi(\tau,\bm x)$ or $\nabla_k v_l(\tau,\bm x)$, and
\be\label{eq:Oij}
  {\cal O}_{ij}(\tau,\bm x;\tau_1,\dots,\tau_{n_{\cal O}}) = \prod_{a=1}^{n_{\cal O}} \, {\cal O}^{(a)}(\tau_a,\bm x_\text{fl}(\tau_a;\tau,\bm x))\,.
\ee
Here each building block ${\cal O}^{(a)}$ is evaluated
at a separate time argument $\tau_a$\footnote{Note that the non-locality in time involves in general an integration over {\it multiple} time arguments $\tau_1,\dots,\tau_{n_{\cal O}}$. A similar EFT structure is well-known {\it e.g.} in soft-collinear effective theory~\cite{Bauer:2000yr,Beneke:2002ph} within collider physics, where the non-locality occurs along the respective light-cone directions in the description of multi-jet events, see {\it e.g.}~\cite{Beneke:2017ztn}.} and a shifted spatial location $\bm x_\text{fl}(\tau_a;\tau,\bm x)$ that corresponds to the Lagrangian trajectory of a fluid element that is located at position $\bm x$ at time $\tau$.
Each distinct possibility to contract the spatial indices of the ${\cal O}^{(a)}$ in the product in Eq.~\eqref{eq:Oij}, and to select two indices $i$ and $j$ that are left open and correspond to the spatial $3\times 3$ matrix structure
of the effective stress tensor Eq.~\eqref{eq:sigmaeff}, is counted as a separate operator ${\cal O}_{ij}$. 

All summands in Eq.~\eqref{eq:sigmaeff}
can be systematically classified by a double series expansion in terms of 
\begin{itemize}
\item the number of fields (perturbative expansion in powers of the initial density contrast $\delta_{k0}$), as well as 
\item the number of spatial derivative operators (gradient expansion in
powers of $k/k_\text{nl}$).
\end{itemize}
In practice, this double series can be terminated at a given, desired order in the perturbative and gradient expansions, respectively. For example, the EFT corrections to the density field
at first~\cite{Carrasco:2012} and second~\cite{BalMerMir1505} order in $\delta_{k0}$ and at leading order ($\propto k^2/k_\text{nl}^2$) in the gradient expansion can be written as corrections to the SPT kernels as\footnote{See also Eq.~(56) and footnote 6 in~\cite{cumPT2} for
a relation to the basis used in~\cite{Baldauf:2021zlt}, that is in turn related to that from~\cite{BalMerMir1505}. See also~\cite{AngForSch1510}.}
\bea\label{eq:F12EFT}
  F_1^\text{EFT}(k,\eta)&=&1-c_s^2(\eta)k^2+\dots\,,\nn\\
  F_2^\text{EFT}(\bm k_1,\bm k_2,\eta)&=&F_2^{\rm SPT}(\bm k_1,\bm k_2) \nn\\
  && {} -\sum_{j=a,b,c,d} c_j(\eta) \times k^2\,\Delta_j(\bm k_1,\bm k_2)  \nn\\
  && {} +\dots\,,
\eea
where $c_s^2$ and $c_j$ are related to Wilson coefficients (``counterterms'')  $c_{\cal O}$ for ``operators'' ${\cal O}$ given by $\nabla_i\nabla_j\Phi$, $\nabla_i\nabla_j\Phi\times \nabla^2\Phi$, $\delta^K_{ij}\nabla_k\nabla_l\Phi\times \nabla_k\nabla_l\Phi$, 
$\nabla_i\nabla_k\Phi\times \nabla_j\nabla_k\Phi$, and the shape functions $\Delta_j$ are obtained from evaluating (linear combinations of) those operators at second order in perturbation theory (here $k^2\equiv (\bm k_1+\bm k_2)^2$),
\bea
  \Delta_a(\bm k_1,\bm k_2) &=& (\bm k_1^2+\bm k_2^2)F_2^{\rm SPT}(\bm k_1,\bm k_2)/k^2\,,\nn\\
  \Delta_b(\bm k_1,\bm k_2) &=& F_2^{\rm SPT}(\bm k_1,\bm k_2)\,,\nn\\
  \Delta_c(\bm k_1,\bm k_2) &=& (\bm k_1^2+\bm k_2^2)K(\bm k_1,\bm k_2)/k^2\,,\nn\\
  \Delta_d(\bm k_1,\bm k_2) &=& K(\bm k_1,\bm k_2)\equiv (\bm k_1\cdot\bm k_2)^2/(\bm k_1^2\bm k_2^2)-1\,.\nn\\
\eea
The ellipsis in Eq.~\eqref{eq:F12EFT} stand for corrections at higher order in the gradient expansion, {\it i.e.} of order $k^4$ and higher, which
would correspond to similar operators as given above, but with additional spatial derivatives.
On the other hand, operators involving a product of three or more building blocks need to be included when considering $F_3$ or higher-order kernels. These two
directions (more building blocks or more gradients) reflect the double expansion in powers of $\delta_{k0}$ as well as in $k/k_\text{nl}$ alluded to above.
All of those EFT corrections are contained in the summation in Eq.~\eqref{eq:sigmaeff}.

The EFT is designed to be able to fit the large-scale density field resulting from {\it any} conceivable dynamics on small scales. This large flexibility is reflected in the freedom to choose the Wilson coefficients (``counterterms'') $c_{\cal O}$.
A given choice corresponds to a particular UV dynamics, leading to these specific Wilson coefficients. A common use of EFT is to marginalize over the Wilson coefficients when fitting perturbative predictions to data or simulations. When including EFT corrections up to a sufficiently high order, this approach is most conservative,
since it is agnostic of the actual dynamics. On the other hand, leaving Wilson coefficients completely free leads to a loss of predictivity and therefore degrades the ability to measure cosmological parameters.

\subsection{Matching of \vpt{} to the EFT: Taylor expansion for $k^2\epsilon\ll 1$}

The Vlasov-Poisson dynamics underlying \vpt{} can be viewed as a particular theory model, based on the assumption of collisionless dynamics. By construction, the EFT has to be able to match the \vpt{} prediction for the density field for some particular choice of the Wilson coefficients (``counterterms'') $c_{\cal O}$, when including
all possible EFT corrections and taking the most general structure of the effective stress tensor Eq.~\eqref{eq:sigmaeff} within the EFT into account. When truncating the EFT expansion at a given order in the gradient expansion, this ``matching'' holds only for sufficiently large scales $k\to 0$. In practice, for the density field, matching \vpt{} to the EFT means that \vpt{} predicts a particular size of {\it e.g.} the
coefficients $c_s^2$ and $c_j$ in Eq.~\eqref{eq:F12EFT}, and analogously for higher orders $n$ and higher gradient corrections, respectively. 

The \vpt{} prediction for these ``counterterms'' can be obtained systematically for each order in the perturbative expansion $n$ by Taylor-expanding the \vpt{} kernels $F_n(\bm k_1,\dots,\bm k_n,\eta)$ in the limit $\bm k_i\to 0$ and matching them to the EFT kernels order by order in the gradient expansion. The {\it zeroth} order in the Taylor expansion is given by the usual SPT kernels at any $n$. Thus, non-trivial EFT matching starts at subleading power in the gradient expansion for the density field.
Due to momentum conservation, the leading correction terms in the gradient expansion arise at order $k^2$. 

How does the Vlasov-Poisson prediction for the ``counterterms'' look like?
Within \vpt{}, the scale associated to corrections to SPT enters via the average cumulants.
Thus, the dimensionless parameter related to the gradient expansion is a Taylor expansion in powers of the ratio $k/k_\sigma$. The leading gradient corrections to the $F_n$ kernels arise at order $k^2/k_\sigma^2=k^2\epsilon$, {\it i.e.}
when formally expanding \vpt{} in the limit $\epsilon\to 0$ and keeping only the linear terms in that expansion. Analytical results for $F_1$ and $F_2$ within \vpt{} have been computed in~\cite{cumPT2} up to linear order in $\epsilon$.
These leading correction terms have a
structure of the form given in Eq.~\eqref{eq:F12EFT}, as expected and as already discussed in Sec.~II$\,$C3 of~\cite{cumPT2}. Matching their prefactors yields the \vpt{} prediction for the leading-order ``counterterms'' $c_s^2$ (for $F_1$) and $c_{a},\dots,c_d$ (for $F_2$), respectively.
For $n=1$, quoting the result from Eq.~(41) in~\cite{cumPT2} gives
\bea\label{eq:cssq}
  c_s^2(\eta)\big|_{\scriptsize\vpt{}} &=& \frac25 \int^\eta d\eta' (1-e^{5(\eta'-\eta)/2})\nn\\
  && {} \times \left(\epsilon(\eta') + 2\int^{\eta'}d\eta'' e^{2(\eta''-\eta')}\epsilon(\eta'')\right)\,.\nn\\
\eea
The sensitivity of $c_s^2(\eta)$ to the past evolution history of the average dispersion $\epsilon(\eta)$ reflects
the non-locality in time within the EFT description.
For $n=2$ the EFT matching yields coefficients $c_j(\eta)=\beta_j^\delta(\eta)$ for $j=a,b,c,d$, with \vpt{} results for the latter
four functions given in Eq.~(58) in~\cite{cumPT2}. These coefficient functions are thus the \vpt{} prediction for
the four ``bispectrum counterterms''.

One may wonder whether these \vpt{} predictions for the EFT ``counterterms'' depend on the truncation order of the cumulant
hierarchy $c_\text{max}$. From the structure of the evolution equation Eq.~\eqref{eq:eomC} for the CGF, one sees that the equation
of motion of the $n$th cumulant contains at most the $(n+1)$th cumulant for $n\geq 1$. Thus the Euler equation does not explicitly contain third-order
cumulant perturbations. Instead, those impact the second-cumulant modes, which in turn enter the Euler equation. Assuming an initially cold system
({\it i.e.} vanishing initial conditions for higher-order cumulant perturbations) as well as a small non-zero average dispersion $\epsilon$, third and
fourth-order cumulants are generated at order $\epsilon^2$ when expanding \vpt{} in the limit $\epsilon\to 0$ (see also Sec.~VII\,C in~\cite{cumPT}).
Even higher cumulants scale as $\epsilon^3$ or more. This implies that the corrections to the $F_n$ kernels at linear order in $\epsilon$
are insensitive to third- and higher cumulants. Thus, the matching coefficients obtained from \vpt{} at leading order in the derivative expansion
({\it i.e.} at linear order in $\epsilon$) are \emph{independent} of $c_\text{max}$ as long as $c_\text{max}\geq 2$. 

This implies that the \vpt{} prediction Eq.~\eqref{eq:cssq} for the $c_s^2$ ``counterterm'' is insensitive to the truncation of the cumulant hierarchy, and thus a genuine
prediction of the Vlasov-Poisson dynamics. The same is true for the ``bispectrum counterterms'' $c_a,\dots,c_d$ entering the leading gradient correction
to the $F_2$ kernel. If one would include the  next-to-leading power corrections of order $k^4$ in Eq.~\eqref{eq:F12EFT}, the corresponding Wilson coefficients
obtained from matching to \vpt{} are of order $\epsilon^2$, and are affected by up to fourth order contributions in the cumulant expansion.

Finally, we stress that the full \vpt{} kernels $F_n$ do {\it not} rely on any gradient expansion, as is the case in the EFT approach. This is essential for the UV screening property
of the kernels observed in \vpt{}, that captures the physically expected decoupling of UV modes when considering full Vlasov-Poisson dynamics, and reflects in the asymptotic scaling of Eq.~\eqref{eq:Fnscaling}. 

In contrast, when performing the EFT expansion, the resulting kernels formally increase for large wavenumber arguments due to the Taylor series expansion up to some finite order.
Actually, this technically leads to an increased UV sensitivity to small scales in loop computations within the EFT approach.
The resulting UV sensitivity needs to be accounted for by artificially splitting the Wilson coefficients into a physical (``finite'') part and a ``counterterm'' contribution. The latter corrects for
the unphysical (and potentially large or even divergent) UV contribution of loop integrals within the SPT/EFT approaches. 

We emphasize that the UV screening captured within full \vpt{} ensures that loop integrals are dominated by modes that are under perturbative control. A further consequence is that the \vpt{} predictions for the Wilson
coefficients, such as Eq.~\eqref{eq:cssq}, correspond only to the ``physical'' contribution. Therefore, the value of $c_s^2$ predicted by the Vlasov-Poisson dynamics
cannot be directly compared to the value of $c_s^2$ obtained when fitting EFT predictions to data or simulations, since the latter includes also the unphysical part
that corrects for the spurious UV dependence of the SPT/EFT loop contributions.

\begin{figure*}[t]
  \begin{center}
  \includegraphics[width=0.48\textwidth]{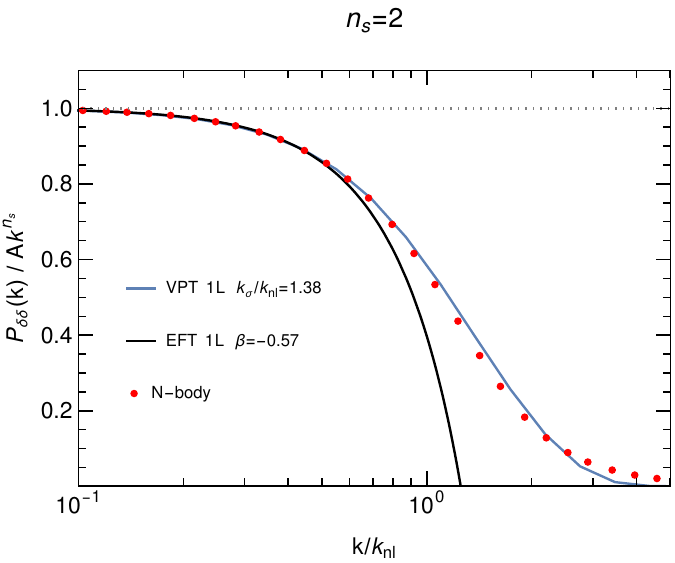}
  \includegraphics[width=0.48\textwidth]{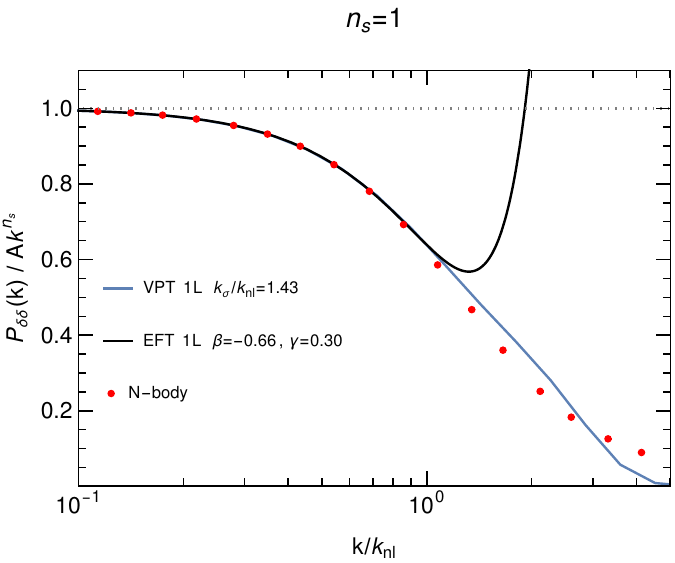}
  \\[1.5ex]
  \includegraphics[width=0.48\textwidth]{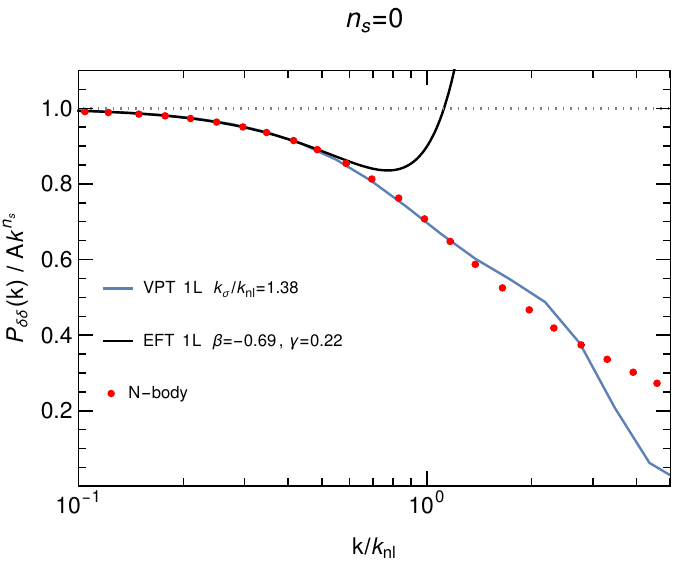}
  \includegraphics[width=0.48\textwidth]{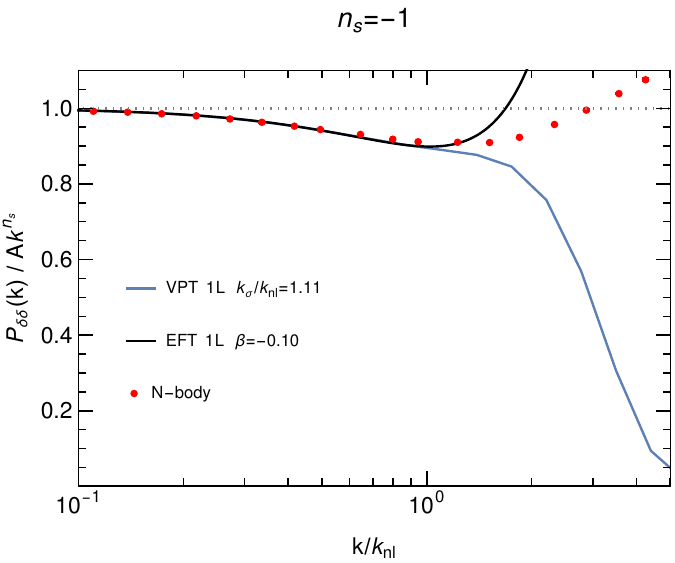}
  \end{center}
  \caption{\label{fig:eft}
 Matter power spectrum $P_{\delta\delta}$ for scale-free initial spectrum $P_0\propto k^{n_s}$, with $n_s=2,1,0,-1$, respectively. The ``standard'' one-loop EFT result including leading-order gradient
 corrections (``$c_s^2$'' as well as $k^4$ noise terms) with two free parameters  (see {\it e.g.}~\cite{Pajer:2013jj} and Eq.~\eqref{eq:EFT}) fitted to $N$-body results~\cite{cumPT2} is compared to the one-loop \vpt{} result as described
 in  Sec.~\ref{eq:mpk}. We note that for $n_s=2$ the two EFT corrections are degenerate, and noise terms are omitted for $n_s=-1$ following~\cite{Pajer:2013jj}. The \vpt{}
 result contains a single parameter $k_\sigma/k_\text{nl}$, related to the average velocity dispersion $\epsilon$, that is adjusted to the $N$-body result. The \vpt{} result is largely insensitive to the truncation order of the
 Vlasov-Poisson cumulant hierarchy  (see Fig.~\ref{fig:Pdd_c234}, we use $c_\text{max}=4$ here).}
\end{figure*}

\subsection{Comparison of $P_{\delta\delta}$}

We now compare the one-loop density power spectrum for power-law initial spectra with $n_s=-1,0,1,2$ within \vpt{} (see Sec.~\ref{eq:mpk}) with the EFT result
obtained when taking the leading EFT correction terms into account~\cite{Pajer:2013jj},
\bea\label{eq:EFT}
  P_{\delta\delta}(k,\eta) &=& P_\text{lin}^\text{SPT}(k,\eta) + P_{1L}^\text{SPT}(k,\eta)\big|_\text{dim\,reg}\nn\\
  && {} + 2c_s^2k^2P_\text{lin}^\text{SPT}(k,\eta) + c_n k^4\,,
\eea
where $c_s^2$ and $c_n$ are free ``counterterm'' parameters related to the leading gradient correction of the $F_1$ kernel in Eq.~\eqref{eq:F12EFT} and to the noise term at leading order in
the gradient expansion, respectively.
The SPT one-loop contribution is UV divergent for $n_s\geq -1$ in three spatial dimensions, but can be computed using dimensional regularization~\cite{Scoccimarro:1996se},
expanding for $d\to 3$ space dimensions, and discarding the pole terms $\propto 1/(d-3)$ that are absorbed in the EFT parameters.
The density power spectrum obtained from fitting both $c_s^2$ and $c_n$ to $N$-body results~\cite{cumPT2} is shown in Fig.~\ref{fig:eft}.
Here we use the EFT result parameterized as given in Eq.~(3.1) of~\cite{Pajer:2013jj}, with dimensionless fit
parameters $\beta\equiv 2c_s^2k_\text{nl}^2$ and $\gamma\equiv 4\pi c_nk_\text{nl}^7$. We note that for $n_s=2$, the contributions from $c_s^2$ and
$c_n$ are degenerate, and we therefore set $\gamma|_{n_s=2}=0$. We also omit $\gamma$ for $n_s=-1$ as argued in~\cite{Pajer:2013jj}.

From Fig.~\ref{fig:eft} we see that one-loop \vpt{} captures the $N$-body result to larger $k$ compared to one-loop EFT with leading gradient corrections.
This can be taken as evidence that higher-gradient contributions that are ``resummed'' when solving for the full \vpt{} kernels provide additional information.
We stress again that the UV screening captured by \vpt{} leads to finite predictions, and also ensures that only modes under perturbative control enter the loop integration.
Of course, the EFT line could be brought closer to the $N$-body result by including additional higher gradient corrections in the EFT fit.\footnote{Formally, the next higher-gradient correction $\propto k^4 P_\text{lin}^\text{SPT}(k,\eta)\propto k^{4+n_s}$ is 
more important than the two-loop contribution $\propto k^{2(3+n_s)}P_\text{lin}^\text{SPT}(k,\eta)\propto k^{6+3n_s}$ in the limit $k\ll k_\text{nl}$ for $n_s>-1$. The even higher $k^6P_\text{lin}^\text{SPT}$ ($k^8P_\text{lin}^\text{SPT}$) gradient corrections formally dominate over the two-loop for $n_s>0$ ($n_s>1$). In \vpt{}, all of these corrections are resummed with only a single free coefficient.} However, this would come at the
price of additional free parameters, and ultimately amount to a Taylor series approximation of the $N$-body result around $k=0$ when done in this way. On the other hand, the \vpt{} prediction
contains only a single parameter ($k_\sigma/k_\text{nl}$) related to the average velocity dispersion that was adjusted in Fig.~\ref{fig:eft}.

\medskip

Let us finally provide an outlook of how the advantages of \vpt{} and EFT approaches could be combined.
In particular, in practice a useful approach can be to employ full \vpt{} as a basis for perturbative predictions, and complement it with additional EFT corrections.
This would correspond to using EFT kernels as in Eq.~\eqref{eq:F12EFT}, but with the SPT kernels entering in there replaced by those in \vpt{} (including $1\mapsto F_1(k,\eta)$ in the first line). This strategy has
the advantage that \vpt{} already captures the bulk of the corrections relative to SPT, and therefore the additional ``counterterm'' corrections are
expected to be numerically much smaller. Moreover, a full tower of physically motivated higher-gradient terms is already contained in \vpt{}. In addition, this approach would combine the advantage of the EFT of being agnostic with respect to UV dynamics (capturing {\it e.g.}  corrections to \vpt{} arising from deviations from collisionless dynamics), and the advantage of \vpt{} of accounting for UV screening and being only sensitive to modes that are under perturbative control.
In this way, tighter priors on counterterms in data analysis could be possible (without leading to unwanted bias in parameter inference), enhancing the predictivity and constraining power.

\begin{figure}[t]
  \begin{center}
  \includegraphics[width=0.48\textwidth]{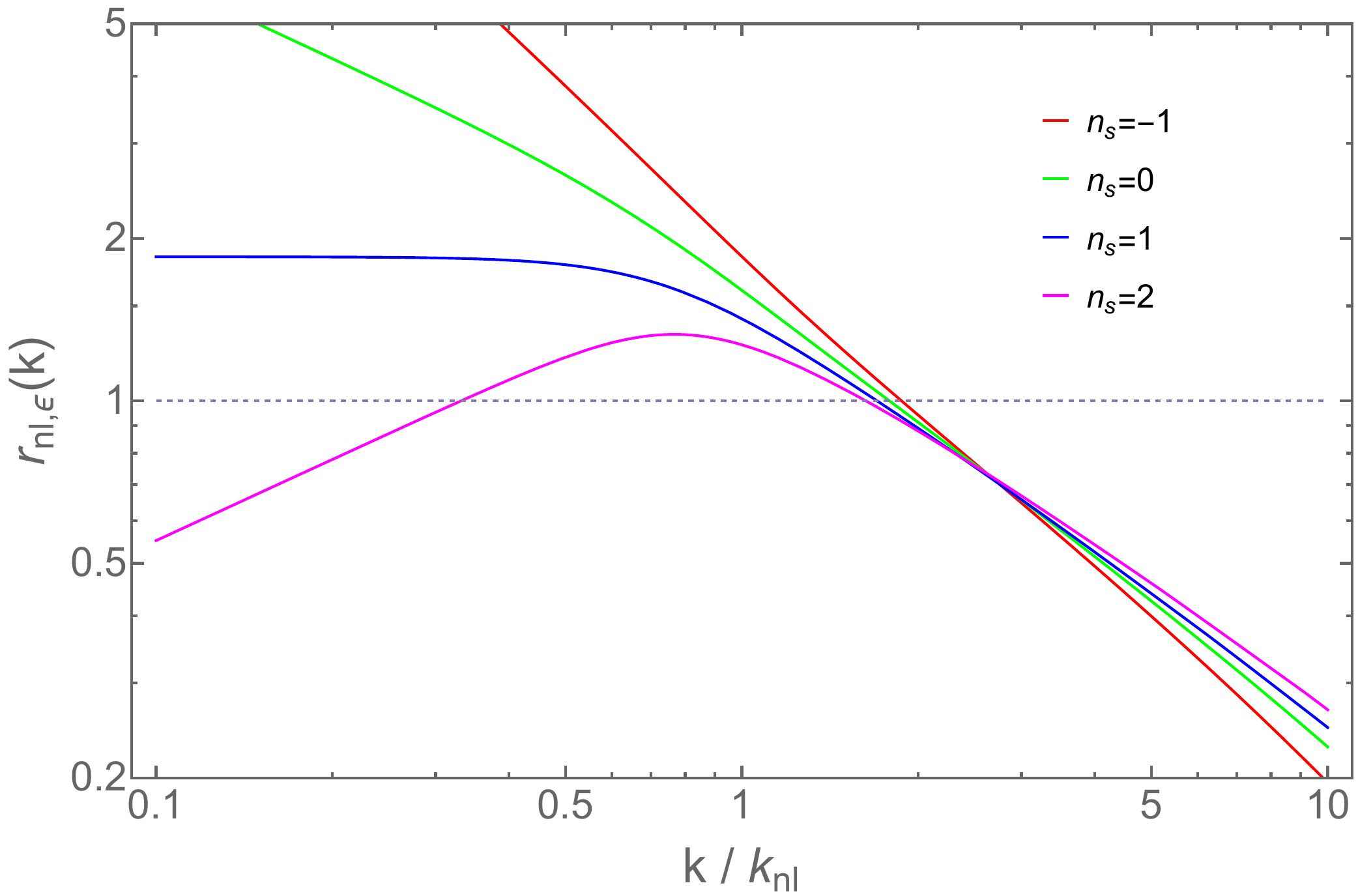}
  \end{center}
  \caption{\label{fig:rnleps}
  Ratio of non-linearity to linear dispersion $r_{{\rm nl},\epsilon}(k)$ as a function of $k/k_{nl}$ (see Eq.~\ref{rsc}) for different spectral indices, as labeled.   }
\end{figure}

\section{Non-linearity vs Dispersion}
\label{sec:NLdisp}

Having studied the robustness of \vpt{} to truncation of the hierarchy, we now step back and look at the \vpt{}  results as a whole  when compared to $N$-body simulations. One lesson that emerges from such comparison (see {\it e.g.} Figs.~\ref{fig:Pdd_c234},~\ref{fig:Pdttt_c234}, ~\ref{fig:Bddd} and~\ref{fig:eft}) is that the reach in $k$ of \vpt{} becomes larger for bluer spectral indices (more positive $n_s$), suggesting that the convergence of perturbation theory {\em improves} for initial conditions with more small-scale power, the opposite of what is expected from SPT (where dispersion is neglected altogether). In order to understand this behavior, it is then worth comparing nonlinearities (present in SPT) to dispersion effects (newly incorporated into \vpt{}). Let us consider the ratio,
\beq
r_{{\rm nl},\epsilon} = {{\rm nonlinearity}\over {\rm dispersion}} \sim  {u_j\nabla_j u_i \over \epsilon \, \nabla_i \delta} \sim {\nabla_i(u_j\nabla_j u_i )\over \epsilon \, \nabla^2 \delta}
\label{scaling}
\eeq 
which compares the nonlinearity present in SPT and the linear dispersion terms in the modified Euler equation. Assuming that  gradients of the bulk velocity $u_i$ scale as the density perturbation $\delta$, and translating the result to Fourier space, we can write
\beq
r_{{\rm nl},\epsilon}(k)  \simeq \Big( {k_\sigma\over k_{nl}}\Big)^2\, {\Delta_{nl}^{1/2}(k/k_{nl}) \over (k/ k_{nl})^2}
\label{rsc}
\eeq
where $\Delta_{nl}\equiv 4\pi k^3 P(k)$ is the dimensionless (nonlinear) power spectrum, {\it i.e.} the contribution to the variance of $\delta$ per logarithmic interval in $k$. The prefactor $k_\sigma/ k_{nl}$ is a number of order unity independent of time (obtained by comparison with simulations as discussed) and thus $r_{{\rm nl},\epsilon}$ is  solely a function of the self-similar variable $k/k_{nl}$.  

Both nonlinearity (as defined in Eq.~\ref{scaling}) and dispersion terms are subdominant to the linear terms in velocity and gravitational potential for $k\ll  k_{nl}$, so we are mainly interested in $r_{{\rm nl},\epsilon}(k)$ close to $k_{nl}$ and as we cross into the nonlinear regime. The assumptions leading to Eq.~(\ref{rsc}) are only approximate in this case, {\it e.g.} we expect bulk velocity gradients to be smaller than density perturbations $\delta$ in the nonlinear regime, but also the linear dispersion term depends more weakly (logarithmically) on $\delta$ (through the $A$ field) so some cancellation may be expected when taking the ratio. Overall, however, as a rough estimate Eq.~(\ref{rsc}) is enough for our purposes here. 

Figure~\ref{fig:rnleps} illustrates this ratio for the spectral indices $n_s=-1,0,1,2$ we have $N$-body simulation measurements for. For simplicity, we have used the nonlinear power spectrum fitting formula from~\cite{Peacock:1996ci} for $\Delta_{nl}(k/k_{nl})$ in Eq.~(\ref{rsc}) and the dispersion scale values $k_\sigma$ obtained by fitting the \vpt{} density power spectrum at two loops to our $N$-body simulations~\cite{cumPT2}. The nonlinear power spectrum fitting formula assumes stable clustering in the nonlinear regime ({\it i.e.} $\Delta_{nl} \simeq (k/ k_{nl})^{3(n_s+3)/(n_s+5)}$), which is consistent with our measurements, see~\cite{cumPT2} for details. This means that in the nonlinear regime, 
\beq
r_{{\rm nl},\epsilon}(k)  \underset{{k\gg k_{nl}}}{\sim} \Big( {k_\sigma\over k_{nl}}\Big)^2\, \Big({k\over k_{nl}}\Big)^{-{(n_s+11)\over 2(n_s+5)}} \ll 1 
\label{rscNL}
\eeq
so linear dispersion always wins over nonlinearity. In addition, we have linear and nonlinear dispersion effects from  $\delta\epsilon_{ij}$ (and its multiplication by $A$) that are probably stronger than the linear dispersion term past the nonlinear scale, {\it e.g.} in linear theory $\delta\epsilon_{ij}\sim {\cal O}(\epsilon\, \theta)$, so one might expect the corresponding result to Eq.~(\ref{rscNL}) to be even smaller.

We see from Fig.~\ref{fig:rnleps} that overall $r_{{\rm nl},\epsilon}$ is smallest for blue indices and drops below unity (dispersion wins over nonlinearity) as we cross into the nonlinear regime. This is consistent with the picture that emerges from the nonlinear power spectrum measured in simulations, where for red spectral indices the power spectrum is enhanced by loop corrections (as for CDM initial conditions) before dispersion dominates and causes a ``virial turnover" (moderation of  the enhancement) at higher $k$. Indeed, from the results in Figs.~\ref{fig:Pdd_c234},~\ref{fig:Pdttt_c234} and~\ref{fig:Bddd}, we see that as we cross past $n_s=-1$ the suppression of the power spectrum and bispectrum happens  closer to the nonlinear scale, erasing altogether the enhancement seen in red spectra, signaling that dispersion increasingly dominates over nonlinearities, as suggested by Fig.~\ref{fig:rnleps}. As a result of this, for blue spectra, dispersion incorporated in linear \vpt{} (or tree-level for the bispectrum) already  gives a reasonable approximation to the full nonlinear answer.

\section{Conclusions}
\label{sec:conclusions}

In this paper we investigated perturbative solutions to the Vlasov-Poisson equations for the phase-space distribution function $f(\tau,{\bm x},{\bm p})$ of collisionless matter coupled to gravity in the context of cosmological large-scale structure formation.
Our work is based on the framework of~\vpt{} introduced in~\cite{cumPT,cumPT2}, obtained by rewriting the Vlasov-Poisson equations in terms of a coupled hierarchy of equations for the cumulants ${\cal C}_{i_1i_2,\dots,i_c}(\tau,{\bm x})$ associated to
the moments $\int d^3p\, p_{i_1}p_{i_2}\cdots p_{i_c}\, f(\tau,{\bm x},{\bm p})$.

In this work, we develop and implement an extension that allows us to systematically study the dependence on the truncation order $c\leq c_\text{max}$ of the cumulant expansion at the non-linear level, which would otherwise be unfeasible with the approach presented in~\cite{cumPT,cumPT2}. While SPT is recovered for $c_\text{max}=1$, \vpt{} is based on truncations with $c_\text{max}\geq 2$. We present numerical solutions up to $c_\text{max}=6$ for non-linear kernels, $c_\text{max}=4$ for one-loop power spectra, and $c_\text{max}=3$
for the one-loop equilateral bispectrum.

Within \vpt{}, all cumulants are split into a homogeneous average part $\langle {\cal C}_{i_1i_2,\dots,i_c}\rangle$ and perturbations $\delta {\cal C}_{i_1i_2,\dots,i_c}(\tau,{\bm x})$. Assuming statistical isotropy, only even cumulants possess an average value. They can be parameterized by time-dependent functions ${\cal E}_2(\eta),{\cal E}_4(\eta),\dots$ for $c=2,4,\dots$, see Eq.~\eqref{eq:Eaverage}. The most relevant is ${\cal E}_2(\eta)\equiv \epsilon(\eta)\equiv 1/k_\sigma^2(\eta)$, providing the velocity dispersion scale $k_\sigma(\eta)$ as a major feature of \vpt{}. The average cumulants encapsulate non-perturbative dynamics related to shell crossing on small scales. In practice, we treat them as an a priori unknown input in \vpt{}, while the $\delta {\cal C}_{i_1i_2,\dots,i_c}(\tau,{\bm x})$ are computed perturbatively for given average values.

Our main findings are:

\begin{itemize}

\item[(i)] We employ an efficient parameterization of all cumulants $\delta {\cal C}_{i_1i_2,\dots,i_c}(\tau,{\bm x})$, taking the increasingly higher-rank tensor structure of order $c$ at the $c$th cumulant order into account by a decomposition
in a suitable tensor basis related to rotated spherical harmonic functions, see Eqs.~(\ref{eq:Ylmdecomp},\ref{eq:Clms}) and Eq.~\eqref{eq:Ylmk}.

\item[(ii)] The perturbation modes are parameterized (in Fourier space) by $\delta{\cal C}_{\ell,m,2s}(\eta,{\bm k})$, where the three indices take values $\ell=0,1,2,\dots$, $m=-\ell,-\ell+1,\dots,\ell$, $2s=0,2,4,\dots$, see Eq.~\eqref{eq:Clms}. The cumulant order is $c=\ell+2s$. In this parameterization scalar modes have $m=0$, vector modes $m=\pm 1$, tensor modes $m=\pm 2$, and so on. A given truncation takes all modes with $\ell+2s\leq c_\text{max}$ into account. We consider in addition truncations with $|m|\leq m_\text{max}$.

\item[(iii)] At the linear level, the cumulant parametrization used in this paper allows us to study the asymptotic behavior of perturbation modes at scales $k\gg k_\sigma$ for vector and tensor modes, generalizing the result for scalar modes given in~\cite{cumPT}. These stability conditions (see Section~\ref{sec:stability}) restrict the non-Gaussianity of the background distribution function through constraints on the kurtosis $\bar{\cal E}_4$, $\bar{\cal E}_6$, etc and guarantee the lack of exponentially growing modes which would otherwise spoil the decoupling of small-scale modes. We find that the stability conditions for modes with a certain $m$ and cumulants up to order $c_\text{max}$ are identical to those derived from scalar modes up to cumulant order $c_\text{max}-|m|$. This means that the stability conditions for scalar modes  are the most constraining of all.

\item[(iv)] We derive a non-linear equation of motion for the cumulant perturbation modes $\delta{\cal C}_{\ell,m,2s}(\eta,{\bm k})$ from the underlying Vlasov-Poisson equations. It becomes exact, \emph{i.e.} fully equivalent to Vlasov-Poisson, in the limit $c_\text{max},m_\text{max}\to\infty$. Collecting all modes in a large vector $\psi_a(\eta,{\bm k})$ with a multi-index $a$ labelling the indices of the various $\ell,m,2s$ modes (and including also the density contrast $\delta$), it takes the form of Eq.~\eqref{eq:eommatrixform} with a linear part characterized by the matrix $\Omega_{ab}(\eta,k)$ and a quadratic non-linearity $\gamma_{abc}({\bm p},{\bm q})$. We find explicit expressions for both, given in Eq.~\eqref{eq:Omegaab} and in Eqs.~(\ref{eq:gammaabcexplicit},\ref{eq:gammaunsymmabcexplicit}), respectively. The average cumulants ${\cal E}_2,{\cal E}_4,\dots$ enter in $\Omega_{ab}(\eta,k)$, giving the backreaction of shell-crossing on linear modes. The ``vertices'' $\gamma_{abc}({\bm p},{\bm q})$ describe mode-coupling, and capture interactions among modes with different $m$-number, leading {\it e.g.} to vorticity generation.

\item[(v)] Expanding the cumulant perturbations $\psi_a(\eta,{\bm k})$ perturbatively in the initial density contrast yields the \vpt{} kernels $F_{n,a}(\bm k_1,\dots,\bm k_n,\eta)$. They replace the usual SPT kernels in perturbative predictions.
We determine them, for a given truncation, by numerically solving recursive equations Eq.~\eqref{eq:kerneleom} which follow from the evolution equations for the $\psi_a$ described above. 

\item[(vi)] The \vpt{} kernels $F_n\equiv F_{n,\delta}$ for the density contrast $\delta$ capture screening of UV modes, meaning they are suppressed relative to SPT for arguments above $k_\sigma$, as already shown in~\cite{cumPT,cumPT2}. Here, we find that the kernels are largely insensitive to the cumulant truncation order $c_\text{max}$ within the regime where the suppression relative to SPT is up to an order of magnitude (see Sec.~\ref{sec:kernels}). Furthermore, we find that any modes with $|m|>1$ have almost no impact on the density kernels. Thus $m_\text{max}=1$, \emph{i.e.} including only scalar and vector modes, is an excellent approximation for the density field. Furthermore, even the lowest truncation order $c_\text{max}=2$ within \vpt{} yields results close to those for $c_\text{max}=6$. We also find similar behavior  for the $G_n$ kernels of the velocity divergence $\theta$ (see Appendix~\ref{app:Gkernels}).

\item[(vii)] We extend the comparison of \vpt{} predictions with $N$-body results from~\cite{cumPT2}, considering truncations with $c_\text{max}=2,3,4$ as well as the dependence on ${\cal E}_4$. We find that both are very mild (see Sec.~\ref{sec:impact}). Specifically, we consider one-loop
predictions for $P_{\delta\delta}$, $P_{\delta\theta}$, $P_{\theta\theta}$ as well as the bispectrum $B(k,k,k)$, and two-loop for the vorticity $P_{ww}$. Notably, there is a noticeable dependence of $P_{\delta\delta}$ on ${\cal E}_4$ but this can be absorbed into a redefinition of the dispersion scale $k_\sigma$ (which is our adjustable parameter), leaving a residual mild dependence as a result (see Fig.~\ref{fig:Pdd_e4}). This residual dependence is more visible in the velocity divergence (see Fig.~\ref{fig:Pdttt_e4}). The dependence on the truncation ($c_\text{max}$ and ${\cal E}_4$) is most pronounced for the vorticity, as might be expected since its generation is sensitive to the details of small-scale shell crossing dynamics~\cite{PicBer9903}. Note that the single free parameter $k_\sigma$ within \vpt{} is determined by fitting $P_{\delta\delta}$ to $N$-body results, and the same value is then used to predict $P_{\delta\theta}$, $P_{\theta\theta}$, $B(k,k,k)$, and $P_{ww}$ without any free parameters.

\item[(viii)] As in~\cite{cumPT}, we focus on
power-law initial spectra $P_0\propto k^{n_s}$, considering $n_s=-1,0,1,2$, for which SPT would be UV divergent. Using the screening property of \vpt{} kernels, we show analytically that \vpt{} power spectra are UV finite at any loop order $L$ and for any
spectral index $-3<n_s<4$ within the regime consistent with hierarchical clustering, see Eq.~\eqref{eq:nsVPT}. For comparison, SPT yields convergent results only for $-3<n_s<-3+2/L$ at loop order $L$.

\item[(ix)] We provide a detailed discussion of the relation between \vpt{} and the EFT approach, and compare one-loop predictions within the standard EFT setup to those in \vpt{}. As expected, for the spectral indices considered here, the \vpt{} one-loop prediction remains close to $N$-body results up to higher $k$ as compared to the  common EFT ansatz with leading-order counterterms, see Sec.~\ref{sec:eft}. This is more pronounced for bluer spectra, we argue that for such cases the effects of dispersion dominate over the standard nonlinearities over most scales (see Fig.~\ref{fig:rnleps}).

\end{itemize}

We conclude that, while third- and higher-order cumulants are generated after shell-crossing, their impact on the density and velocity divergence fields on weakly non-linear scales is rather small. Including the velocity dispersion tensor $\sigma_{ij}$, \emph{i.e.} the second cumulant, in the perturbative description yields the major change with respect to SPT. Even higher cumulants lead to small quantitative changes. We attribute this \emph{(a)} to the fact that UV screening (being the dominant qualitative effect captured by \vpt{} but not by SPT) arises already for $c_\text{max}=2$ and \emph{(b)} the backreaction of cumulants with $c>2$ on the density field is suppressed on sufficiently large length scales. The combination of both effects implies that within \vpt{} loop integrals are largely dominated by wavenumbers within the regime of validity of the perturbative approach, and that the impact of third and even higher-order cumulants is correspondingly small. This qualifies \vpt{} as a promising and practicable setup for overcoming the well-known shortcomings of SPT within a predictive approach based on a first-principle description of collisionless dynamics. In addition, as we argue in Sec.~\ref{sec:eft}, this can be complemented with additional EFT corrections arising from deviation from collisionless dynamics. Since a full tower of physically motivated gradient terms is already contained in \vpt{}, tighter priors on counterterms in data analysis should be possible, enhancing overall predictivity and constraining power.

\vspace*{2em}
\acknowledgments

We acknowledge support by the Excellence Cluster ORIGINS, which is funded by the Deutsche Forschungsgemeinschaft (DFG, German Research
Foundation) under Germany's Excellence Strategy - EXC-2094 - 390783311.

\begin{widetext}

\appendix

\section{Manifestly real formulation}
\label{app:real}

For the numerical implementation, it is convenient to use a formulation that is manifestly real.
This can be achieved by expanding the CGF in \emph{real spherical harmonics} ${\cal Y}_{\ell m}$ (instead of the usual complex ones as in Eq.~\ref{eq:Ylmdecomp}),
\bea
  {\cal Y}_{\ell m} &\equiv& \left\{\begin{array}{ll}
    \frac{1}{\sqrt{2}}\left((-1)^mY_{\ell m}+Y_{\ell, -m}\right) = \frac{(-1)^m}{\sqrt{2}}\left(Y_{\ell m}+Y_{\ell m}^*\right), & m>0\,,\\
    Y_{\ell 0}, & m=0\,,\\
    \frac{1}{\sqrt{2}i}\left((-1)^mY_{\ell |m|}-Y_{\ell, -|m|}\right) =  \frac{(-1)^m}{\sqrt{2}i}\left(Y_{\ell |m|}-Y_{\ell, |m|}^*\right), & m<0    \,.
  \end{array}\right.
\eea
As the complex ones, they satisfy the orthogonality condition
\be\label{eq:YRlmortho}
  \int d^2\hat L\, {\cal Y}_{\ell m}(\hat L) {\cal Y}_{\ell'm'}(\hat L) =\delta_{\ell\ell'}^K\delta_{mm'}^K\,,
\ee
where $m,m'$ in this relation can be positive, negative or zero. Under a rotation of the coordinate system specified by an orthogonal tranformation matrix $R$, they transform
with real Wigner matrices ${\cal D}^{m'}_{\ell m}(R)$,
\be
  {\cal Y}_{\ell m}(\theta',\phi')={\cal Y}_{\ell m'}(\theta,\phi){\cal D}^{m'}_{\ell m}(R)\,.
\ee
The corresponding Wigner matrix $D^{m'}_{\ell m}(R)$ for the complex spherical harmonics, parameterized by Euler angles $\alpha,\beta,\gamma$ in the $z-y-z$ convention, can be written as
\be
  D^{m'}_{\ell m}(\alpha,\beta,\gamma) = e^{-i m'\alpha} d^{m'}_{\ell m}(\beta) e^{-i m\gamma}\,,
\ee
with real-valued $d^{m'}_{\ell m}(\beta)$ that satisfy $d^{m'}_{\ell m}(\beta)=(-1)^{m'-m}d^{-m'}_{\ell ,-m}(\beta)$.
From this one can derive (here $m,m'>0$, and negative values are denoted by $-m$ or $-m'$, respectively)
\be
\left(\begin{array}{ccc}
  {\cal D}^{m'}_{\ell m} & {\cal D}^{m'}_{\ell 0} & {\cal D}^{m'}_{\ell,- m} \\
  {\cal D}^{0}_{\ell m} & {\cal D}^{0}_{\ell 0} & {\cal D}^{0}_{\ell, -m} \\
  {\cal D}^{-m'}_{\ell m} & {\cal D}^{-m'}_{\ell 0} & {\cal D}^{-m'}_{\ell, -m} 
  \end{array}\right) =
  \left(\begin{array}{ccc}
 (-1)^m (\cos(\gamma m - \alpha m') d^{-m'}_{\ell m} \atop + (-1)^{m'} \cos(\gamma m + \alpha m') d^{m'}_{\ell m} ) &
  \sqrt{2} \cos(\alpha m') d^{-m'}_{\ell 0}  &
   -(-1)^m (d^{-m'}_{\ell m} \sin(\gamma m - \alpha mp) \atop + (-1)^{m'} d^{m'}_{\ell m}\sin(\gamma m + \alpha mp)) \\[1.5ex]
 (-1)^m \sqrt{2} \cos(\gamma m) d^0_{\ell m} & d^0_{\ell 0} & -(-1)^m \sqrt{2} d^0_{\ell m }\sin(\gamma m) \\[1.5ex]
 -(-1)^m (d^{-m'}_{\ell m}\sin(\gamma m - \alpha m') \atop - (-1)^{m'} d^{m'}_{\ell m}\sin(\gamma m + \alpha m')) &
  \sqrt{2} d^{-m'}_{\ell 0}\sin(   \alpha m') &
   (-1)^m (-\cos(\gamma m - \alpha m') d^{-m'}_{\ell m} \atop + (-1)^{m'} \cos(\gamma m + \alpha m') d^{m'}_{\ell m})
  \end{array}\right) \,.  
\ee
The Gaunt integral for real spherical harmonics reads
\bea\label{eq:GauntR}
  \int d^2\hat L\, {\cal Y}_{\ell_1 m_1}(\hat L){\cal Y}_{\ell_2 m_2}(\hat L){\cal Y}_{\ell_3 m_3}(\hat L) &=& \sqrt{\frac{(2\ell_1+1)(2\ell_2+1)(2\ell_3+1)}{4\pi}}\left(\begin{array}{ccc}\ell_1&\ell_2&\ell_3\\ 0&0&0\end{array}\right)\nn\\
 && {} \times {\cal J}^{\ell_1\ell_2\ell_3}_{m_1m_2m_3}\,.
\eea
Denoting the usual Wigner 3j symbols entering the Gaunt integral of complex spherical harmonics by
\be
  J^{\ell_1\ell_2\ell_3}_{m_1m_2m_3} \equiv \left(\begin{array}{ccc}\ell_1&\ell_2&\ell_3\\ m_1&m_2&m_3\end{array}\right)\,,\nn\\
\ee
we find (for $m_i>0$, and denoting negative values by $-m_i$)
\bea
  {\cal J}^{\ell_1\ell_2\ell_3}_{m_1m_2m_3} &=& \frac{1}{\sqrt{2}} \left( (-1)^{m_1} J^{\ell_1\ell_2\ell_3}_{-m_1m_2m_3} + (-1)^{m_2} J^{\ell_1\ell_2\ell_3}_{m_1-m_2m_3} + (-1)^{m_3} J^{\ell_1\ell_2\ell_3}_{m_1m_2-m_3} \right)\,,\nn\\
  {\cal J}^{\ell_1\ell_2\ell_3}_{m_1-m_2-m_3} &=& \frac{1}{\sqrt{2}} \left( - (-1)^{m_1} J^{\ell_1\ell_2\ell_3}_{-m_1m_2m_3} + (-1)^{m_2} J^{\ell_1\ell_2\ell_3}_{m_1-m_2m_3} + (-1)^{m_3} J^{\ell_1\ell_2\ell_3}_{m_1m_2-m_3} \right)\,,\nn\\
  {\cal J}^{\ell_1\ell_2\ell_3}_{m_1m_20} &=&  (-1)^{m_1} J^{\ell_1\ell_2\ell_3}_{m_1-m_20} \quad \text{(non-zero only for} \ m_1=m_2)\,,\nn\\
  {\cal J}^{\ell_1\ell_2\ell_3}_{-m_1-m_20} &=&  (-1)^{m_1} J^{\ell_1\ell_2\ell_3}_{m_1-m_20} \quad \text{(non-zero only for} \ m_1=m_2)\,,\nn\\
  {\cal J}^{\ell_1\ell_2\ell_3}_{000} &=&   J^{\ell_1\ell_2\ell_3}_{000} \,.
\eea
The cases $-+-,\ --+$ and $+0+,\ 0++$ as well as $-0-,\ --0$ can be obtained from the $+--$, $++0$ and $--0$ configurations given in the second, third and fourth line, respectively, using that the Gaunt integral is non-zero only for $\ell_1+\ell_2+\ell_3=$even,
and that the $J^{\ell_1\ell_2\ell_3}_{m_1m_2m_3}$ are totally symmetric under permuting the $(\ell_1m_1),(\ell_2m_2),(\ell_3m_3)$ in that case. This property is inherited by the ${\cal J}^{\ell_1\ell_2\ell_3}_{m_1m_2m_3}$.
The cases $---,\ -++,\ -00,\ -+0$ and $+00$ are zero, as well as the corresponding permutations of the three arguments. In addition, ${\cal J}^{\ell_1\ell_2\ell_3}_{m_1m_2m_3}$ (for any sign of the $m_i$) is non-zero only if $|m_1|=|m_2|+|m_3|$ or $|m_2|=|m_1|+|m_3|$ or $|m_3|=|m_1|+|m_2|$.

We define rotated real spherical harmonics in analogy to the complex case Eq.~\eqref{eq:Ylmk} by 
\be\label{eq:YRlmk}
  {\cal Y}^{\bm k}_{\ell m}(\theta,\phi)\equiv {\cal Y}_{\ell m'}(\theta,\phi){\cal D}^{m'}_{\ell m}(R^{\bm k})\,.
\ee
Expanding the CGF in those instead of the complex ones in Eq.~\eqref{eq:Ylmdecomp}, yields evolution equations for the cumulant
perturbation modes that only differ in the non-linear vertex Eq.~\eqref{eq:gammaunsymmabcexplicit} by the replacements
\bea
 (-1)^{m'}[D^{m'}_{\ell m}((R^{\bm p})^{-1}R^{\bm k})]^* D^{m_2'}_{\ell_2 m_2}((R^{\bm p})^{-1}R^{\bm q}) &\mapsto&
 {\cal D}^{m'}_{\ell m}((R^{\bm p})^{-1}R^{\bm k}) {\cal D}^{m_2'}_{\ell_2 m_2}((R^{\bm p})^{-1}R^{\bm q}) \,,
 \nn\\
  \left(\begin{array}{ccc}\ell&\ell_1&\ell_2\pm 1\\ -m'&m_1&m_2'\end{array}\right) &\mapsto& {\cal J}^{\ell,\ell_1,\ell_2\pm 1}_{m'm_1m_2'}\,.
\eea
The linear part of the evolution equations Eq.~\eqref{eq:eommatrixform} is identical when expanding in real spherical harmonics since the $\Omega$-matrix Eq.~\eqref{eq:Omegaab}
is diagonal with respect to the $m$-number, and since its entries depend only on $m^2$, \emph{i.e.} they are identical for $m$ and $-m$. Note that expanding in real instead of complex spherical harmonics affects
the form of the mapping to the cartesian basis discussed in Sec.~\ref{sec:kart}. However, for scalar modes (with $m=0$) there is no difference. Overall, the expansion in real spherical harmonics allows us
to avoid the appearance of any complex numbers in the numerical computation of arbitrary non-linear kernels within \vpt{}.

\section{Useful relations}
\label{app:relations}

Here we collect some useful relations for the derivations in the main text.
For the derivation of the evolution equation Eq.~\eqref{eq:Clm_raw_eq_of_motion}
we use (assuming ${\bm k}$ in $z$-direction for the moment)
\bea
{\bm L}\cdot \nabla_L &=& L\frac{\partial}{\partial L}=\ell+L^{\ell+1} \frac{\partial}{\partial L} L^{-\ell}\,,\nn\\
{\bm L}\cdot i{\bm k} &=& ikL_z = ikL\cos\theta\,,\nn\\
\nabla_L\cdot i{\bm k} &=& ik\frac{\partial}{\partial L_z} = ik\left[\cos\theta\frac{\partial}{\partial L} - \frac{\sin\theta}{L}\frac{\partial}{\partial\theta}\right]
= ik\left[\cos\theta L^\ell\frac{\partial}{\partial L}L^{-\ell} +\frac{1}{L}\left(\ell\cos\theta -\sin\theta\frac{\partial}{\partial\theta}\right)\right]\,,
\eea
and
\bea
\cos\theta Y_{\ell m}(\hat L) &=& \sqrt{\frac{(\ell+1)^2-m^2}{(2\ell+1)(2\ell+3)}}Y_{\ell+1,m}(\hat L) +\sqrt{\frac{\ell^2-m^2}{(2\ell-1)(2\ell+1)}}Y_{\ell-1,m}(\hat L)\,,\nn\\
\left(\ell\cos\theta -\sin\theta\frac{\partial}{\partial\theta}\right) Y_{\ell m}(\hat L) &=& (2\ell+1)\sqrt{\frac{\ell^2-m^2}{(2\ell-1)(2\ell+1)}}Y_{\ell-1,m}(\hat L)\,.
\eea
For deriving Eq.~\eqref{eq:Clm_raw_eq_of_motion} we encounter \emph{e.g.} expressions of the form $i{\bm k}\cdot\nabla_L  \, Y_{\ell m}^{\bm k}(\hat L)(-iL)^\ell f(L^2)$, with some generic function $f(L^2)$.
Using the relations from above, its dependence on the angles (contained in the unit vector $\hat L$) can be expressed as a linear combination of $Y_{\ell+1, m}^{\bm k}(\hat L)$
and $Y_{\ell-1, m}^{\bm k}(\hat L)$. In a first step, this relation can be derived in the frame in which ${\bm k}$ points in the $z$-direction, such that in this frame the usual spherical harmonics $Y_{\ell m}(\vartheta',\phi')$ appear
and the equations from above, evaluated in the ``primed'' frame, can be used. In a second step, 
using $Y_{\ell m}^{\bm k}(\vartheta,\varphi)=Y_{\ell m}(\vartheta',\varphi')$ (see Eq.~\ref{eq:Ylmk}) and that the scalar product $i{\bm k}\cdot\nabla_L$ is invariant under rotations, the relation can be trivially generalized to a generic frame.

For the derivation of the non-linear interaction vertex  Eq.~\eqref{eq:gammaunsymmabcexplicit} starting from Eq.~\eqref{eq:Clm_raw_eq_of_motion}, we encounter also
expressions of the form $i{\bm p}\cdot\nabla_L  \, Y_{\ell m}^{\bm q}(\hat L)(-iL)^\ell f(L^2)$ with two different vectors ${\bm p}$ and ${\bm q}$. In this case we first express $Y_{\ell m}^{\bm q}(\hat L)$ in terms of a linear combination of
$Y_{\ell m'}^{\bm p}(\hat L)$ with $m'=-\ell,\dots,\ell$
using Eq.~\eqref{eq:Ylmk},
\be\label{eq:Ylmpqtrafo}
  Y_{\ell m}^{\bm q}(\hat L) = Y_{\ell m'}^{\bm p}(\hat L)D^{m'}_{\ell m''}((R^{\bm p})^{-1})D^{m''}_{\ell m}(R^{\bm q})
  =Y_{\ell m'}^{\bm p}(\hat L)D^{m'}_{\ell m}((R^{\bm p})^{-1}R^{\bm q})\,.
\ee
Here we used the group property for the Wigner representation matrices in the last step.
We then rewrite $i{\bm p}\cdot\nabla_L  \, Y_{\ell m'}^{\bm p}(\hat L)(-iL)^\ell f(L^2)$ as a linear combination of $Y_{\ell+1, m'}^{\bm p}(\hat L)$
and $Y_{\ell-1, m'}^{\bm p}(\hat L)$ analogously as described above.

\section{Impact of truncation on velocity kernels}
\label{app:Gkernels}

Here we present the dependence of \vpt{} kernels $G_n$ for the velocity divergence $\theta$
on the order of the cumulant truncation, showing analogous figures as in Sec.~\ref{sec:kernels}
for the density kernels $F_n$. We note that for the configuration of wave-numbers for $G_3(\bm k,\bm q,-\bm q)$, the linear scaling of the velocity divergence kernels
observed in~\cite{cumPT2} is absent.

\begin{figure*}[t]
  \begin{center}
  \includegraphics[width=0.48\textwidth]{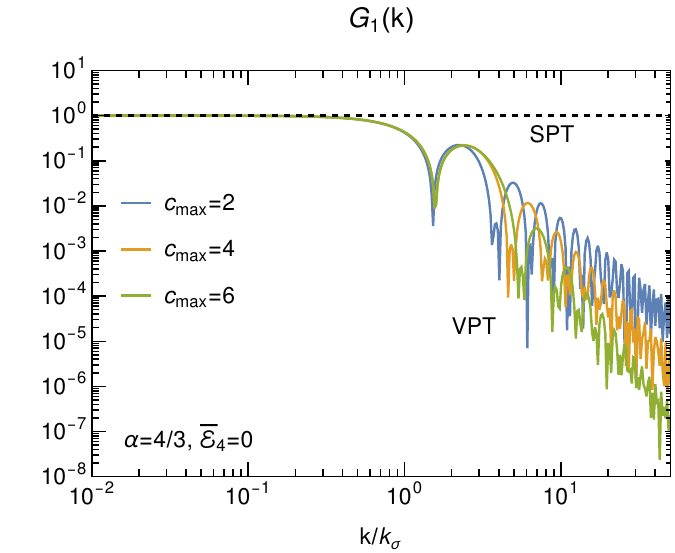}
  \includegraphics[width=0.48\textwidth]{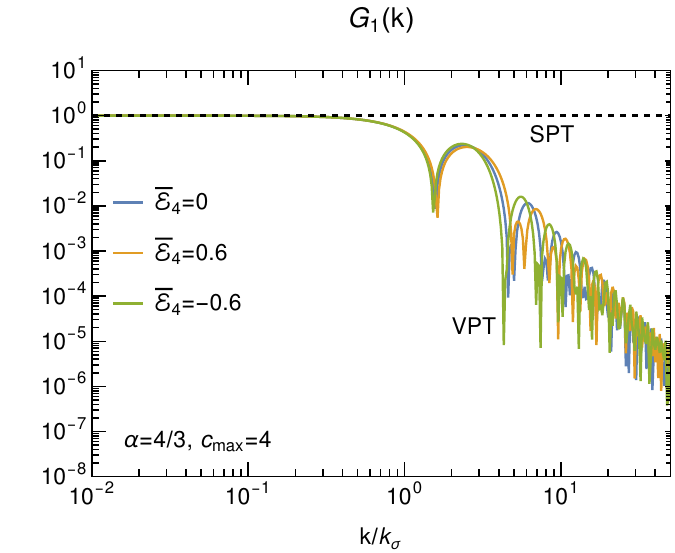}
  \end{center}
  \caption{\label{fig:G1}
  As Fig.~\ref{fig:F1}, but for the velocity divergence kernel $G_1(k,\eta=0)$.
  }
\end{figure*}

\begin{figure*}[t]
  \begin{center}
  \includegraphics[width=0.48\textwidth]{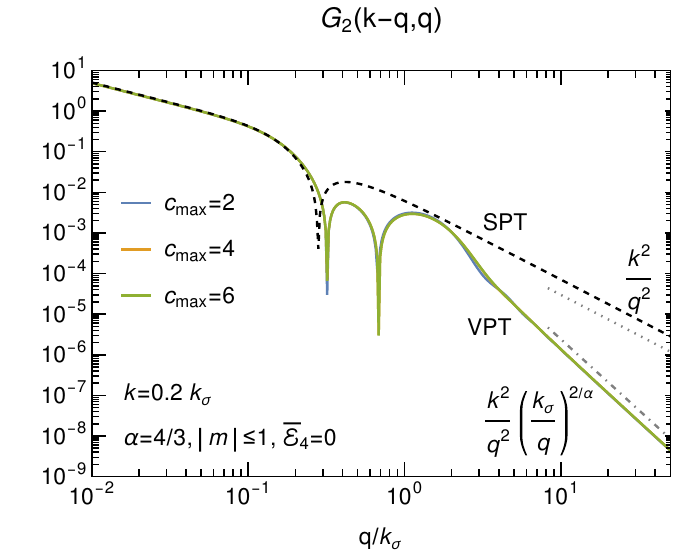}
  \includegraphics[width=0.48\textwidth]{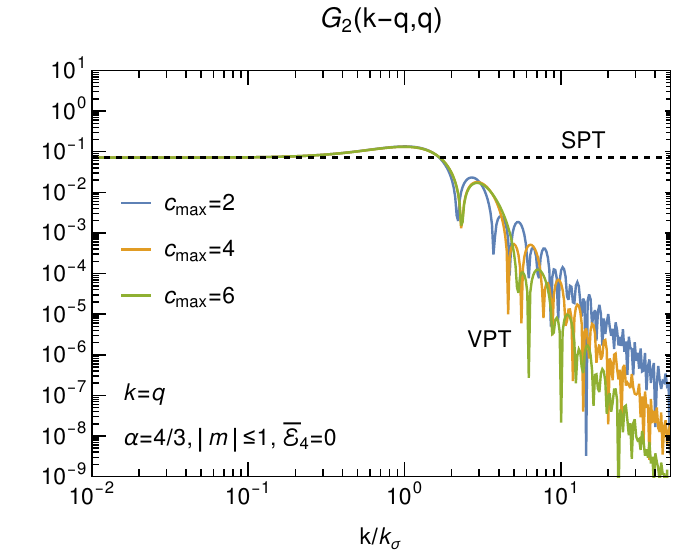}
  \\[1.5ex]
  \includegraphics[width=0.48\textwidth]{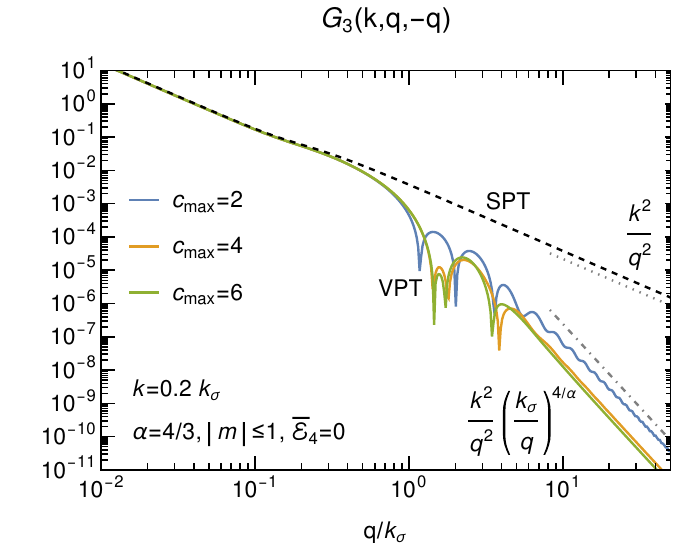}
  \includegraphics[width=0.48\textwidth]{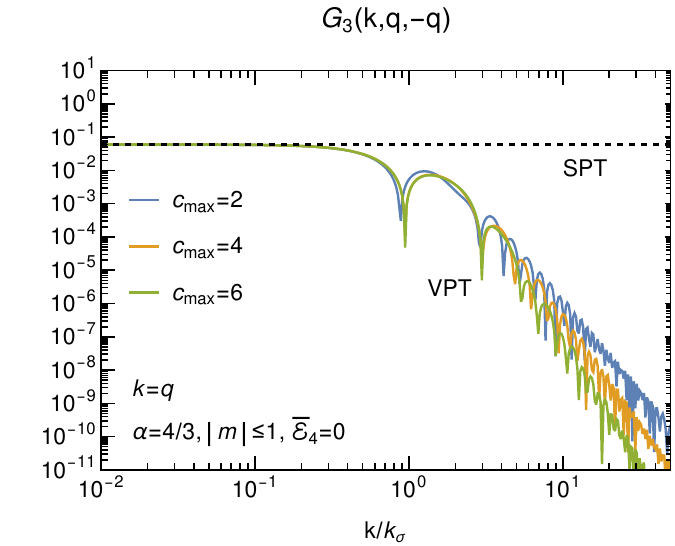}
  \end{center}
  \caption{\label{fig:G23_c246}
  As Fig.~\ref{fig:F23_c246}, but for the velocity divergence kernels $G_2$ and $G_3$, all at $\eta=0$.
  }
\end{figure*}

\begin{figure*}[t]
  \begin{center}
  \includegraphics[width=0.48\textwidth]{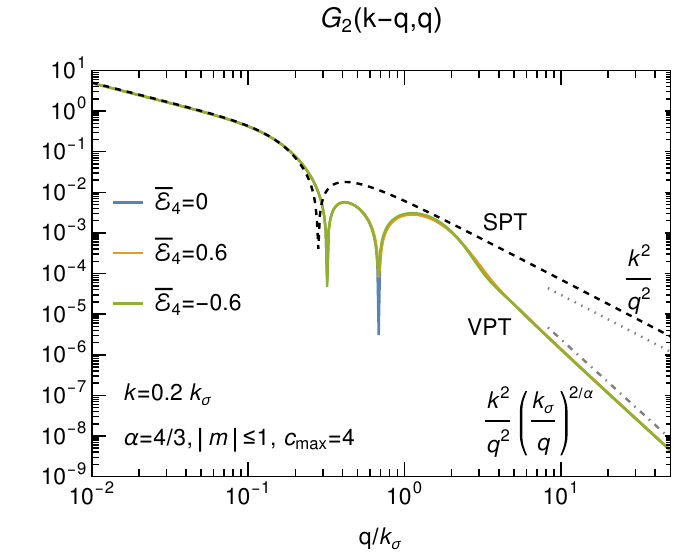}
  \includegraphics[width=0.48\textwidth]{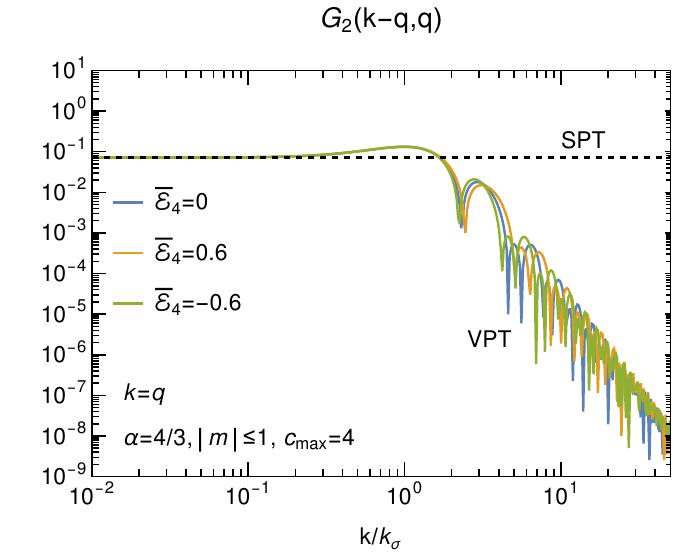}
  \\[1.5ex]
  \includegraphics[width=0.48\textwidth]{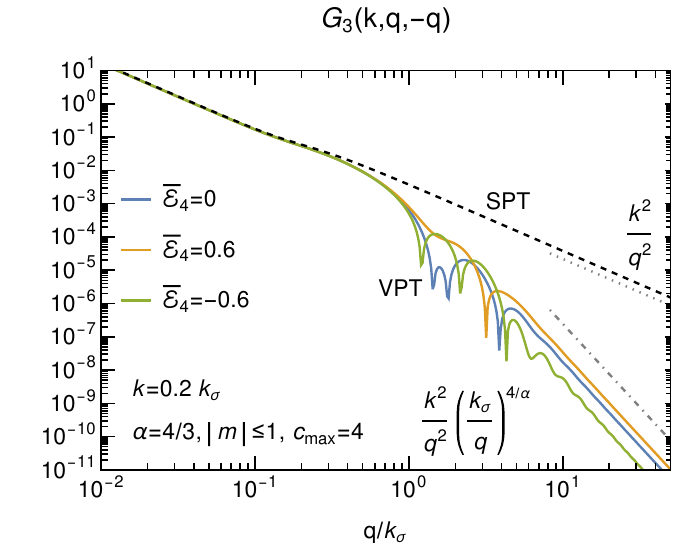}
  \includegraphics[width=0.48\textwidth]{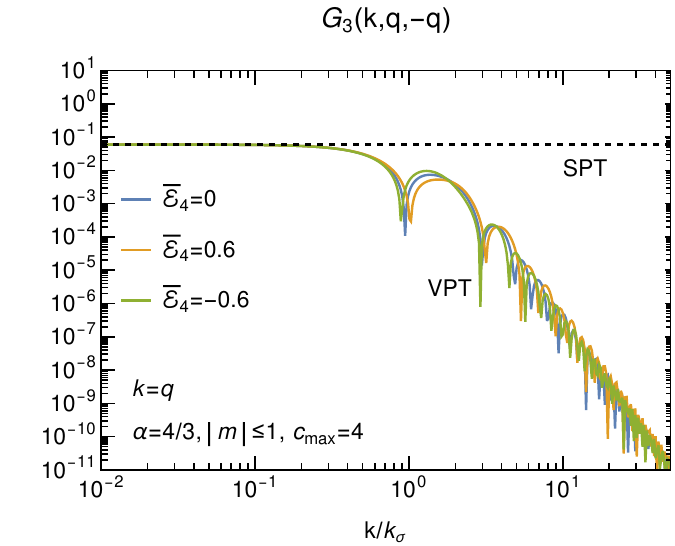}
  \end{center}
  \caption{\label{fig:G23_e4}
  As Fig.~\ref{fig:F23_e4}, but for the velocity divergence kernels $G_2$ and $G_3$, all at $\eta=0$.
  }
\end{figure*}

\begin{figure*}[t]
  \begin{center}
  \includegraphics[width=0.48\textwidth]{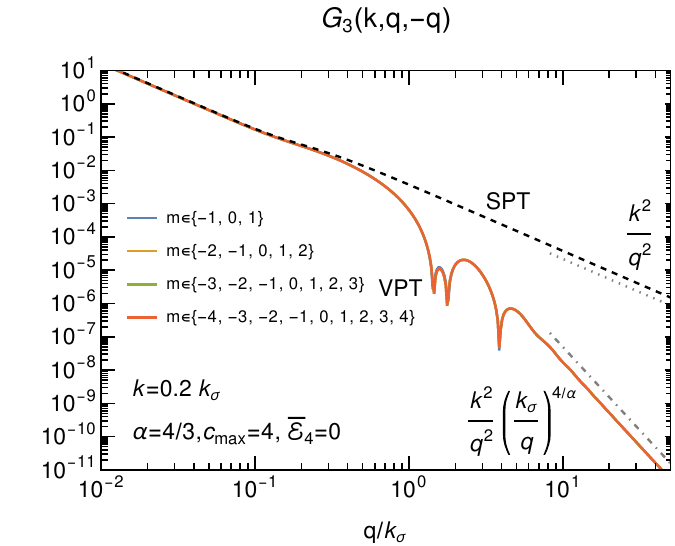}
  \includegraphics[width=0.48\textwidth]{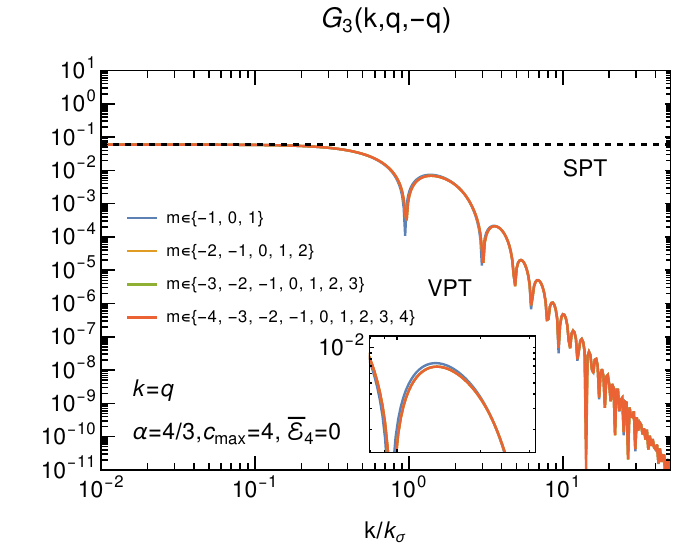}
  \end{center}
  \caption{\label{fig:G3_m}
  As Fig.~\ref{fig:F3_m}, but for the velocity divergence kernel $G_3$, at $\eta=0$.
  }
\end{figure*}

\end{widetext}


\begin{thebibliography}{48}%
\makeatletter
\providecommand \@ifxundefined [1]{%
 \@ifx{#1\undefined}
}%
\providecommand \@ifnum [1]{%
 \ifnum #1\expandafter \@firstoftwo
 \else \expandafter \@secondoftwo
 \fi
}%
\providecommand \@ifx [1]{%
 \ifx #1\expandafter \@firstoftwo
 \else \expandafter \@secondoftwo
 \fi
}%
\providecommand \natexlab [1]{#1}%
\providecommand \enquote  [1]{``#1''}%
\providecommand \bibnamefont  [1]{#1}%
\providecommand \bibfnamefont [1]{#1}%
\providecommand \citenamefont [1]{#1}%
\providecommand \href@noop [0]{\@secondoftwo}%
\providecommand \href [0]{\begingroup \@sanitize@url \@href}%
\providecommand \@href[1]{\@@startlink{#1}\@@href}%
\providecommand \@@href[1]{\endgroup#1\@@endlink}%
\providecommand \@sanitize@url [0]{\catcode `\\12\catcode `\$12\catcode
  `\&12\catcode `\#12\catcode `\^12\catcode `\_12\catcode `\%12\relax}%
\providecommand \@@startlink[1]{}%
\providecommand \@@endlink[0]{}%
\providecommand \url  [0]{\begingroup\@sanitize@url \@url }%
\providecommand \@url [1]{\endgroup\@href {#1}{\urlprefix }}%
\providecommand \urlprefix  [0]{URL }%
\providecommand \Eprint [0]{\href }%
\providecommand \doibase [0]{http://dx.doi.org/}%
\providecommand \selectlanguage [0]{\@gobble}%
\providecommand \bibinfo  [0]{\@secondoftwo}%
\providecommand \bibfield  [0]{\@secondoftwo}%
\providecommand \translation [1]{[#1]}%
\providecommand \BibitemOpen [0]{}%
\providecommand \bibitemStop [0]{}%
\providecommand \bibitemNoStop [0]{.\EOS\space}%
\providecommand \EOS [0]{\spacefactor3000\relax}%
\providecommand \BibitemShut  [1]{\csname bibitem#1\endcsname}%
\let\auto@bib@innerbib\@empty
\bibitem [{\citenamefont {Pueblas}\ and\ \citenamefont
  {Scoccimarro}(2009)}]{Pueblas:2008uv}%
  \BibitemOpen
  \bibfield  {author} {\bibinfo {author} {\bibfnamefont {S.}~\bibnamefont
  {Pueblas}}\ and\ \bibinfo {author} {\bibfnamefont {R.}~\bibnamefont
  {Scoccimarro}},\ }\href {\doibase 10.1103/PhysRevD.80.043504} {\bibfield
  {journal} {\bibinfo  {journal} {Phys. Rev. D}\ }\textbf {\bibinfo {volume}
  {80}},\ \bibinfo {pages} {043504} (\bibinfo {year} {2009})},\ \Eprint
  {http://arxiv.org/abs/0809.4606} {arXiv:0809.4606 [astro-ph]} \BibitemShut
  {NoStop}%
\bibitem [{\citenamefont {Bernardeau}\ \emph {et~al.}(2002)\citenamefont
  {Bernardeau}, \citenamefont {Colombi}, \citenamefont {Gaztanaga},\ and\
  \citenamefont {Scoccimarro}}]{Bernardeau:2001qr}%
  \BibitemOpen
  \bibfield  {author} {\bibinfo {author} {\bibfnamefont {F.}~\bibnamefont
  {Bernardeau}}, \bibinfo {author} {\bibfnamefont {S.}~\bibnamefont {Colombi}},
  \bibinfo {author} {\bibfnamefont {E.}~\bibnamefont {Gaztanaga}}, \ and\
  \bibinfo {author} {\bibfnamefont {R.}~\bibnamefont {Scoccimarro}},\ }\href
  {\doibase 10.1016/S0370-1573(02)00135-7} {\bibfield  {journal} {\bibinfo
  {journal} {Phys. Rept.}\ }\textbf {\bibinfo {volume} {367}},\ \bibinfo
  {pages} {1} (\bibinfo {year} {2002})},\ \Eprint
  {http://arxiv.org/abs/astro-ph/0112551} {arXiv:astro-ph/0112551} \BibitemShut
  {NoStop}%
\bibitem [{\citenamefont {Blas}\ \emph {et~al.}(2014)\citenamefont {Blas},
  \citenamefont {Garny},\ and\ \citenamefont {Konstandin}}]{BlaGarKon1309}%
  \BibitemOpen
  \bibfield  {author} {\bibinfo {author} {\bibfnamefont {D.}~\bibnamefont
  {Blas}}, \bibinfo {author} {\bibfnamefont {M.}~\bibnamefont {Garny}}, \ and\
  \bibinfo {author} {\bibfnamefont {T.}~\bibnamefont {Konstandin}},\ }\href
  {\doibase 10.1088/1475-7516/2014/01/010} {\bibfield  {journal} {\bibinfo
  {journal} {JCAP}\ }\textbf {\bibinfo {volume} {01}},\ \bibinfo {pages} {010}
  (\bibinfo {year} {2014})},\ \Eprint {http://arxiv.org/abs/1309.3308}
  {arXiv:1309.3308 [astro-ph.CO]} \BibitemShut {NoStop}%
\bibitem [{\citenamefont {Peebles}(1980)}]{peebles1980large}%
  \BibitemOpen
  \bibfield  {author} {\bibinfo {author} {\bibfnamefont {P.~J.~E.}\
  \bibnamefont {Peebles}},\ }\href@noop {} {\emph {\bibinfo {title} {The
  large-scale structure of the universe}}}\ (\bibinfo  {publisher} {Princeton
  university press},\ \bibinfo {year} {1980})\BibitemShut {NoStop}%
\bibitem [{\citenamefont {{Nishimichi}}\ \emph {et~al.}(2016)\citenamefont
  {{Nishimichi}}, \citenamefont {{Bernardeau}},\ and\ \citenamefont
  {{Taruya}}}]{NisBerTar1611}%
  \BibitemOpen
  \bibfield  {author} {\bibinfo {author} {\bibfnamefont {T.}~\bibnamefont
  {{Nishimichi}}}, \bibinfo {author} {\bibfnamefont {F.}~\bibnamefont
  {{Bernardeau}}}, \ and\ \bibinfo {author} {\bibfnamefont {A.}~\bibnamefont
  {{Taruya}}},\ }\href {\doibase 10.1016/j.physletb.2016.09.035} {\bibfield
  {journal} {\bibinfo  {journal} {Physics Letters B}\ }\textbf {\bibinfo
  {volume} {762}},\ \bibinfo {pages} {247} (\bibinfo {year} {2016})},\ \Eprint
  {http://arxiv.org/abs/1411.2970} {arXiv:1411.2970} \BibitemShut {NoStop}%
\bibitem [{\citenamefont {{Nishimichi}}\ \emph {et~al.}(2017)\citenamefont
  {{Nishimichi}}, \citenamefont {{Bernardeau}},\ and\ \citenamefont
  {{Taruya}}}]{NisBerTar1712}%
  \BibitemOpen
  \bibfield  {author} {\bibinfo {author} {\bibfnamefont {T.}~\bibnamefont
  {{Nishimichi}}}, \bibinfo {author} {\bibfnamefont {F.}~\bibnamefont
  {{Bernardeau}}}, \ and\ \bibinfo {author} {\bibfnamefont {A.}~\bibnamefont
  {{Taruya}}},\ }\href {\doibase 10.1103/PhysRevD.96.123515} {\bibfield
  {journal} {\bibinfo  {journal} {\prd}\ }\textbf {\bibinfo {volume} {96}},\
  \bibinfo {eid} {123515} (\bibinfo {year} {2017})},\ \Eprint
  {http://arxiv.org/abs/1708.08946} {arXiv:1708.08946 [astro-ph.CO]}
  \BibitemShut {NoStop}%
\bibitem [{\citenamefont {Aghamousa}\ \emph {et~al.}(2016)\citenamefont
  {Aghamousa} \emph {et~al.}}]{DESI:2016fyo}%
  \BibitemOpen
  \bibfield  {author} {\bibinfo {author} {\bibfnamefont {A.}~\bibnamefont
  {Aghamousa}} \emph {et~al.} (\bibinfo {collaboration} {DESI}),\ }\href@noop
  {} {\  (\bibinfo {year} {2016})},\ \Eprint {http://arxiv.org/abs/1611.00036}
  {arXiv:1611.00036 [astro-ph.IM]} \BibitemShut {NoStop}%
\bibitem [{\citenamefont {Mellier}\ \emph {et~al.}(2024)\citenamefont {Mellier}
  \emph {et~al.}}]{Euclid:2024yrr}%
  \BibitemOpen
  \bibfield  {author} {\bibinfo {author} {\bibfnamefont {Y.}~\bibnamefont
  {Mellier}} \emph {et~al.} (\bibinfo {collaboration} {Euclid}),\ }\href@noop
  {} {\  (\bibinfo {year} {2024})},\ \Eprint {http://arxiv.org/abs/2405.13491}
  {arXiv:2405.13491 [astro-ph.CO]} \BibitemShut {NoStop}%
\bibitem [{\citenamefont {{Baumann}}\ \emph {et~al.}(2012)\citenamefont
  {{Baumann}}, \citenamefont {{Nicolis}}, \citenamefont {{Senatore}},\ and\
  \citenamefont {{Zaldarriaga}}}]{BauNicSen1207}%
  \BibitemOpen
  \bibfield  {author} {\bibinfo {author} {\bibfnamefont {D.}~\bibnamefont
  {{Baumann}}}, \bibinfo {author} {\bibfnamefont {A.}~\bibnamefont
  {{Nicolis}}}, \bibinfo {author} {\bibfnamefont {L.}~\bibnamefont
  {{Senatore}}}, \ and\ \bibinfo {author} {\bibfnamefont {M.}~\bibnamefont
  {{Zaldarriaga}}},\ }\href {\doibase 10.1088/1475-7516/2012/07/051} {\bibfield
   {journal} {\bibinfo  {journal} {\jcap}\ }\textbf {\bibinfo {volume} {7}},\
  \bibinfo {eid} {051} (\bibinfo {year} {2012})},\ \Eprint
  {http://arxiv.org/abs/1004.2488} {arXiv:1004.2488 [astro-ph.CO]} \BibitemShut
  {NoStop}%
\bibitem [{\citenamefont {Garny}\ \emph
  {et~al.}(2023{\natexlab{a}})\citenamefont {Garny}, \citenamefont {Laxhuber},\
  and\ \citenamefont {Scoccimarro}}]{cumPT}%
  \BibitemOpen
  \bibfield  {author} {\bibinfo {author} {\bibfnamefont {M.}~\bibnamefont
  {Garny}}, \bibinfo {author} {\bibfnamefont {D.}~\bibnamefont {Laxhuber}}, \
  and\ \bibinfo {author} {\bibfnamefont {R.}~\bibnamefont {Scoccimarro}},\
  }\href {\doibase 10.1103/PhysRevD.107.063539} {\bibfield  {journal} {\bibinfo
   {journal} {Phys. Rev. D}\ }\textbf {\bibinfo {volume} {107}},\ \bibinfo
  {pages} {063539} (\bibinfo {year} {2023}{\natexlab{a}})},\ \Eprint
  {http://arxiv.org/abs/2210.08088} {arXiv:2210.08088 [astro-ph.CO]}
  \BibitemShut {NoStop}%
\bibitem [{\citenamefont {Garny}\ \emph
  {et~al.}(2023{\natexlab{b}})\citenamefont {Garny}, \citenamefont {Laxhuber},\
  and\ \citenamefont {Scoccimarro}}]{cumPT2}%
  \BibitemOpen
  \bibfield  {author} {\bibinfo {author} {\bibfnamefont {M.}~\bibnamefont
  {Garny}}, \bibinfo {author} {\bibfnamefont {D.}~\bibnamefont {Laxhuber}}, \
  and\ \bibinfo {author} {\bibfnamefont {R.}~\bibnamefont {Scoccimarro}},\
  }\href {\doibase 10.1103/PhysRevD.107.063540} {\bibfield  {journal} {\bibinfo
   {journal} {Phys. Rev. D}\ }\textbf {\bibinfo {volume} {107}},\ \bibinfo
  {pages} {063540} (\bibinfo {year} {2023}{\natexlab{b}})},\ \Eprint
  {http://arxiv.org/abs/2210.08089} {arXiv:2210.08089 [astro-ph.CO]}
  \BibitemShut {NoStop}%
\bibitem [{\citenamefont {{McDonald}}(2011)}]{McD1104}%
  \BibitemOpen
  \bibfield  {author} {\bibinfo {author} {\bibfnamefont {P.}~\bibnamefont
  {{McDonald}}},\ }\href {\doibase 10.1088/1475-7516/2011/04/032} {\bibfield
  {journal} {\bibinfo  {journal} {\jcap}\ }\textbf {\bibinfo {volume} {2011}},\
  \bibinfo {eid} {032} (\bibinfo {year} {2011})},\ \Eprint
  {http://arxiv.org/abs/0910.1002} {arXiv:0910.1002 [astro-ph.CO]} \BibitemShut
  {NoStop}%
\bibitem [{\citenamefont {Aviles}(2016)}]{Aviles_2016}%
  \BibitemOpen
  \bibfield  {author} {\bibinfo {author} {\bibfnamefont {A.}~\bibnamefont
  {Aviles}},\ }\href {\doibase 10.1103/physrevd.93.063517} {\bibfield
  {journal} {\bibinfo  {journal} {Physical Review D}\ }\textbf {\bibinfo
  {volume} {93}} (\bibinfo {year} {2016}),\
  10.1103/physrevd.93.063517}\BibitemShut {NoStop}%
\bibitem [{\citenamefont {{McDonald}}\ and\ \citenamefont
  {{Vlah}}(2018)}]{McDVla1801}%
  \BibitemOpen
  \bibfield  {author} {\bibinfo {author} {\bibfnamefont {P.}~\bibnamefont
  {{McDonald}}}\ and\ \bibinfo {author} {\bibfnamefont {Z.}~\bibnamefont
  {{Vlah}}},\ }\href {\doibase 10.1103/PhysRevD.97.023508} {\bibfield
  {journal} {\bibinfo  {journal} {\prd}\ }\textbf {\bibinfo {volume} {97}},\
  \bibinfo {eid} {023508} (\bibinfo {year} {2018})},\ \Eprint
  {http://arxiv.org/abs/1709.02834} {arXiv:1709.02834 [astro-ph.CO]}
  \BibitemShut {NoStop}%
\bibitem [{\citenamefont {{Erschfeld}}\ and\ \citenamefont
  {{Floerchinger}}(2019)}]{ErsFlo1906}%
  \BibitemOpen
  \bibfield  {author} {\bibinfo {author} {\bibfnamefont {A.}~\bibnamefont
  {{Erschfeld}}}\ and\ \bibinfo {author} {\bibfnamefont {S.}~\bibnamefont
  {{Floerchinger}}},\ }\href {\doibase 10.1088/1475-7516/2019/06/039}
  {\bibfield  {journal} {\bibinfo  {journal} {\jcap}\ }\textbf {\bibinfo
  {volume} {2019}},\ \bibinfo {eid} {039} (\bibinfo {year} {2019})},\ \Eprint
  {http://arxiv.org/abs/1812.06891} {arXiv:1812.06891 [astro-ph.CO]}
  \BibitemShut {NoStop}%
\bibitem [{\citenamefont {Cusin}\ \emph {et~al.}(2017)\citenamefont {Cusin},
  \citenamefont {Tansella},\ and\ \citenamefont {Durrer}}]{Cusin:2016zvu}%
  \BibitemOpen
  \bibfield  {author} {\bibinfo {author} {\bibfnamefont {G.}~\bibnamefont
  {Cusin}}, \bibinfo {author} {\bibfnamefont {V.}~\bibnamefont {Tansella}}, \
  and\ \bibinfo {author} {\bibfnamefont {R.}~\bibnamefont {Durrer}},\ }\href
  {\doibase 10.1103/PhysRevD.95.063527} {\bibfield  {journal} {\bibinfo
  {journal} {Phys. Rev. D}\ }\textbf {\bibinfo {volume} {95}},\ \bibinfo
  {pages} {063527} (\bibinfo {year} {2017})},\ \Eprint
  {http://arxiv.org/abs/1612.00783} {arXiv:1612.00783 [astro-ph.CO]}
  \BibitemShut {NoStop}%
\bibitem [{\citenamefont {Erschfeld}\ and\ \citenamefont
  {Floerchinger}(2024)}]{Erschfeld:2023aqr}%
  \BibitemOpen
  \bibfield  {author} {\bibinfo {author} {\bibfnamefont {A.}~\bibnamefont
  {Erschfeld}}\ and\ \bibinfo {author} {\bibfnamefont {S.}~\bibnamefont
  {Floerchinger}},\ }\href {\doibase 10.1088/1475-7516/2024/02/053} {\bibfield
  {journal} {\bibinfo  {journal} {JCAP}\ }\textbf {\bibinfo {volume} {02}},\
  \bibinfo {pages} {053} (\bibinfo {year} {2024})},\ \Eprint
  {http://arxiv.org/abs/2305.18517} {arXiv:2305.18517 [astro-ph.CO]}
  \BibitemShut {NoStop}%
\bibitem [{\citenamefont {Uhlemann}(2018)}]{Uhlemann:2018olp}%
  \BibitemOpen
  \bibfield  {author} {\bibinfo {author} {\bibfnamefont {C.}~\bibnamefont
  {Uhlemann}},\ }\href {\doibase 10.1088/1475-7516/2018/10/030} {\bibfield
  {journal} {\bibinfo  {journal} {JCAP}\ }\textbf {\bibinfo {volume} {10}},\
  \bibinfo {pages} {030} (\bibinfo {year} {2018})},\ \Eprint
  {http://arxiv.org/abs/1807.07274} {arXiv:1807.07274 [astro-ph.CO]}
  \BibitemShut {NoStop}%
\bibitem [{\citenamefont {Nascimento}\ and\ \citenamefont
  {Loverde}(2024)}]{Nascimento:2024hle}%
  \BibitemOpen
  \bibfield  {author} {\bibinfo {author} {\bibfnamefont {C.}~\bibnamefont
  {Nascimento}}\ and\ \bibinfo {author} {\bibfnamefont {M.}~\bibnamefont
  {Loverde}},\ }\href@noop {} {\  (\bibinfo {year} {2024})},\ \Eprint
  {http://arxiv.org/abs/2410.05389} {arXiv:2410.05389 [astro-ph.CO]}
  \BibitemShut {NoStop}%
\bibitem [{\citenamefont {Seljak}\ and\ \citenamefont
  {McDonald}(2011)}]{Seljak:2011tx}%
  \BibitemOpen
  \bibfield  {author} {\bibinfo {author} {\bibfnamefont {U.}~\bibnamefont
  {Seljak}}\ and\ \bibinfo {author} {\bibfnamefont {P.}~\bibnamefont
  {McDonald}},\ }\href {\doibase 10.1088/1475-7516/2011/11/039} {\bibfield
  {journal} {\bibinfo  {journal} {JCAP}\ }\textbf {\bibinfo {volume} {11}},\
  \bibinfo {pages} {039} (\bibinfo {year} {2011})},\ \Eprint
  {http://arxiv.org/abs/1109.1888} {arXiv:1109.1888 [astro-ph.CO]} \BibitemShut
  {NoStop}%
\bibitem [{\citenamefont {Okumura}\ \emph
  {et~al.}(2012{\natexlab{a}})\citenamefont {Okumura}, \citenamefont {Seljak},
  \citenamefont {McDonald},\ and\ \citenamefont {Desjacques}}]{Okumura:2011pb}%
  \BibitemOpen
  \bibfield  {author} {\bibinfo {author} {\bibfnamefont {T.}~\bibnamefont
  {Okumura}}, \bibinfo {author} {\bibfnamefont {U.}~\bibnamefont {Seljak}},
  \bibinfo {author} {\bibfnamefont {P.}~\bibnamefont {McDonald}}, \ and\
  \bibinfo {author} {\bibfnamefont {V.}~\bibnamefont {Desjacques}},\ }\href
  {\doibase 10.1088/1475-7516/2012/02/010} {\bibfield  {journal} {\bibinfo
  {journal} {JCAP}\ }\textbf {\bibinfo {volume} {02}},\ \bibinfo {pages} {010}
  (\bibinfo {year} {2012}{\natexlab{a}})},\ \Eprint
  {http://arxiv.org/abs/1109.1609} {arXiv:1109.1609 [astro-ph.CO]} \BibitemShut
  {NoStop}%
\bibitem [{\citenamefont {Okumura}\ \emph
  {et~al.}(2012{\natexlab{b}})\citenamefont {Okumura}, \citenamefont {Seljak},\
  and\ \citenamefont {Desjacques}}]{Okumura:2012xh}%
  \BibitemOpen
  \bibfield  {author} {\bibinfo {author} {\bibfnamefont {T.}~\bibnamefont
  {Okumura}}, \bibinfo {author} {\bibfnamefont {U.}~\bibnamefont {Seljak}}, \
  and\ \bibinfo {author} {\bibfnamefont {V.}~\bibnamefont {Desjacques}},\
  }\href {\doibase 10.1088/1475-7516/2012/11/014} {\bibfield  {journal}
  {\bibinfo  {journal} {JCAP}\ }\textbf {\bibinfo {volume} {11}},\ \bibinfo
  {pages} {014} (\bibinfo {year} {2012}{\natexlab{b}})},\ \Eprint
  {http://arxiv.org/abs/1206.4070} {arXiv:1206.4070 [astro-ph.CO]} \BibitemShut
  {NoStop}%
\bibitem [{\citenamefont {Vlah}\ \emph {et~al.}(2012)\citenamefont {Vlah},
  \citenamefont {Seljak}, \citenamefont {McDonald}, \citenamefont {Okumura},\
  and\ \citenamefont {Baldauf}}]{Vlah:2012ni}%
  \BibitemOpen
  \bibfield  {author} {\bibinfo {author} {\bibfnamefont {Z.}~\bibnamefont
  {Vlah}}, \bibinfo {author} {\bibfnamefont {U.}~\bibnamefont {Seljak}},
  \bibinfo {author} {\bibfnamefont {P.}~\bibnamefont {McDonald}}, \bibinfo
  {author} {\bibfnamefont {T.}~\bibnamefont {Okumura}}, \ and\ \bibinfo
  {author} {\bibfnamefont {T.}~\bibnamefont {Baldauf}},\ }\href {\doibase
  10.1088/1475-7516/2012/11/009} {\bibfield  {journal} {\bibinfo  {journal}
  {JCAP}\ }\textbf {\bibinfo {volume} {11}},\ \bibinfo {pages} {009} (\bibinfo
  {year} {2012})},\ \Eprint {http://arxiv.org/abs/1207.0839} {arXiv:1207.0839
  [astro-ph.CO]} \BibitemShut {NoStop}%
\bibitem [{\citenamefont {Vlah}\ \emph {et~al.}(2013)\citenamefont {Vlah},
  \citenamefont {Seljak}, \citenamefont {Okumura},\ and\ \citenamefont
  {Desjacques}}]{Vlah:2013lia}%
  \BibitemOpen
  \bibfield  {author} {\bibinfo {author} {\bibfnamefont {Z.}~\bibnamefont
  {Vlah}}, \bibinfo {author} {\bibfnamefont {U.}~\bibnamefont {Seljak}},
  \bibinfo {author} {\bibfnamefont {T.}~\bibnamefont {Okumura}}, \ and\
  \bibinfo {author} {\bibfnamefont {V.}~\bibnamefont {Desjacques}},\ }\href
  {\doibase 10.1088/1475-7516/2013/10/053} {\bibfield  {journal} {\bibinfo
  {journal} {JCAP}\ }\textbf {\bibinfo {volume} {10}},\ \bibinfo {pages} {053}
  (\bibinfo {year} {2013})},\ \Eprint {http://arxiv.org/abs/1308.6294}
  {arXiv:1308.6294 [astro-ph.CO]} \BibitemShut {NoStop}%
\bibitem [{\citenamefont {Vlah}\ \emph {et~al.}(2020)\citenamefont {Vlah},
  \citenamefont {Chisari},\ and\ \citenamefont {Schmidt}}]{Vlah:2019byq}%
  \BibitemOpen
  \bibfield  {author} {\bibinfo {author} {\bibfnamefont {Z.}~\bibnamefont
  {Vlah}}, \bibinfo {author} {\bibfnamefont {N.~E.}\ \bibnamefont {Chisari}}, \
  and\ \bibinfo {author} {\bibfnamefont {F.}~\bibnamefont {Schmidt}},\ }\href
  {\doibase 10.1088/1475-7516/2020/01/025} {\bibfield  {journal} {\bibinfo
  {journal} {JCAP}\ }\textbf {\bibinfo {volume} {01}},\ \bibinfo {pages} {025}
  (\bibinfo {year} {2020})},\ \Eprint {http://arxiv.org/abs/1910.08085}
  {arXiv:1910.08085 [astro-ph.CO]} \BibitemShut {NoStop}%
\bibitem [{\citenamefont {Vlah}\ \emph {et~al.}(2021)\citenamefont {Vlah},
  \citenamefont {Chisari},\ and\ \citenamefont {Schmidt}}]{Vlah:2020ovg}%
  \BibitemOpen
  \bibfield  {author} {\bibinfo {author} {\bibfnamefont {Z.}~\bibnamefont
  {Vlah}}, \bibinfo {author} {\bibfnamefont {N.~E.}\ \bibnamefont {Chisari}}, \
  and\ \bibinfo {author} {\bibfnamefont {F.}~\bibnamefont {Schmidt}},\ }\href
  {\doibase 10.1088/1475-7516/2021/05/061} {\bibfield  {journal} {\bibinfo
  {journal} {JCAP}\ }\textbf {\bibinfo {volume} {05}},\ \bibinfo {pages} {061}
  (\bibinfo {year} {2021})},\ \Eprint {http://arxiv.org/abs/2012.04114}
  {arXiv:2012.04114 [astro-ph.CO]} \BibitemShut {NoStop}%
\bibitem [{\citenamefont {Matsubara}(2024)}]{Matsubara:2022ohx}%
  \BibitemOpen
  \bibfield  {author} {\bibinfo {author} {\bibfnamefont {T.}~\bibnamefont
  {Matsubara}},\ }\href {\doibase 10.1103/PhysRevD.110.063543} {\bibfield
  {journal} {\bibinfo  {journal} {Phys. Rev. D}\ }\textbf {\bibinfo {volume}
  {110}},\ \bibinfo {pages} {063543} (\bibinfo {year} {2024})},\ \Eprint
  {http://arxiv.org/abs/2210.10435} {arXiv:2210.10435 [astro-ph.CO]}
  \BibitemShut {NoStop}%
\bibitem [{\citenamefont {Bakx}\ \emph {et~al.}(2023)\citenamefont {Bakx},
  \citenamefont {Kurita}, \citenamefont {Chisari}, \citenamefont {Vlah},\ and\
  \citenamefont {Schmidt}}]{Bakx:2023mld}%
  \BibitemOpen
  \bibfield  {author} {\bibinfo {author} {\bibfnamefont {T.}~\bibnamefont
  {Bakx}}, \bibinfo {author} {\bibfnamefont {T.}~\bibnamefont {Kurita}},
  \bibinfo {author} {\bibfnamefont {N.~E.}\ \bibnamefont {Chisari}}, \bibinfo
  {author} {\bibfnamefont {Z.}~\bibnamefont {Vlah}}, \ and\ \bibinfo {author}
  {\bibfnamefont {F.}~\bibnamefont {Schmidt}},\ }\href {\doibase
  10.1088/1475-7516/2023/10/005} {\bibfield  {journal} {\bibinfo  {journal}
  {JCAP}\ }\textbf {\bibinfo {volume} {10}},\ \bibinfo {pages} {005} (\bibinfo
  {year} {2023})},\ \Eprint {http://arxiv.org/abs/2303.15565} {arXiv:2303.15565
  [astro-ph.CO]} \BibitemShut {NoStop}%
\bibitem [{\citenamefont {Slepian}\ and\ \citenamefont
  {Eisenstein}(2015)}]{Slepian:2015qza}%
  \BibitemOpen
  \bibfield  {author} {\bibinfo {author} {\bibfnamefont {Z.}~\bibnamefont
  {Slepian}}\ and\ \bibinfo {author} {\bibfnamefont {D.~J.}\ \bibnamefont
  {Eisenstein}},\ }\href {\doibase 10.1093/mnras/stv2119} {\bibfield  {journal}
  {\bibinfo  {journal} {Mon. Not. Roy. Astron. Soc.}\ }\textbf {\bibinfo
  {volume} {454}},\ \bibinfo {pages} {4142} (\bibinfo {year} {2015})},\ \Eprint
  {http://arxiv.org/abs/1506.02040} {arXiv:1506.02040 [astro-ph.CO]}
  \BibitemShut {NoStop}%
\bibitem [{\citenamefont {Ma}\ and\ \citenamefont
  {Bertschinger}(1995)}]{Ma:1995ey}%
  \BibitemOpen
  \bibfield  {author} {\bibinfo {author} {\bibfnamefont {C.-P.}\ \bibnamefont
  {Ma}}\ and\ \bibinfo {author} {\bibfnamefont {E.}~\bibnamefont
  {Bertschinger}},\ }\href {\doibase 10.1086/176550} {\bibfield  {journal}
  {\bibinfo  {journal} {Astrophys. J.}\ }\textbf {\bibinfo {volume} {455}},\
  \bibinfo {pages} {7} (\bibinfo {year} {1995})},\ \Eprint
  {http://arxiv.org/abs/astro-ph/9506072} {arXiv:astro-ph/9506072} \BibitemShut
  {NoStop}%
\bibitem [{\citenamefont {{Scoccimarro}}\ and\ \citenamefont
  {{Frieman}}(1996)}]{ScoFri9607}%
  \BibitemOpen
  \bibfield  {author} {\bibinfo {author} {\bibfnamefont {R.}~\bibnamefont
  {{Scoccimarro}}}\ and\ \bibinfo {author} {\bibfnamefont {J.}~\bibnamefont
  {{Frieman}}},\ }\href {\doibase 10.1086/192306} {\bibfield  {journal}
  {\bibinfo  {journal} {\apjs}\ }\textbf {\bibinfo {volume} {105}},\ \bibinfo
  {pages} {37} (\bibinfo {year} {1996})},\ \Eprint
  {http://arxiv.org/abs/arXiv:astro-ph/9509047} {arXiv:astro-ph/9509047}
  \BibitemShut {NoStop}%
\bibitem [{\citenamefont {Blas}\ \emph {et~al.}(2013)\citenamefont {Blas},
  \citenamefont {Garny},\ and\ \citenamefont {Konstandin}}]{Blas:2013bpa}%
  \BibitemOpen
  \bibfield  {author} {\bibinfo {author} {\bibfnamefont {D.}~\bibnamefont
  {Blas}}, \bibinfo {author} {\bibfnamefont {M.}~\bibnamefont {Garny}}, \ and\
  \bibinfo {author} {\bibfnamefont {T.}~\bibnamefont {Konstandin}},\ }\href
  {\doibase 10.1088/1475-7516/2013/09/024} {\bibfield  {journal} {\bibinfo
  {journal} {JCAP}\ }\textbf {\bibinfo {volume} {09}},\ \bibinfo {pages} {024}
  (\bibinfo {year} {2013})},\ \Eprint {http://arxiv.org/abs/1304.1546}
  {arXiv:1304.1546 [astro-ph.CO]} \BibitemShut {NoStop}%
\bibitem [{\citenamefont {{Pichon}}\ and\ \citenamefont
  {{Bernardeau}}(1999)}]{PicBer9903}%
  \BibitemOpen
  \bibfield  {author} {\bibinfo {author} {\bibfnamefont {C.}~\bibnamefont
  {{Pichon}}}\ and\ \bibinfo {author} {\bibfnamefont {F.}~\bibnamefont
  {{Bernardeau}}},\ }\href@noop {} {\bibfield  {journal} {\bibinfo  {journal}
  {\aap}\ }\textbf {\bibinfo {volume} {343}},\ \bibinfo {pages} {663} (\bibinfo
  {year} {1999})},\ \Eprint {http://arxiv.org/abs/astro-ph/9902142}
  {arXiv:astro-ph/9902142 [astro-ph]} \BibitemShut {NoStop}%
\bibitem [{\citenamefont {Jelic-Cizmek}\ \emph {et~al.}(2018)\citenamefont
  {Jelic-Cizmek}, \citenamefont {Lepori}, \citenamefont {Adamek},\ and\
  \citenamefont {Durrer}}]{Jelic-Cizmek:2018gdp}%
  \BibitemOpen
  \bibfield  {author} {\bibinfo {author} {\bibfnamefont {G.}~\bibnamefont
  {Jelic-Cizmek}}, \bibinfo {author} {\bibfnamefont {F.}~\bibnamefont
  {Lepori}}, \bibinfo {author} {\bibfnamefont {J.}~\bibnamefont {Adamek}}, \
  and\ \bibinfo {author} {\bibfnamefont {R.}~\bibnamefont {Durrer}},\ }\href
  {\doibase 10.1088/1475-7516/2018/09/006} {\bibfield  {journal} {\bibinfo
  {journal} {JCAP}\ }\textbf {\bibinfo {volume} {09}},\ \bibinfo {pages} {006}
  (\bibinfo {year} {2018})},\ \Eprint {http://arxiv.org/abs/1806.05146}
  {arXiv:1806.05146 [astro-ph.CO]} \BibitemShut {NoStop}%
\bibitem [{\citenamefont {Pajer}\ and\ \citenamefont
  {Zaldarriaga}(2013)}]{Pajer:2013jj}%
  \BibitemOpen
  \bibfield  {author} {\bibinfo {author} {\bibfnamefont {E.}~\bibnamefont
  {Pajer}}\ and\ \bibinfo {author} {\bibfnamefont {M.}~\bibnamefont
  {Zaldarriaga}},\ }\href {\doibase 10.1088/1475-7516/2013/08/037} {\bibfield
  {journal} {\bibinfo  {journal} {JCAP}\ }\textbf {\bibinfo {volume} {08}},\
  \bibinfo {pages} {037} (\bibinfo {year} {2013})},\ \Eprint
  {http://arxiv.org/abs/1301.7182} {arXiv:1301.7182 [astro-ph.CO]} \BibitemShut
  {NoStop}%
\bibitem [{\citenamefont {{Mercolli}}\ and\ \citenamefont
  {{Pajer}}(2014)}]{MerPaj1403}%
  \BibitemOpen
  \bibfield  {author} {\bibinfo {author} {\bibfnamefont {L.}~\bibnamefont
  {{Mercolli}}}\ and\ \bibinfo {author} {\bibfnamefont {E.}~\bibnamefont
  {{Pajer}}},\ }\href {\doibase 10.1088/1475-7516/2014/03/006} {\bibfield
  {journal} {\bibinfo  {journal} {\jcap}\ }\textbf {\bibinfo {volume} {2014}},\
  \bibinfo {eid} {006} (\bibinfo {year} {2014})},\ \Eprint
  {http://arxiv.org/abs/1307.3220} {arXiv:1307.3220 [astro-ph.CO]} \BibitemShut
  {NoStop}%
\bibitem [{\citenamefont {{Mirbabayi}}\ \emph {et~al.}(2015)\citenamefont
  {{Mirbabayi}}, \citenamefont {{Schmidt}},\ and\ \citenamefont
  {{Zaldarriaga}}}]{Mirbabayi:2015}%
  \BibitemOpen
  \bibfield  {author} {\bibinfo {author} {\bibfnamefont {M.}~\bibnamefont
  {{Mirbabayi}}}, \bibinfo {author} {\bibfnamefont {F.}~\bibnamefont
  {{Schmidt}}}, \ and\ \bibinfo {author} {\bibfnamefont {M.}~\bibnamefont
  {{Zaldarriaga}}},\ }\href {\doibase 10.1088/1475-7516/2015/07/030} {\bibfield
   {journal} {\bibinfo  {journal} {\jcap}\ }\textbf {\bibinfo {volume} {7}},\
  \bibinfo {eid} {030} (\bibinfo {year} {2015})},\ \Eprint
  {http://arxiv.org/abs/1412.5169} {arXiv:1412.5169} \BibitemShut {NoStop}%
\bibitem [{\citenamefont {{Akbar Abolhasani}}\ \emph
  {et~al.}(2016)\citenamefont {{Akbar Abolhasani}}, \citenamefont
  {{Mirbabayi}},\ and\ \citenamefont {{Pajer}}}]{2016JCAP...05..063A}%
  \BibitemOpen
  \bibfield  {author} {\bibinfo {author} {\bibfnamefont {A.}~\bibnamefont
  {{Akbar Abolhasani}}}, \bibinfo {author} {\bibfnamefont {M.}~\bibnamefont
  {{Mirbabayi}}}, \ and\ \bibinfo {author} {\bibfnamefont {E.}~\bibnamefont
  {{Pajer}}},\ }\href {\doibase 10.1088/1475-7516/2016/05/063} {\bibfield
  {journal} {\bibinfo  {journal} {\jcap}\ }\textbf {\bibinfo {volume} {2016}},\
  \bibinfo {eid} {063} (\bibinfo {year} {2016})},\ \Eprint
  {http://arxiv.org/abs/1509.07886} {arXiv:1509.07886 [hep-th]} \BibitemShut
  {NoStop}%
\bibitem [{\citenamefont {Baldauf}\ \emph {et~al.}(2021)\citenamefont
  {Baldauf}, \citenamefont {Garny}, \citenamefont {Taule},\ and\ \citenamefont
  {Steele}}]{Baldauf:2021zlt}%
  \BibitemOpen
  \bibfield  {author} {\bibinfo {author} {\bibfnamefont {T.}~\bibnamefont
  {Baldauf}}, \bibinfo {author} {\bibfnamefont {M.}~\bibnamefont {Garny}},
  \bibinfo {author} {\bibfnamefont {P.}~\bibnamefont {Taule}}, \ and\ \bibinfo
  {author} {\bibfnamefont {T.}~\bibnamefont {Steele}},\ }\href {\doibase
  10.1103/PhysRevD.104.123551} {\bibfield  {journal} {\bibinfo  {journal}
  {Phys. Rev. D}\ }\textbf {\bibinfo {volume} {104}},\ \bibinfo {pages}
  {123551} (\bibinfo {year} {2021})},\ \Eprint
  {http://arxiv.org/abs/2110.13930} {arXiv:2110.13930 [astro-ph.CO]}
  \BibitemShut {NoStop}%
\bibitem [{\citenamefont {{Desjacques}}\ \emph {et~al.}(2018)\citenamefont
  {{Desjacques}}, \citenamefont {{Jeong}},\ and\ \citenamefont
  {{Schmidt}}}]{Desjacques:2018}%
  \BibitemOpen
  \bibfield  {author} {\bibinfo {author} {\bibfnamefont {V.}~\bibnamefont
  {{Desjacques}}}, \bibinfo {author} {\bibfnamefont {D.}~\bibnamefont
  {{Jeong}}}, \ and\ \bibinfo {author} {\bibfnamefont {F.}~\bibnamefont
  {{Schmidt}}},\ }\href {\doibase 10.1016/j.physrep.2017.12.002} {\bibfield
  {journal} {\bibinfo  {journal} {\physrep}\ }\textbf {\bibinfo {volume}
  {733}},\ \bibinfo {pages} {1} (\bibinfo {year} {2018})}\BibitemShut {NoStop}%
\bibitem [{\citenamefont {Bauer}\ \emph {et~al.}(2001)\citenamefont {Bauer},
  \citenamefont {Fleming}, \citenamefont {Pirjol},\ and\ \citenamefont
  {Stewart}}]{Bauer:2000yr}%
  \BibitemOpen
  \bibfield  {author} {\bibinfo {author} {\bibfnamefont {C.~W.}\ \bibnamefont
  {Bauer}}, \bibinfo {author} {\bibfnamefont {S.}~\bibnamefont {Fleming}},
  \bibinfo {author} {\bibfnamefont {D.}~\bibnamefont {Pirjol}}, \ and\ \bibinfo
  {author} {\bibfnamefont {I.~W.}\ \bibnamefont {Stewart}},\ }\href {\doibase
  10.1103/PhysRevD.63.114020} {\bibfield  {journal} {\bibinfo  {journal} {Phys.
  Rev. D}\ }\textbf {\bibinfo {volume} {63}},\ \bibinfo {pages} {114020}
  (\bibinfo {year} {2001})},\ \Eprint {http://arxiv.org/abs/hep-ph/0011336}
  {arXiv:hep-ph/0011336} \BibitemShut {NoStop}%
\bibitem [{\citenamefont {Beneke}\ \emph {et~al.}(2002)\citenamefont {Beneke},
  \citenamefont {Chapovsky}, \citenamefont {Diehl},\ and\ \citenamefont
  {Feldmann}}]{Beneke:2002ph}%
  \BibitemOpen
  \bibfield  {author} {\bibinfo {author} {\bibfnamefont {M.}~\bibnamefont
  {Beneke}}, \bibinfo {author} {\bibfnamefont {A.~P.}\ \bibnamefont
  {Chapovsky}}, \bibinfo {author} {\bibfnamefont {M.}~\bibnamefont {Diehl}}, \
  and\ \bibinfo {author} {\bibfnamefont {T.}~\bibnamefont {Feldmann}},\ }\href
  {\doibase 10.1016/S0550-3213(02)00687-9} {\bibfield  {journal} {\bibinfo
  {journal} {Nucl. Phys. B}\ }\textbf {\bibinfo {volume} {643}},\ \bibinfo
  {pages} {431} (\bibinfo {year} {2002})},\ \Eprint
  {http://arxiv.org/abs/hep-ph/0206152} {arXiv:hep-ph/0206152} \BibitemShut
  {NoStop}%
\bibitem [{\citenamefont {Beneke}\ \emph {et~al.}(2018)\citenamefont {Beneke},
  \citenamefont {Garny}, \citenamefont {Szafron},\ and\ \citenamefont
  {Wang}}]{Beneke:2017ztn}%
  \BibitemOpen
  \bibfield  {author} {\bibinfo {author} {\bibfnamefont {M.}~\bibnamefont
  {Beneke}}, \bibinfo {author} {\bibfnamefont {M.}~\bibnamefont {Garny}},
  \bibinfo {author} {\bibfnamefont {R.}~\bibnamefont {Szafron}}, \ and\
  \bibinfo {author} {\bibfnamefont {J.}~\bibnamefont {Wang}},\ }\href {\doibase
  10.1007/JHEP03(2018)001} {\bibfield  {journal} {\bibinfo  {journal} {JHEP}\
  }\textbf {\bibinfo {volume} {03}},\ \bibinfo {pages} {001} (\bibinfo {year}
  {2018})},\ \Eprint {http://arxiv.org/abs/1712.04416} {arXiv:1712.04416
  [hep-ph]} \BibitemShut {NoStop}%
\bibitem [{\citenamefont {{Carrasco}}\ \emph {et~al.}(2012)\citenamefont
  {{Carrasco}}, \citenamefont {{Hertzberg}},\ and\ \citenamefont
  {{Senatore}}}]{Carrasco:2012}%
  \BibitemOpen
  \bibfield  {author} {\bibinfo {author} {\bibfnamefont {J.~J.~M.}\
  \bibnamefont {{Carrasco}}}, \bibinfo {author} {\bibfnamefont {M.~P.}\
  \bibnamefont {{Hertzberg}}}, \ and\ \bibinfo {author} {\bibfnamefont
  {L.}~\bibnamefont {{Senatore}}},\ }\href {\doibase 10.1007/JHEP09(2012)082}
  {\bibfield  {journal} {\bibinfo  {journal} {Journal of High Energy Physics}\
  }\textbf {\bibinfo {volume} {2012}},\ \bibinfo {eid} {82} (\bibinfo {year}
  {2012})}\BibitemShut {NoStop}%
\bibitem [{\citenamefont {{Baldauf}}\ \emph {et~al.}(2015)\citenamefont
  {{Baldauf}}, \citenamefont {{Mercolli}}, \citenamefont {{Mirbabayi}},\ and\
  \citenamefont {{Pajer}}}]{BalMerMir1505}%
  \BibitemOpen
  \bibfield  {author} {\bibinfo {author} {\bibfnamefont {T.}~\bibnamefont
  {{Baldauf}}}, \bibinfo {author} {\bibfnamefont {L.}~\bibnamefont
  {{Mercolli}}}, \bibinfo {author} {\bibfnamefont {M.}~\bibnamefont
  {{Mirbabayi}}}, \ and\ \bibinfo {author} {\bibfnamefont {E.}~\bibnamefont
  {{Pajer}}},\ }\href {\doibase 10.1088/1475-7516/2015/05/007} {\bibfield
  {journal} {\bibinfo  {journal} {\jcap}\ }\textbf {\bibinfo {volume} {2015}},\
  \bibinfo {eid} {007} (\bibinfo {year} {2015})},\ \Eprint
  {http://arxiv.org/abs/1406.4135} {arXiv:1406.4135 [astro-ph.CO]} \BibitemShut
  {NoStop}%
\bibitem [{\citenamefont {{Angulo}}\ \emph {et~al.}(2015)\citenamefont
  {{Angulo}}, \citenamefont {{Foreman}}, \citenamefont {{Schmittfull}},\ and\
  \citenamefont {{Senatore}}}]{AngForSch1510}%
  \BibitemOpen
  \bibfield  {author} {\bibinfo {author} {\bibfnamefont {R.~E.}\ \bibnamefont
  {{Angulo}}}, \bibinfo {author} {\bibfnamefont {S.}~\bibnamefont {{Foreman}}},
  \bibinfo {author} {\bibfnamefont {M.}~\bibnamefont {{Schmittfull}}}, \ and\
  \bibinfo {author} {\bibfnamefont {L.}~\bibnamefont {{Senatore}}},\ }\href
  {\doibase 10.1088/1475-7516/2015/10/039} {\bibfield  {journal} {\bibinfo
  {journal} {Journal of Cosmology and Astro-Particle Physics}\ }\textbf
  {\bibinfo {volume} {2015}},\ \bibinfo {eid} {039} (\bibinfo {year} {2015})},\
  \Eprint {http://arxiv.org/abs/1406.4143} {arXiv:1406.4143 [astro-ph.CO]}
  \BibitemShut {NoStop}%
\bibitem [{\citenamefont {Scoccimarro}\ and\ \citenamefont
  {Frieman}(1996)}]{Scoccimarro:1996se}%
  \BibitemOpen
  \bibfield  {author} {\bibinfo {author} {\bibfnamefont {R.}~\bibnamefont
  {Scoccimarro}}\ and\ \bibinfo {author} {\bibfnamefont {J.}~\bibnamefont
  {Frieman}},\ }\href {\doibase 10.1086/178177} {\bibfield  {journal} {\bibinfo
   {journal} {Astrophys. J.}\ }\textbf {\bibinfo {volume} {473}},\ \bibinfo
  {pages} {620} (\bibinfo {year} {1996})},\ \Eprint
  {http://arxiv.org/abs/astro-ph/9602070} {arXiv:astro-ph/9602070} \BibitemShut
  {NoStop}%
\bibitem [{\citenamefont {Peacock}\ and\ \citenamefont
  {Dodds}(1996)}]{Peacock:1996ci}%
  \BibitemOpen
  \bibfield  {author} {\bibinfo {author} {\bibfnamefont {J.~A.}\ \bibnamefont
  {Peacock}}\ and\ \bibinfo {author} {\bibfnamefont {S.~J.}\ \bibnamefont
  {Dodds}},\ }\href {\doibase 10.1093/mnras/280.3.L19} {\bibfield  {journal}
  {\bibinfo  {journal} {Mon. Not. Roy. Astron. Soc.}\ }\textbf {\bibinfo
  {volume} {280}},\ \bibinfo {pages} {L19} (\bibinfo {year} {1996})},\ \Eprint
  {http://arxiv.org/abs/astro-ph/9603031} {arXiv:astro-ph/9603031} \BibitemShut
  {NoStop}%
\end{thebibliography}
%

\end{document}